\documentclass[nofootinbib,aps,amssymb,floatfix,superscriptaddress,preprintnumbers]{revtex4}

\usepackage{graphicx}
\usepackage{bm}
\usepackage{amsmath}
\usepackage{amssymb}
\usepackage{amsfonts}
\usepackage{float}
\usepackage{hyperref}
\usepackage{dsfont}  
\usepackage{slashed}  
\usepackage{booktabs}
\usepackage{multirow}%https://www.overleaf.com/project/5f75d0158ffdb40001394bbb
\usepackage{subfigure}
\usepackage[sort&compress]{natbib}
\usepackage{xcolor}
\usepackage{ulem}

\newcommand{\be}{\begin{equation}}  
\newcommand{\ee}{\end{equation}}  
\newcommand{\beq}{\begin{eqnarray}} 
\newcommand{\eeq}{\end{eqnarray}}

\newcommand{\bea}{\begin{eqnarray}}
\newcommand{\eea}{\end{eqnarray}}

\newcommand{\MSb}{{\overline{\rm MS}}}
\newcommand{\MMSb}{{\rm{M}\overline{\rm MS}}}

\begin{document}

\title{Parton distribution functions beyond leading twist from lattice QCD: \\ The $h_L(x)$ case}

\author{Shohini Bhattacharya}
\affiliation{Department of Physics,  Temple University,  Philadelphia,  PA 19122 - 1801,  USA}
\author{Krzysztof Cichy}
\affiliation{Faculty of Physics, Adam Mickiewicz University, ul.\ Uniwersytetu Pozna\'nskiego 2, 61-614 Pozna\'{n}, Poland}
\author{Martha Constantinou}
\affiliation{Department of Physics,  Temple University,  Philadelphia,  PA 19122 - 1801,  USA}
\author{Andreas Metz}
\affiliation{Department of Physics,  Temple University,  Philadelphia,  PA 19122 - 1801,  USA}
\author{Aurora Scapellato}
\affiliation{Department of Physics,  Temple University,  Philadelphia,  PA 19122 - 1801,  USA}
\author{Fernanda Steffens}
\affiliation{Institut f\"ur Strahlen- und Kernphysik, Rheinische
  Friedrich-Wilhelms-Universit\"at Bonn, Nussallee 14-16, 53115 Bonn}

\newcommand{\mcc}[1]{{\color{magenta} {#1}}}
\newcommand{\mc}[1]{{\color{magenta}(\textbf{MC}: {#1})}}
\newcommand{\am}[1]{{\color{cyan} {#1}}}
\newcommand{\ams}[2]{\textcolor{cyan}{\sout{#1} #2}}
\newcommand{\Shohini}[1]{{\color{red}\textbf{SB}: {#1}}}

\begin{abstract}
We report the first-ever calculation of the isovector flavor combination of the chiral-odd twist-3 parton distribution $h_L(x)$ for the proton from lattice QCD. 
We employ gauge configurations with two degenerate light, a strange and a charm quark ($N_f=2+1+1$) of maximally twisted mass fermions with a clover improvement. 
The lattice has a spatial extent of 3~fm and lattice spacing of 0.093~fm. 
The values of the quark masses lead to a pion mass of $260$~MeV.  
We use a source-sink time separation of 1.12~fm to control contamination from excited states. 
Our calculation is based on the quasi-distribution approach, with three values for the proton momentum: 0.83~GeV, 1.25~GeV, and 1.67~GeV. 
The lattice data are renormalized non-perturbatively using the RI$'$ scheme, and the final result for $h_L(x)$ is presented in the $\MSb$ scheme at the scale of 2~GeV. 
Furthermore, we compute in the same setup the transversity distribution, $h_1(x)$, which allows us, in particular, to compare $h_L(x)$ to its Wandzura-Wilczek approximation. We also combine results for the isovector and isoscalar flavor combinations to disentangle the individual quark contributions for $h_1(x)$ and $h_L(x)$, and address the Wandzura-Wilczek approximation in that case as well.
\end{abstract}

\maketitle

%%%%%%%%%%%%%%%%%%%%%%%%%%%%%%%%%%%%%%%%%%%%%%%%%%%%%%%%%%%%%%%%%%
\section{Introduction}
\label{sec:introduction}
%%%%%%%%%%%%%%%%%%%%%%%%%%%%%%%%%%%%%%%%%%%%%%%%%%%%%%%%%%%%%%%%%%
Parton distribution functions (PDFs) are the most important quantities characterizing the structure of strongly interacting systems such as the nucleon in terms of quarks and gluons, the elementary degrees of freedom of quantum chromodynamics (QCD)~\cite{Collins:1981uw, Collins:2011zzd}. QCD factorization theorems allow one to extract (the non-perturbative) PDFs from cross section measurements for high-energy processes~\cite{Collins:1989gx, Collins:2011zzd}. 
An important property of PDFs is their twist, which specifies the order in $1/Q$ at which they enter a factorization formula for a given observable, where $Q$ indicates the large scale of the process~\cite{Jaffe:1996zw}. 
The leading-order PDFs, also denoted as twist-2 PDFs, can be considered probability densities for finding a parton with momentum fraction $x$ inside a hadron.
Twist-2 PDFs have been studied for decades, and in the meantime the community has gathered a wealth of information about those quantities.
In contrast, twist-3 PDFs are presently poorly known.
At the same time, for a number of reasons they are important too.
First, they are typically as large as twist-2 PDFs.
Second, they contain information about quark-gluon correlations inside hadrons~\cite{Balitsky:1987bk,Kanazawa:2015ajw} and as such characterize the parton structure of hadrons in new ways.
Third, twist-3 PDFs appear in QCD factorization theorems for various observables, where arguably the best known example is the structure function $g_2^{\rm s.f.}$ in inclusive deep-inelastic scattering (DIS).
Forth, certain twist-3 PDFs are related to the transverse-momentum-dependent PDFs~\cite{Boer:2003cm, Accardi:2009au, Gamberg:2017jha, Cammarota:2020qcw}, which are important for understanding the three-dimensional structure of hadrons. 
Fifth, some twist-3 PDFs have a semi-classical relation to the average force experienced by partons inside the nucleon~\cite{Burkardt:2008ps}.  

For a spin-$\frac{1}{2}$ hadron, three (collinear) twist-3 quark PDFs can be identified, which are defined through quark-antiquark matrix elements: $g_T(x)$, $e(x)$, $h_L(x)$~\cite{Jaffe:1991kp, Jaffe:1991ra}. 
The PDF $g_T(x)$ can be measured through the aforementioned DIS structure function $g_2^{\rm s.f.}$; see Refs.~\cite{Flay:2016wie, Armstrong:2018xgk} for recent related experiments.
On the other hand, since both $e(x)$ and $h_L(x)$ are chiral-odd and therefore decouple from the ``simple'' DIS process, there exists hardly any experimental information about these quantities.
In fact, the function $h_L(x)$, on which we concentrate in the present work and which, for instance, could be measured through the double-polarized Drell-Yan process~\cite{Jaffe:1991kp, Jaffe:1991ra, Koike:2008du} or single-inclusive particle production in proton-proton collisions~\cite{Koike:2016ura}, has never been addressed in an experiment.
Also model calculations for $h_L(x)$ of the nucleon are sparse~\cite{Jakob:1997wg, Bastami:2020rxn}.
Generally, it is fair to say that $h_{L}(x)$ is the most elusive of the three twist-3 PDF.

Here we present the first-ever lattice-QCD calculation of the isovector flavor combination $h_L^{u - d}(x)$ for the proton, which represents an extension of our previous study of $g_T^{u - d}(x)$~\cite{Bhattacharya:2020cen}. 
To this end, we employ the so-called quasi-PDF approach suggested by X.~Ji~\cite{Ji:2013dva,Ji:2014gla}. 
While standard (light-cone) PDFs are given by light-cone correlation functions, quasi-PDFs and related quantities~\cite{Braun:2007wv, Radyushkin:2017cyf, Ma:2017pxb} are defined through spatial correlation functions accessible in lattice QCD.
Recent years have seen a surge of studies of (spatial) Euclidean correlators which provide access to the $x$-dependent parton structure of hadrons; see, e.g., Refs.~\cite{Lin:2014zya,Alexandrou:2015rja,Chen:2016utp,Alexandrou:2016jqi,Chambers:2017dov,Alexandrou:2017huk,Orginos:2017kos,Ishikawa:2017faj,Ji:2017oey,Radyushkin:2018cvn,Alexandrou:2018pbm,Chen:2018fwa,Alexandrou:2018eet,Liu:2018uuj,Karpie:2018zaz,Zhang:2018diq,Bhattacharya:2018zxi,Li:2018tpe,Sufian:2019bol,Karpie:2019eiq,Alexandrou:2019lfo,Izubuchi:2019lyk,Cichy:2019ebf,Joo:2019jct,Radyushkin:2019owq,Joo:2019bzr,Chai:2020nxw,Ji:2020baz,Braun:2020ymy,Bhat:2020ktg,Alexandrou:2020zbe,Alexandrou:2020uyt,Bringewatt:2020ixn,Liu:2020rqi,DelDebbio:2020rgv,Alexandrou:2020qtt,Liu:2020krc,Huo:2021rpe,Detmold:2021uru,Karpie:2021pap,Alexandrou:2021oih} and the recent reviews in Refs.~\cite{Cichy:2018mum,Ji:2020ect,Constantinou:2020pek}. 
Because quasi-PDFs and light-cone PDFs share the same infrared (non-perturbative) physics~\cite{Ji:2013dva,Ji:2014gla, Briceno:2017cpo}, they can be related via a matching procedure in perturbative QCD~\cite{Ji:2013dva, Xiong:2013bka, Ma:2014jla, Radyushkin:2017cyf, Wang:2017qyg, Stewart:2017tvs, Izubuchi:2018srq, Balitsky:2019krf, Bhattacharya:2020xlt, Bhattacharya:2020jfj, Li:2020xml, Chen:2020ody, Braun:2021aon}. 
In the present work we will use the one-loop matching result derived in Ref.~\cite{Bhattacharya:2020jfj}.

The paper is organized as follows:
In Sec.~\ref{sec:lattice_setup}, we present some details of the lattice setup and show our results for the relevant matrix elements in position space. In Sec.~\ref{sec:renormalization}, we discuss the key ingredients that are needed for the renormalization of the lattice data, while Sec.~\ref{sec:reco} contains information on how we obtain the $x$-dependent results.
In Sec.~\ref{sec:matching}, we recall the matching kernel in the $\MSb$ scheme from Ref.~\cite{Bhattacharya:2020jfj} and transform that kernel to the so-called modified $\MSb$ (M$\MSb$) scheme~\cite{Alexandrou:2019lfo}.
A scheme change of this type is needed in order to avoid a divergence in the light-cone PDF.
We also highlight and discuss a specific term in the matching kernel which has its origin in singular zero-mode contributions in the twist-3 light-cone and quasi-PDFs.
Generally, such contributions proportional to $\delta(x)$ can arise in model-independent analyses and in model calculations of twist-3 PDFs~\cite{Burkardt:1995ts, Burkardt:2001iy, Efremov:2002qh, Wakamatsu:2003uu, Pasquini:2018oyz, Aslan:2018zzk, Aslan:2018tff, Bhattacharya:2020xlt, Bhattacharya:2020jfj, Bhattacharya:2021boh}.
We present the main numerical results for the light-cone PDF $h_L(x)$ in Sec.~\ref{sec:numerics}.
This includes a discussion of the numerical impact of the zero-mode term in the matching kernel and, in particular, a study of the so-called Wandzura-Wilczek approximation for $h_L(x)$~\cite{Wandzura:1977qf, Jaffe:1991ra}.
In this approximation, $h_L(x)$ is entirely determined through the twist-2 transversity PDF $h_1(x)$~\cite{Ralston:1979ys}, which we have computed as well in the same lattice setup. 
In Sec.~\ref{sec:up_down_PDFs}, we discuss the individual quark contributions by combining the isovector ($u-d$) and isoscalar ($u+d$) flavor combinations. 
We summarize the findings of our work in Sec.~\ref{sec:summary}.

%%%%%%%%%%%%%%%%%%%%%%%%%%%%%%%%%%%%%%%%%%%%%%%%%%%%%%%%%%%%%%%%%%
\section{Lattice setup}
\label{sec:lattice_setup}
%%%%%%%%%%%%%%%%%%%%%%%%%%%%%%%%%%%%%%%%%%%%%%%%%%%%%%%%%%%%%%%%%%
We use one $N_f=2+1+1$ ensemble of two dynamical degenerate light quarks, reproducing a pion mass of 260 MeV and a dynamical strange and charm quark with masses near to the physical ones. The gauge configurations were generated by the ETM collaboration (ETMC)~\cite{Alexandrou:2021gqw}, using the Iwasaki improved gauge action~\cite{Iwasaki:1985we} and Wilson fermions at maximal twist with clover improvement~\cite{Sheikholeslami:1985ij}. The clover parameter is denoted by $c_{\rm SW}$. The lattice spacing is $a\simeq 0.093$ fm and the lattice volume is $32^3\times 64$ ($L\approx 3$ fm). The parameters of the ensemble are given in Table~\ref{tab:ensemble}.
\begin{table}[h!]
\begin{center}
\renewcommand{\arraystretch}{1.4}
\renewcommand{\tabcolsep}{6pt}
\begin{tabular}{c|c c c c c c }
\hline
Name & $\beta$ & $N_f$ & $L^3\times L_T$ & $a$ [fm] & $M_\pi$ & $m_\pi L$ \\
\hline 
cA211.32 & $1.726$ & $u,d,s,c$ & $32^3\times 64$  & 0.093 & 260 MeV & 4 \\
\hline
\end{tabular}
\begin{minipage}{15cm}
\caption{Parameters of the ensemble used in this work: $\beta$ is the bare coupling and $L,L_T$ are the size of the lattice along the spatial and temporal directions.}
\label{tab:ensemble}
\end{minipage}
\end{center}
\end{table}

The proton isovector $h_L(x)$ distributions are extracted through the quasi-PDF formalism, that involves the calculation of the following non-local matrix elements:
\begin{equation}
{\mathcal M}_{h_L}(z,P)\,=\,\langle P \,\vert\, \overline{\psi}(0,\vec{0})\,\sigma_{jk}\tau_3 W(0,\vec{z})\,\psi(0,\vec{z})\,\vert P\rangle\,,
    %{\mathcal M}_{h_L}(P,z)\,=\,\langle P\vert \, \overline{\psi}(0,z)\,\sigma_{jk}\, \tau_3 W(z)\,\psi(0,0)\,\vert P\rangle\,.
\label{eq:matrix_elements}
\end{equation}
where $\vert P\rangle$ denotes a proton state with four-momentum $P=(iE,0,0,P_3)$ and $W$ is a straight Wilson line in the direction of the boost. The fermion fields, $\psi$ and $\bar{\psi}$, are here a doublet of up and down quarks separated by a space-like distance, and $\tau_3$ is the third Pauli matrix, selecting the isovector combination $u-d$. 
Unlike the twist-2 transversity PDF, the twist-3 distribution $h_L(x)$ is extracted from a tensor structure whose indices are perpendicular to the boost direction. In this work, we always set the nucleon momentum to be along the $+z$-direction and therefore $\sigma_{ij}$ in the operator is taken to be $\sigma_{12}$.  
%The Pauli matrix $\tau_3$ allow to isolate the isovector combination $u-d$, that receives contribution only from the connected diagram of the type of Fig.~\ref{fig:diagram}. 
In fact, the desired matrix element for the proton ground state, $F_{h_L}(z,P_3)$, can be obtained through the following continuum decomposition in the Euclidean space:
\be
F_{h_L}(z,P_3)=-i\,\epsilon_{ij30}\,\frac{E}{m} M_{h_L}(z,P_3)\,,
\ee
where $m$ is the proton mass, $E=\sqrt{m^2+P_3^2}$ is the energy of the state with momentum boost $P_3$. Also, the indices $i$ and $j$ are in the transverse spatial plane ($i,\,j=1,2$).

\begin{figure}[h!]
    \begin{center} 
    \includegraphics[scale=0.8]{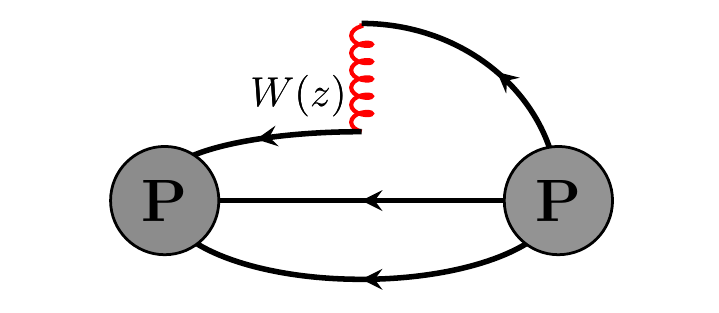}
    \end{center}
    \vspace{-0.4cm}
    \caption{Schematic representation of the connected diagram entering the calculation of the distribution $h_L(x)$. The curly line in red denotes the Wilson line $W(z)$.}
    \label{fig:diagram}
\end{figure}

The details of the lattice calculation follow those of Ref.~\cite{Bhattacharya:2020cen}, where the first investigation of the twist-3 $g_T(x)$ distribution is presented. The nucleon interpolating fields are defined using the momentum smearing technique~\cite{Bali:2016lva},
which has been proven to be very advantageous in reducing the exponential increase of the gauge noise as the energy of the particle increases. The momentum smearing is performed on APE-smeared~\cite{Albanese:1987ds} gauge links. To the Wilson line in the insertion operator we instead apply stout smearing~\cite{Morningstar:2003gk}, which helps in reducing statistical uncertainties in gluonic~\cite{Alexandrou:2016ekb,Alexandrou:2020sml} and non-local matrix elements~\cite{Alexandrou:2019lfo}. The matrix elements of Eq.~(\ref{eq:matrix_elements}) are accessed through computation of proton two-point functions and three-point functions, whose connected part is schematically represented in Fig.~\ref{fig:diagram}. 
This diagram requires the evaluation of an all-to-all propagator (from the spatial positions of the final proton state to the insertion points of the non-local operator), that is computed using the sequential method with the fixed sink approach. The time-slice of the sink is set to $T_{sink}=12a\simeq 1.12$ fm, a value at which excited-states contaminations are assumed to be sufficiently suppressed within the achieved statistical uncertainties and range of proton boosts considered in this work; see Ref.~\cite{Alexandrou:2019lfo} for a detailed study of excited-states effects on PDF matrix elements. The ground state matrix element is then extracted by seeking for the region where the ratios between the three-point functions and the two-point functions are independent of the insertion time of the operator (\textit{plateau method}).

To investigate the momentum dependence on $h_L(x)$, we perform the lattice calculation using three values of the nucleon boost, namely $P_3=4\pi/L$, $6\pi/L$ and $8\pi/L$, corresponding in physical units to $0.83$, $1.25$ and $1.67$~GeV. For each momentum, separate inversions of the Dirac operator have to be carried out, because the quark propagators depend on the $P_3$ value, which enters both in the momentum smearing phase and in the construction of the sequential source. To keep the statistical uncertainties under control, we perform a different number of measurements, reported in Table~\ref{tab:statistics}, where it can be seen that around 60 times larger statistics has been employed at $P_3=1.67$~GeV compared to the statistics at the lowest boost, $P_3=0.83$~GeV.
\begin{table}[h!]
\begin{center}
\renewcommand{\arraystretch}{1.4}
\begin{tabular}{c c c c}
\hline
$P_3 \; [\frac{2\pi}{L}]$ & $\,\,P_3$ [GeV] \,\,& \,\,$N_{conf}$ \,\,& $N_{meas}$\\
\hline 
$2$ & $0.83$ & $194$ & $1552$ \\
$3$ & $1.25$ & $731$ & $23392$ \\
$4$ & $1.67$ & $1644$ & $105216$ \\
\hline
\end{tabular}
\begin{minipage}{16cm}
\caption{Statistics used in this work, at $T_{sink}=1.12$~fm. We report the nucleon momentum in lattice and in physical units, the number of analyzed configurations, $N_{conf}$, and the total number of measurements $N_{meas}$.}
\label{tab:statistics}
\end{minipage}
\end{center}
\end{table}

\begin{figure}[h!]
    \begin{center} 
    \includegraphics[scale=0.575]{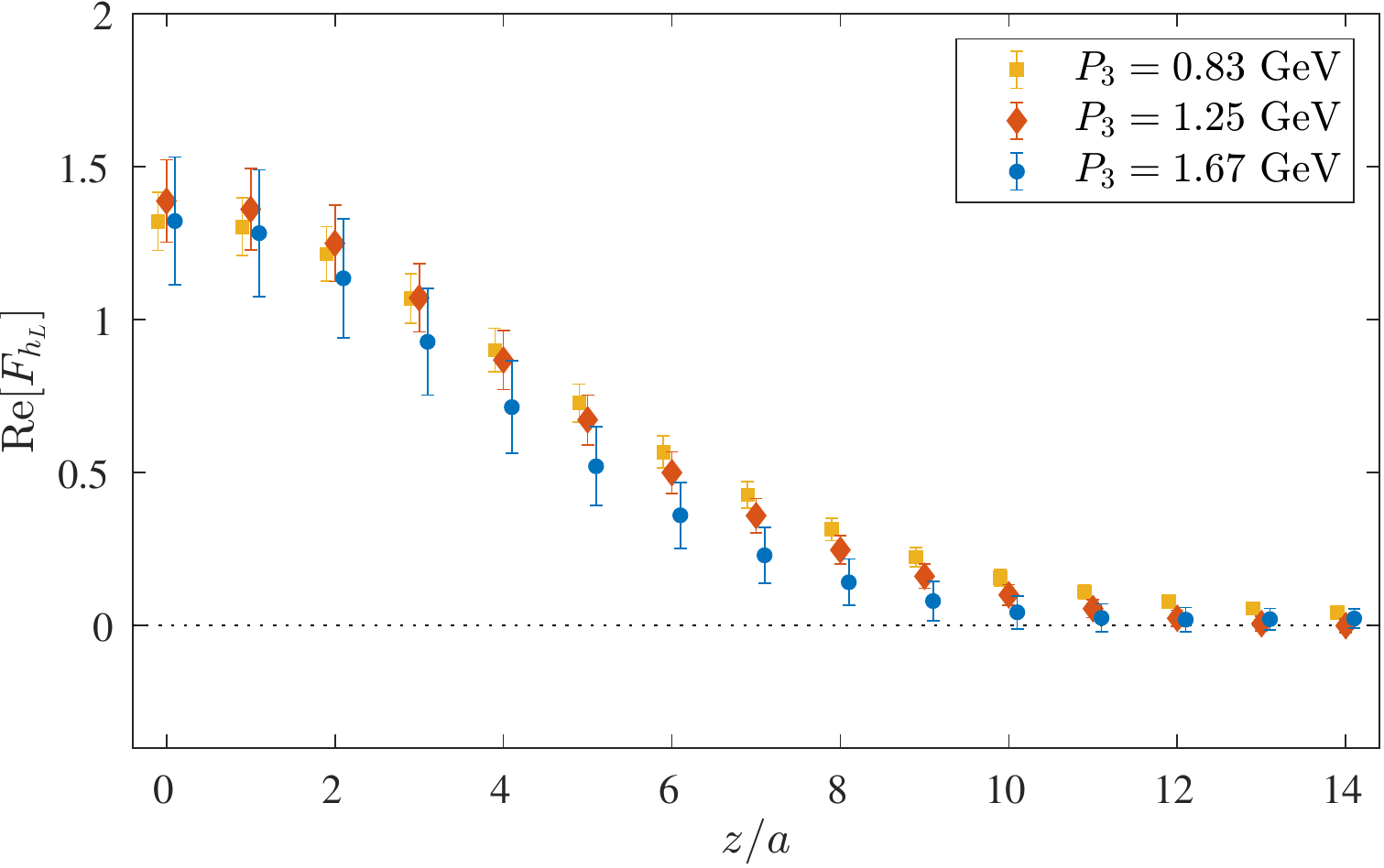}
    \includegraphics[scale=0.575]{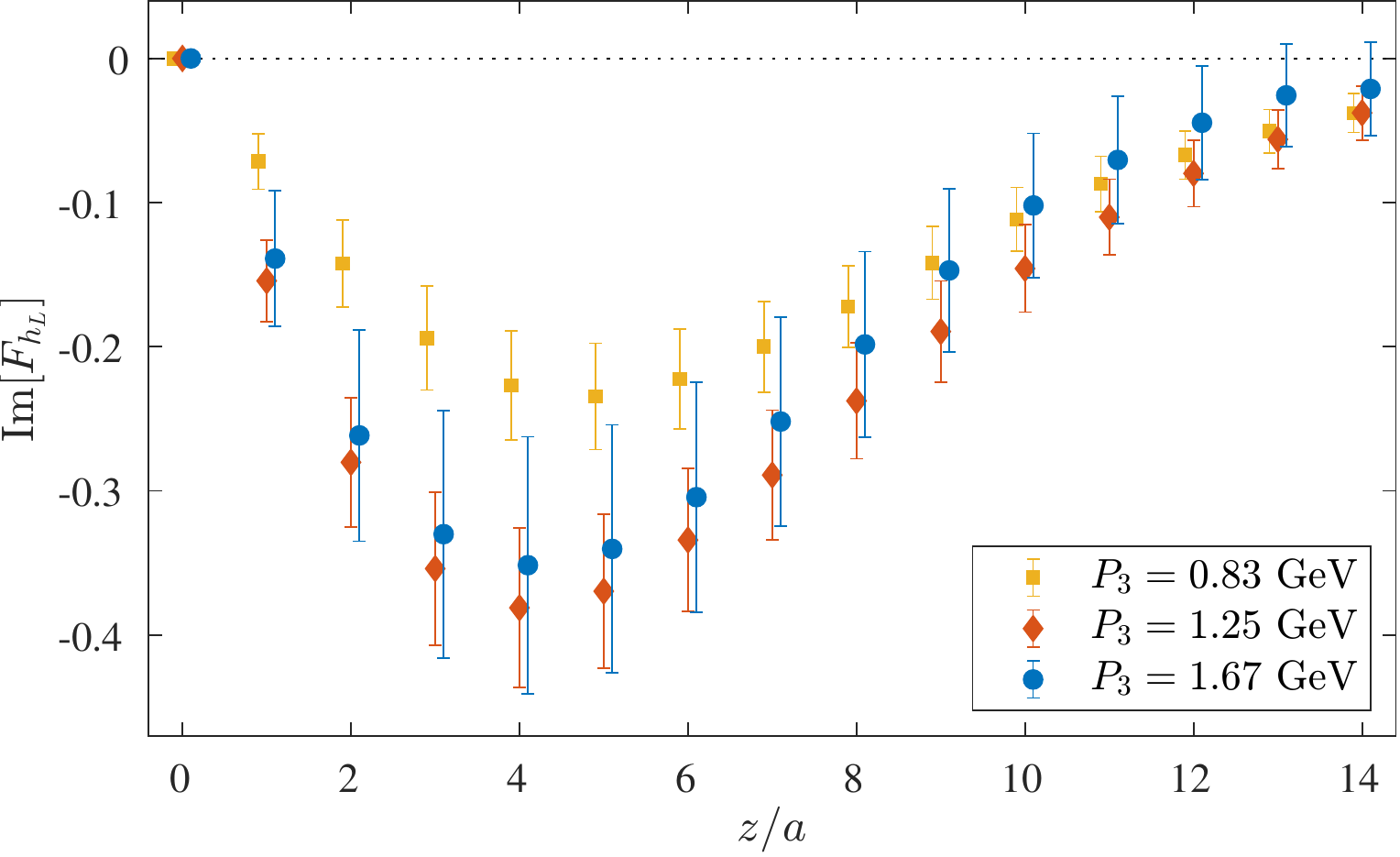}
    \end{center}
    \vspace*{-0.4cm}
    \caption{Real (left) and imaginary (right) part of the bare matrix elements for momenta $0.83$~GeV (yellow squares), $1.25$~GeV (red diamonds) and $1.67$~GeV (blue circles), extracted using the statistics of Table~\ref{tab:statistics}.}
    \label{fig:bare_Fhl}
\end{figure}

The resulting matrix elements $F_{h_L}$ as a function of the Wilson line length $z/a$ are shown in  Fig.~\ref{fig:bare_Fhl}. We find that with increasing momentum, the real part of the matrix elements decay faster and convergence for all $z$-values is obtained at the two largest boosts.

%%%%%%%%%%%%%%%%%%%%%%%%%%%%%%%%%%%%%%%%%%%%%%%%%%%%%%%%%%%%%%%%%%
\section{Non-perturbative Renormalization}
\label{sec:renormalization}
%%%%%%%%%%%%%%%%%%%%%%%%%%%%%%%%%%%%%%%%%%%%%%%%%%%%%%%%%%%%%%%%%%
The matrix elements of Eq.~(\ref{eq:matrix_elements}) are renormalized non-perturbatively, following the procedure developed and implemented in Refs.~\cite{Constantinou:2017sej,Alexandrou:2017huk}. We calculate the vertex function $G_{h_L}(p,z)$ of the tensor non-local operator within quark states, that is $\bar{\psi}_u\, \sigma_{12} \, W(z)\, \psi_d$, where $\psi_u$ ($\psi_d$) is the up-quark (down-quark) field. The vertex functions with momentum $p$ are amputated using the up-quark and down-quark propagator in momentum space, that is,
\begin{equation}
   {\cal V}_{h_L}(p,z) = (S_u(p))^{-1}\, G_{h_L}(p,z)\, (S_d(p))^{-1} \, .
\label{vertexfunction}
\end{equation}
The amputated vertex function is matched with its tree-level value in an RI$'$-type scheme~\cite{Martinelli:1994ty}, where the vertex momentum is set equal to the renormalization scale. The appropriate condition for the renormalization functions, $Z_{h_L}$, is 
\be
\label{renormZO}
Z_q^{-1}Z_{h_L}(z)\,\frac{1}{12} {\rm Tr} \left[{\cal V}_{h_L}(p,z) \left({\cal V}_{h_L}^{\rm Born}(p,z)\right)^{-1}\right] \Bigr|_{p^2{=}\bar\mu_0^2} {=} 1\, ,
\ee
where the quark field renormalization, $Z_q$, is given by
\be
\label{renormZq}
Z_q \, \frac{1}{12} {\rm Tr} \left[(S(p))^{-1}\, S^{\rm Born}(p)\right] \Bigr|_{p^2=\bar\mu_0^2}\,.
\ee
Eq.~(\ref{renormZO}) is a generalization of the condition used for local operators; here it is applied at each value of $z$ separately. ${\cal V}(p,z)$ ($S(p)$) is the amputated vertex function of the operator (fermion propagator) and ${\cal V}^{{\rm Born}}$ ($S^{{\rm Born}}(p)$) is its tree-level value. 

We calculate $Z_{h_L}$ using five ensembles with all quarks degenerate ($N_f=4$) and at different values of the pion mass. The relevant parameters are given in Table~\ref{Table:Z_ensembles}. Gauge configurations with all quark flavors degenerate is necessary for the calculation of the renormalization functions. This is because RI-type schemes are mass-independent schemes and a chiral extrapolation is needed. Therefore, the $N_f=2+1+1$ ensemble used to extract the proton matrix elements cannot be used for $Z_{h_L}$, as the strange and charm quarks are fixed to their physical value. For the $N_f=4$ ensembles to produce the appropriate value of $Z_{h_L}$, they have the same lattice formulations and with the same lattice spacing as the $N_f=2+1+1$ ensemble used for the extraction of $h_L(x)$, which is the case for the ones we use here.
\begin{table}[h]
\begin{center}
\renewcommand{\arraystretch}{1.5}
\renewcommand{\tabcolsep}{5.5pt}
\begin{tabular}{ccc}
\hline 
$\beta=1.726$ & $c_{\rm SW} = 1.74$ & $a=0.093$~fm \\
\hline\hline \\[-3ex]
{$24^3\times 48$}  & {$\,\,a\mu = 0.0060$}  & $\,\,m_\pi = 357.84$~MeV     \\
\hline
{$24^3\times 48$}  & $\,\,a\mu = 0.0080$     & $\,\,m_\pi = 408.11$~MeV     \\
\hline
{$24^3\times 48$}  & $\,\,a\mu = 0.0100$    & $\,\,m_\pi = 453.48$~MeV    \\
\hline
{$24^3\times 48$}  & $\,\,a\mu = 0.0115$    & $\,\,m_\pi = 488.41$~MeV    \\
\hline
{$24^3\times 48$}  & $\,\,a\mu = 0.0130$    & $\,\,m_\pi = 518.02$~MeV    \\
\hline
\end{tabular}
\vspace*{-0.25cm}
\begin{center}
\caption{\small{Parameters of the $N_f=4$ ensembles used for the calculation of the renormalization function $Z_{h_L}$}.}
\label{Table:Z_ensembles}
\end{center}
\end{center}
\vspace*{-0.2cm}
\end{table} 

The scale $\bar\mu_0$ of Eq.~(\ref{renormZO}) (RI$'$ renormalization scale) is chosen so that it has reduced discretization effects; see, e.g., Ref.~\cite{Alexandrou:2015sea}). We choose several values, and the corresponding estimates of $Z_{h_L}$ in the $\overline{\rm MS}$ scheme are fitted to eliminate residual $(a\,\bar\mu_0)^2$ dependence. In particular, we choose the momentum of the vertex function to have the same spatial components, $p=(p_0,p_1,p_1,p_1)$, leading to suppresses Lorentz non-invariant contributions~\cite{Constantinou:2010gr}. In practice, the ratio $\frac{p^4}{(p^2)^2}$ is less than 0.35 to control unwanted discretization effects. We use 17 different values of $\bar\mu_0$ within $(a\,\bar\mu_0)^2 \in [0.7,2.6]$, and apply a chiral extrapolation of the form
\begin{equation}
\label{eq:Zchiral_fit}
Z^{\rm RI}_{h_L}(z,\bar\mu_0,m_\pi) = {Z}^{\rm RI}_{h_L,0}(z,\mu_0) + m_\pi^2 \,{Z}^{\rm RI}_{gT,1}(z,\mu_0) \,.
\end{equation}
The fit is used for every value of $\bar\mu_0$ to eliminate any pion mass dependence. We find that the mass dependence is negligible for $z\le 5\,a$, and very small for the remaining $z$ values used in our analysis for the reconstruction of the $x$-dependence of $h_L$. The desirable mass-independent estimate, ${Z}^{\rm RI}_{h_L,0}(z,\mu_0)$, is then converted to the $\overline{\rm MS}$ scheme and evolved to $\mu{=}2$ GeV using the results of Ref.~\cite{Constantinou:2017sej}. Since the perturbative expressions are only known to the one-loop level, and due to present discretization effects, we extrapolate $(a\,\bar\mu_0)^2 {\to} 0$ using a linear fit and data, which gives the final estimates ${Z}^{\overline{\rm MS}}_{h_L,0}(z,2 \,{\rm GeV})$. 

\begin{figure}[h!]
    %\centering
    \includegraphics[scale=0.57]{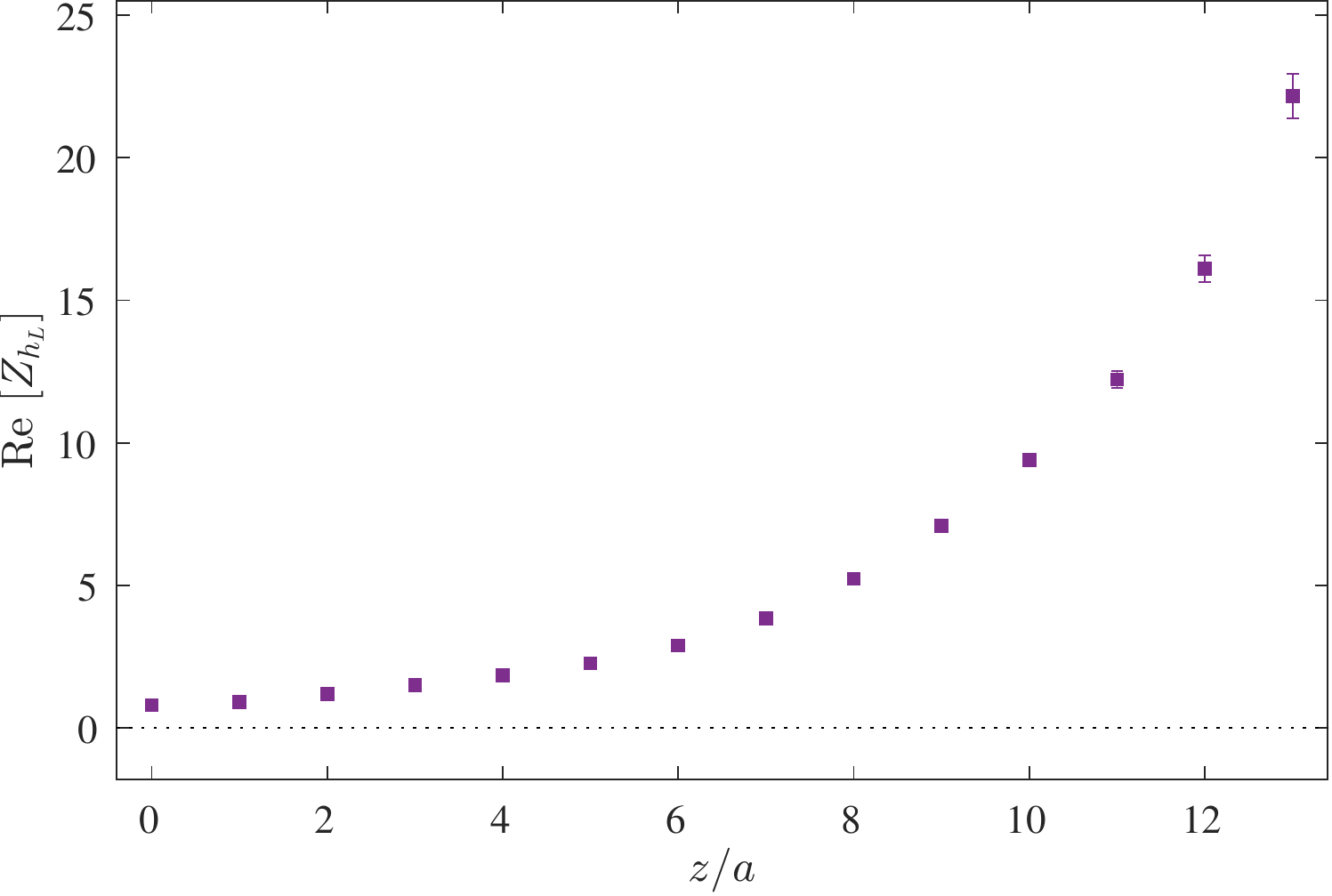}
    \includegraphics[scale=0.57]{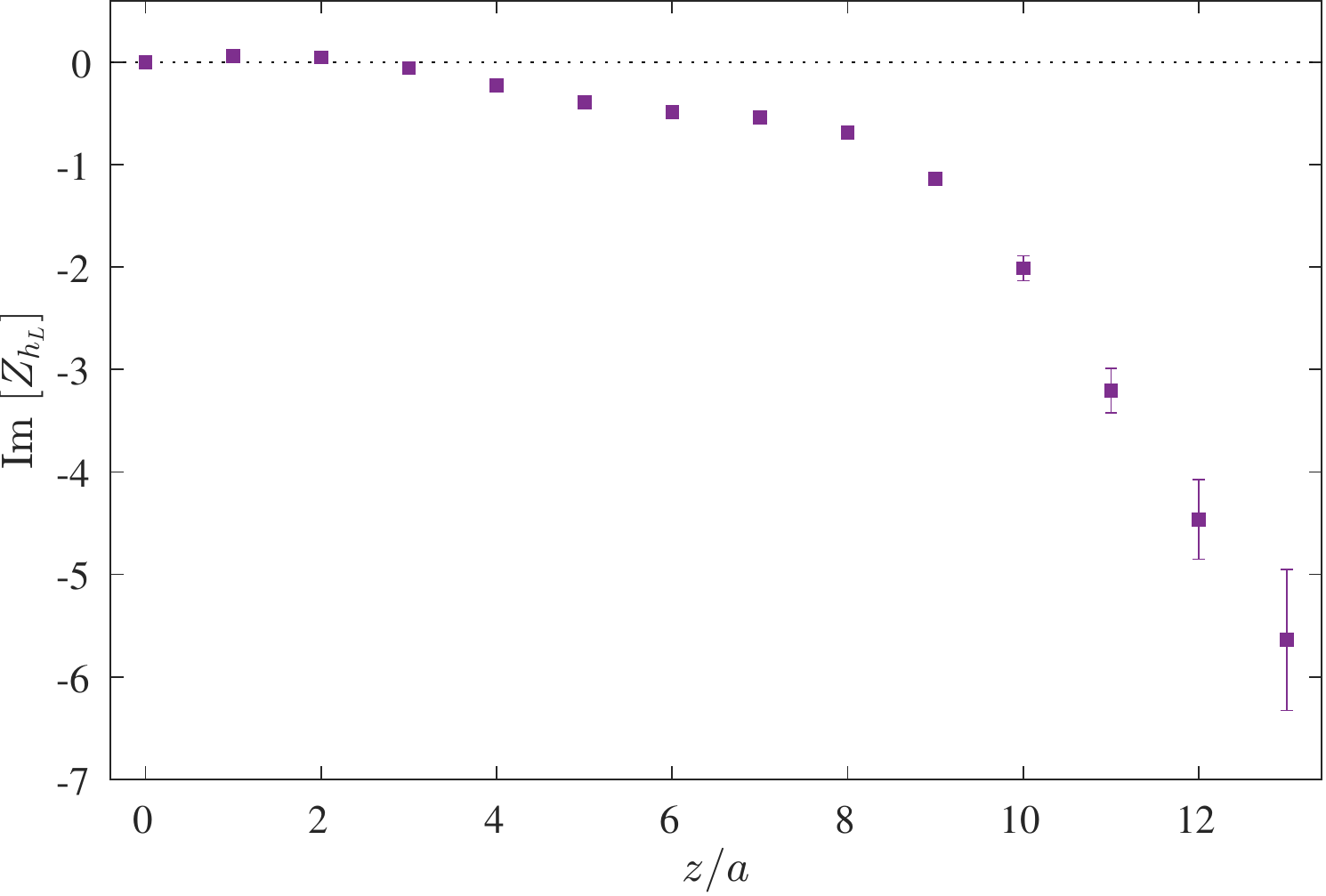}
    \caption{Renormalization function ${Z}^{{\rm M}\overline{\rm MS}}_{h_L,0}(z,2 \, {\rm GeV})$ after the chiral extrapolation, the conversion to the ${\rm M}{\overline{\rm MS}}$ scheme, evolution to 2 GeV, and the extrapolation $(a\,\bar\mu_0)^2 \to 0$.}
    \label{fig:Zfac}
\end{figure}

For the renormalization of the matrix elements of non-local operators we use a modified $\overline{\rm MS}$ scheme (${\rm M}\overline{\rm MS}$). This scheme was developed in Ref.~\cite{Alexandrou:2019lfo} as the matching in the $\overline{\rm MS}$ scheme does not preserve the norm of the light-cone PDFs and can even lead to a divergence in those quantities. 
To bring the renormalization function to the ${\rm M}\overline{\rm MS}$ scheme, one needs an additional conversion factor, that is,
\begin{equation}
\label{eq:ZMMS}
\displaystyle Z ^{\MMSb}_{{h_L},0}(z,\bar\mu) = Z ^{\MSb}_{{h_L},0}(z,\bar\mu)\,  {\cal C}^{\MSb,{\rm M\overline{MS}}}\,.
\end{equation}
We computed this additional finite factor in this work (see Sec.~\ref{sec:matching}), and the result in momentum space can be found in Eq.~(\ref{e:conv_mom}). 
In position space we find
\begin{eqnarray}
\label{eq:CMStoMMS}
\hspace*{-0.45cm}
{\cal C}_{h_L}^{\overline{\rm MS}, {\rm M\overline{MS}}} &=& 1 + \frac{\alpha_{s} C_F}{2\pi} e^{i z \mu_F}   \left( -2 \ln\left(\frac{1}{4}\right)\right) \nonumber \\[0.15cm]
&+& \frac{\alpha_{s}C_F}{2\pi} \bigg ( - 2{\rm Ci}(z \mu_F) + 2\ln(z \mu_F) - 2\ln(|z \mu_F|) \bigg ) \nonumber \\[0.15cm]
&+& \frac{\alpha_{s}C_F}{2\pi}\left(i\pi \frac{|z \mu_F|}{2 z \mu_F} - {\rm Ci}(z \mu_F) + \ln(z \mu_F) - \ln(|z \mu_F|)- i {\rm Si}(z \mu_F)\right) \nonumber \\[0.15cm]
&+& \frac{\alpha_{s}C_F}{2\pi} \big ( - e^{i z \mu_F} \big ) \left(\frac{2 {\rm Ei}(-i z \mu_F)- \ln (-i z \mu_F) + \ln (i z \mu_F) + i \pi {\rm Sign}(z \mu_F)}{2} \right) \, ,
\label{e:ZMMS_pos_space}
\end{eqnarray}
where $\mu_F$ is the factorization scale chosen to 2 GeV. In Eq.~(\ref{e:ZMMS_pos_space}), ${\rm Ci}$, ${\rm Si}$ and ${\rm Ei}$ are the special functions cosine integral, sine integral and exponential integral, respectively. Also, ${\rm Sign}$ is the sign function. After multiplying with the above conversion factor we extract ${Z}^{{\rm M}\overline{\rm MS}}_{h_L,0}(z,2 {\rm GeV})$, which is shown in Fig.~\ref{fig:Zfac}.

\begin{figure}[h!]
    \includegraphics[scale=0.57]{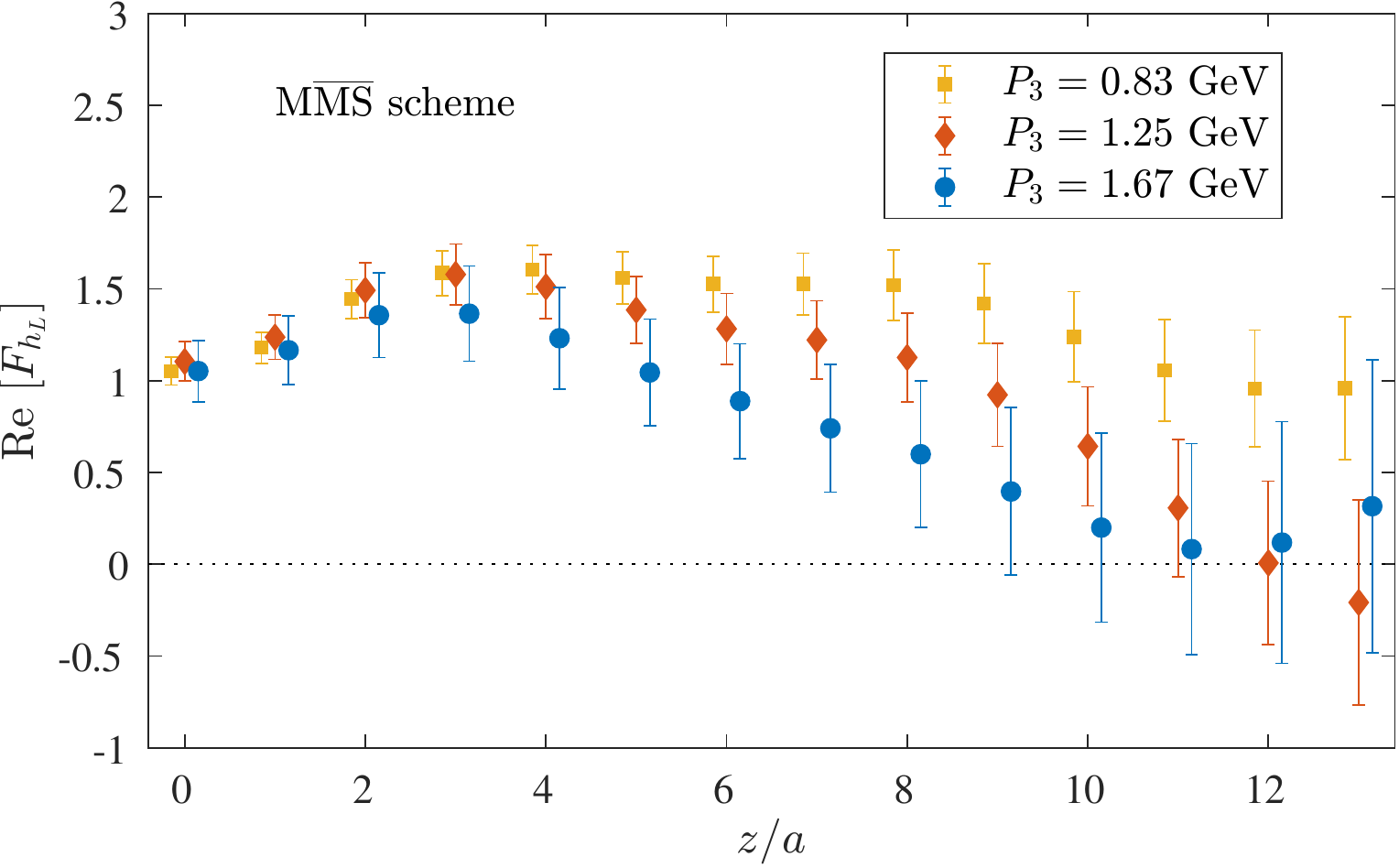}
    \includegraphics[scale=0.57]{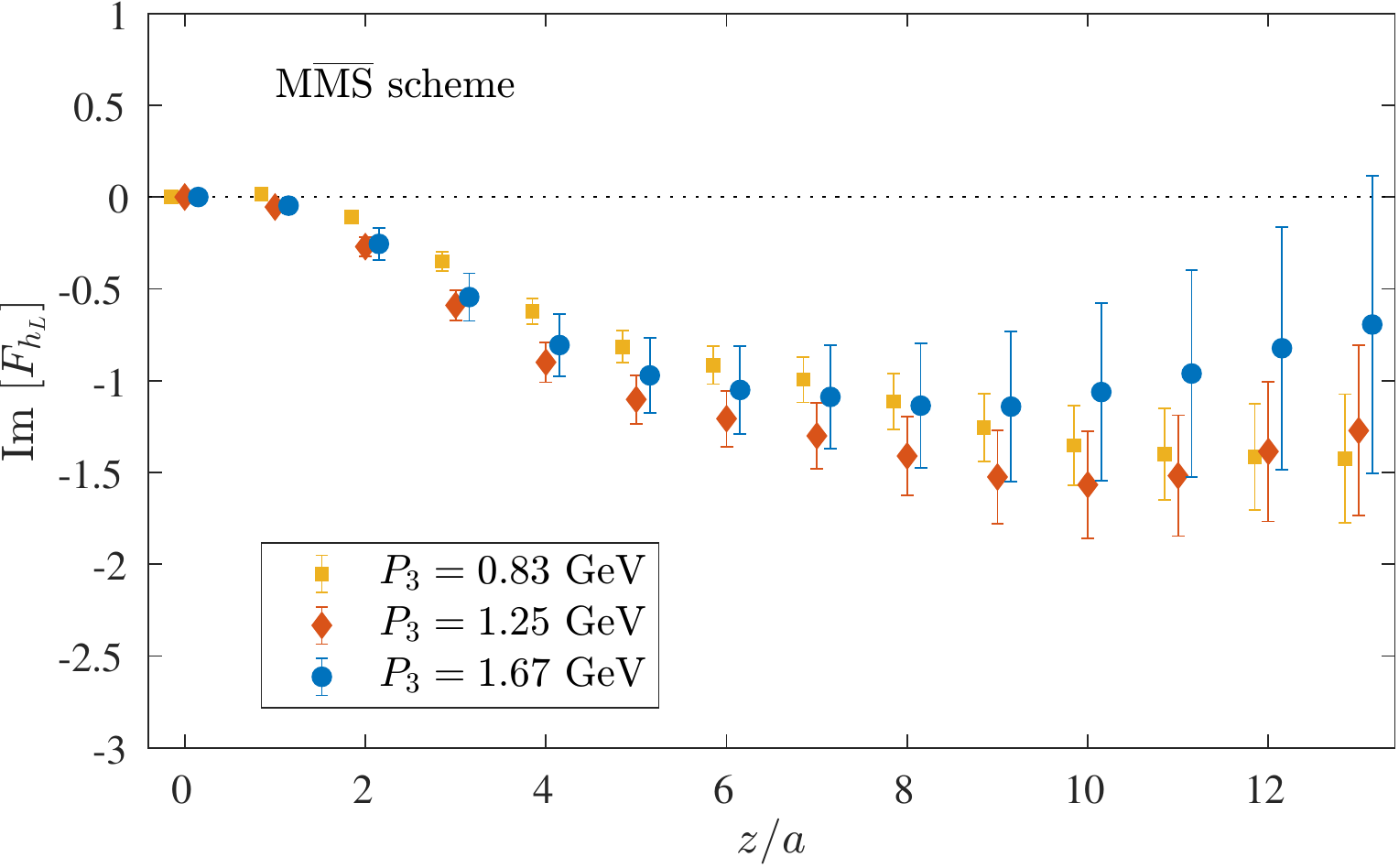}
    \caption{Renormalized matrix elements in ${\rm M}\overline{\rm MS}$ scheme at $2$~GeV, at the nucleon momenta $0.83$~GeV (yellow squares), $1.25$~GeV (red diamonds) and $1.67$~GeV (blue circles).}
    \label{fig:ME_R}
\end{figure}

${Z}^{{\rm M}\overline{\rm MS}}_{h_L,0}(z,2 \, {\rm GeV})$ is applied multiplicatively on the bare matrix elements of Eq.~(\ref{eq:matrix_elements}), and the resulting matrix element is shown in Fig.~\ref{fig:ME_R}.

%%%%%%%%%%%%%%%%%%%%%%%%%%%%%%%%%%%%%%%%%%%%%%%%%%%%%%%%%%%%%%%%%%
\section{Reconstruction of the $x$-dependence}
\label{sec:reco}
%%%%%%%%%%%%%%%%%%%%%%%%%%%%%%%%%%%%%%%%%%%%%%%%%%%%%%%%%%%%%%%%%%
The renormalized matrix elements of Fig.~\ref{fig:ME_R} are used to extract the $x$-dependence of the quasi-distribution, $\tilde{h}_L(x,P_3)$, by performing a Fourier transform. 
Such a reconstruction procedure is subject to an inverse problem, since one is attempting to obtain a continuous distribution from a finite and truncated set of lattice data, see Ref.~\cite{Karpie:2018zaz} for a detailed discussion of the issue and the proposed solutions.
One of the latter, that we utilize in this work, is the Backus-Gilbert method~\cite{BackusGilbert} with Tikhonov regularization~\cite{Tikhonov:1963}. 
This approach does not introduce any modeling for the quasi-PDFs and contains only three free parameters: the regularization parameter $\rho$ related to the resolution of the method, the maximum value of $x$ for which the quasi-distribution is taken to be nonzero, called $x_c$ here, and the maximal length $z_{max}/a$ included in the reconstruction procedure. 
For a detailed description of our implementation of the Backus-Gilbert method for quasi-PDFs we refer to Ref.~\cite{Alexandrou:2020zbe}. 
The results presented below are obtained using $\rho=10^{-3}$, which leads to a reasonable resolution avoiding bias in the final distributions, and $|x_c|=2$. 
The latter is justifiable by the fact that quasi-PDFs are not bound to vanish beyond the canonical support $x\in [-1,1]$ and, therefore, the reconstruction of such distributions needs to be extended outside of this interval. 
We also find that other choices of the $\rho$-parameter down to $10^{-5}$, as well as different values of $x_c>1$, do not lead to any significant difference in the final PDFs, and they are not taken into account in the uncertainty budget. 
Instead, more important is the amount of input data that is used in the reconstruction process, because the matrix elements do not decay to zero fast enough within the attained separations (see Fig.~\ref{fig:ME_R}) and, in most cases, they do not remain compatible with zero as the length of the Wilson line increases. 
Our criterion is to include lattice data up to the value $z_{max}/a$ at which the real or the imaginary part is compatible with zero. 
While it is possible to meet this condition for $P_3=1.25$, $1.67$ GeV, we note that at the lowest boost no such value can be found. 
In practice, to compute $\tilde{h}_L(x,P_3)$, we use matrix elements up to $z_{max}=\lbrace 13, 12, 10\rbrace a$ for $P_3=0.83$, $1.25$ and $1.67$ GeV, respectively. 
Moreover, we estimate the systematic uncertainty from this choice of the cutoff by varying $z_{max}$ up to three lattice units, 
\begin{equation}
\Delta_{syst.}(x,P_3)=\frac{\vert \tilde{h}_L(x,P_3)_{z_{max}+3} -  \tilde{h}_L(x,P_3)_{z_{max}-3}\vert }{2} \,,
\label{eq:syst_err}
\end{equation}
and analogously for the matched distributions $h_L(x)$.
Finally, we estimate the total error by summing in quadrature $\Delta_{syst.}(x,P_3)$ and the statistical uncertainty. The error bands of all numerical results below include this combined uncertainty.

%%%%%%%%%%%%%%%%%%%%%%%%%%%%%%%%%%%%%%%%%%%%%%%%%%%%%%%%%%%%%%%%%%
\section{Matching to the light-cone PDF $h_{L}(x)$}
\label{sec:matching}
%%%%%%%%%%%%%%%%%%%%%%%%%%%%%%%%%%%%%%%%%%%%%%%%%%%%%%%%%%%%%%%%%%
A perturbative matching procedure relates the quasi-PDFs to the light-cone PDFs of interest.
While the matching has been discussed in quite some detail for twist-2 PDFs~\cite{Ji:2013dva, Xiong:2013bka, Ma:2014jla, Radyushkin:2017cyf, Wang:2017qyg, Stewart:2017tvs, Izubuchi:2018srq}, only recently the twist-3 case was considered for the first time~\cite{Bhattacharya:2020cen, Bhattacharya:2020xlt, Bhattacharya:2020jfj, Braun:2021aon}.
Here we take as starting point the $\MSb$ matching result obtained in Ref.~\cite{Bhattacharya:2020jfj}, according to which the quasi-PDF and light-cone PDF are connected via
%%%%%%%%%%%%%%%%%%%%%%%
\begin{eqnarray}
h_{L}(x, \mu) &=& \int^{\infty}_{-\infty} \dfrac{d\xi}{|\xi|} \, C_{\overline{\mathrm{MS}}} \left ( \xi, \dfrac{\mu^{2}}{p^{2}_{3}} \right ) \tilde{h}_{L} \left (\dfrac{x}{\xi}, \mu, P_3 \right ) \, , 
\label{e:def_matching}
\end{eqnarray}
%%%%%%%%%%%%%%%%%%%%%%%
where $C$ is the perturbatively calculable matching coefficient, and $p_3 = (x/\xi)P_3$ the quark momentum.
In Ref.~\cite{Braun:2021aon} it was shown in the context of $g_T(x)$ that, generally, for twist-3 PDFs the matching formula actually has a more complicated structure than the one in Eq.~(\ref{e:def_matching}).
In particular, our matching does not take into account quark-gluon-quark correlations.
Nevertheless, for several reasons we believe that, at present, our approach is justified.
First, a matching formula along the lines discussed in Ref.~\cite{Braun:2021aon} is currently not available for $h_L(x)$.
Second, the results in Ref.~\cite{Braun:2021aon} indicate that still an approximation will be needed to extract the light-cone $h_L(x)$ with a complete one-loop matching formula.
Third, our previous numerical results for $g_T(x)$~\cite{Bhattacharya:2020cen}, and the results for $h_L(x)$ discussed in the present work, already look encouraging.  

The matching coefficient for $h_{L}(x)$ extracted in Ref.~\cite{Bhattacharya:2020jfj} takes the form
%%%%%%%%%%%%%%%%%%%%%%%
\begin{eqnarray}
C_{\overline{\mathrm{MS}}} \bigg (\xi, \dfrac{\mu^{2}}{p^{2}_{3}} \bigg ) & = & \delta (1-\xi) + C^{(\rm{s})}_{\overline{\mathrm{MS}}} \bigg (\xi, \dfrac{\mu^{2}}{p^{2}_{3}} \bigg ) + C^{(\rm{c})}_{\overline{\mathrm{MS}}} \bigg (\xi, \dfrac{\mu^{2}}{p^{2}_{3}} \bigg ) \, ,
\label{e:structure_CMS}
\end{eqnarray}
%%%%%%%%%%%%%%%%%%%%%%%
where the first term represents the (trivial) leading-order contribution, while the second and third terms are the one-loop results for which we distinguish between a (singular) term caused by a zero-mode contribution and a canonical term.
The singular term, which has no counterpart at twist-2, is given by
%%%%%%%%%%%%%%%%%%%%%%
\begin{eqnarray}
C^{(\rm{s})}_{\overline{\mathrm{MS}}} \bigg (\xi, \dfrac{\mu^{2}}{p^{2}_{3}} \bigg ) = \dfrac{\alpha_{s} C_{F}}{2\pi}
\begin{cases}
- \dfrac{1}{\xi}
& \quad \xi > 1  \\[0.5cm]
- \delta(\xi) \bigg ( \ln \dfrac{4p^{2}_{3}}{\mu^{2}} + 1 \bigg ) \, - \, {\rm{R}}_{0}(|\xi|) 
& \quad -1 < \xi < 1 \\[0.5cm]
\phantom{+} \dfrac{1}{\xi}
& \quad \xi < -1 \, .
\end{cases}
\label{e:matching_e_sing_MSbar}
\end{eqnarray}
%%%%%%%%%%%%%%%%%%%%%%
In this equation, $R_{0} (|\xi|)$ is a plus-function at $\xi =0$, defined as
%%%%%%%%%%%%%%%%%%%%%%
\begin{eqnarray}
\label{eq:R0_def}
{\rm{R}}_{0}(|\xi|) &\equiv \bigg [ \dfrac{1}{|\xi|} \bigg ]_{+[0]}  = \theta(|\xi|) \, \theta(1-|\xi|) \lim_{\beta \rightarrow 0} \bigg [ \dfrac{\theta(|\xi| - \beta)}{|\xi|} + \delta(|\xi| - \beta) \ln \beta \bigg ] \, ,
\label{e:R0_def}
\end{eqnarray}
%%%%%%%%%%%%%%%%%%%%%%
while the canonical term reads
%%%%%%%%%%%%%%%%%%%%%%
\begin{eqnarray}
C^{(\rm{c})}_{\overline{\mathrm{MS}}} \bigg (\xi, \dfrac{\mu^{2}}{p^{2}_{3}} \bigg ) & = &  \dfrac{\alpha_{s} C_{F}}{2\pi}
\begin{cases}
\bigg [ \dfrac{2}{1-\xi} \ln \dfrac{\xi}{\xi - 1} + \dfrac{1}{1-\xi} + \dfrac{1}{\xi} \bigg ]_{+} - \dfrac{1}{\xi}
& \quad \xi > 1  \\[0.5cm]
\bigg [ \dfrac{2}{1-\xi} \ln \dfrac{4\xi (1-\xi) p^{2}_{3}}{\mu^{2}} + 2(1-\xi) - \dfrac{1}{1-\xi} \bigg ]_{+} 
& \quad 0 < \xi < 1 \\[0.5cm]
\bigg [ \dfrac{2}{1-\xi} \ln \dfrac{\xi -1}{\xi} - \dfrac{1}{1-\xi} + \dfrac{1}{1-\xi} \bigg ]_{+} - \dfrac{1}{1- \xi}
& \quad \xi < 0 
\end{cases}
\nonumber \\[0.5cm]
& + &  \dfrac{\alpha_{s}C_{F}}{2\pi} \delta (1-\xi) \, \bigg ( 1 + \ln \dfrac{\mu^{2}}{4p^{2}_{3}} \bigg ) \,,
\label{e:canonical_matching_MSbar}
\end{eqnarray}
%%%%%%%%%%%%%%%%%%%%%%
where the plus-prescription $[...]_{+}$ for the canonical terms have been defined at $\xi=1$.

The problem of matching in the $\MSb$ scheme is that it leads to a divergent norm for the resulting light-cone PDF.
To overcome this issue, we employ the so-called modified $\MSb$ ($\MMSb$) scheme, in which we perform an extra subtraction of terms in the regions $\xi >1$ and $\xi <0$ which give rise to (logarithmic) divergences~\cite{Alexandrou:2019lfo}. 
Similar to Eq.~(\ref{e:structure_CMS}), the structure for the one-loop matching coefficient in the $\MMSb$ scheme is
%%%%%%%%%%%%%%%%%%%%%%%
\begin{eqnarray}
C_{\rm{M}\overline{\mathrm{MS}}} \bigg (\xi, \dfrac{\mu^{2}}{p^{2}_{3}} \bigg ) & = & \delta (1-\xi) + C^{(\rm{s})}_{\rm{M}\overline{\mathrm{MS}}} \bigg (\xi, \dfrac{\mu^{2}}{p^{2}_{3}} \bigg ) + C^{(\rm{c})}_{\rm{M}\overline{\mathrm{MS}}} \bigg (\xi, \dfrac{\mu^{2}}{p^{2}_{3}} \bigg ) \,,
\label{eq:matching_kernel}
\end{eqnarray}
%%%%%%%%%%%%%%%%%%%%%%%
where the individual terms are
%%%%%%%%%%%%%%%%%%%%%%
\begin{eqnarray}
\label{eq:singular_part}
C^{(\rm{s})}_{\rm{M}\overline{\mathrm{MS}}} \bigg (\xi, \dfrac{\mu^{2}}{p^{2}_{3}} \bigg ) &=& \dfrac{\alpha_{s} C_{F}}{2\pi}
\begin{cases}
\phantom{+} \delta (1 - \xi) \bigg ( \dfrac{1}{2} - \dfrac{1}{2} \ln \dfrac{\mu^{2}}{4p^{2}_{3}} \bigg )
& \quad \xi > 1  \\[0.5cm]
- \delta(\xi) \bigg ( \ln \dfrac{4p^{2}_{3}}{\mu^{2}} + 1 \bigg ) \, - \, {\rm{R}}_{0}(|\xi|) 
& \quad -1 < \xi < 1 \\[0.5cm]
\phantom{+} \delta (1 + \xi) \bigg ( \dfrac{1}{2} - \dfrac{1}{2} \ln \dfrac{\mu^{2}}{4p^{2}_{3}} \bigg )
& \quad \xi < -1 \,,
\end{cases}
\end{eqnarray}
%%%%%%%%%%%%%%%%%%%%%%
and
%%%%%%%%%%%%%%%%%%%%%%
\begin{eqnarray}
\label{eq:canonical_part}
C^{(\rm{c})}_{{\rm M}\overline{\mathrm{MS}}} \bigg (\xi, \dfrac{\mu^{2}}{p^{2}_{3}} \bigg ) & = &  \dfrac{\alpha_{s} C_{F}}{2\pi}
\begin{cases}
\bigg [ \dfrac{2}{1-\xi} \ln \dfrac{\xi}{\xi - 1} + \dfrac{1}{1-\xi} + \dfrac{1}{\xi} \bigg ]_{+} 
& \quad \xi > 1  \\[0.5cm]
\bigg [ \dfrac{2}{1-\xi} \ln \dfrac{4\xi (1-\xi) p^{2}_{3}}{\mu^{2}} + 2(1-\xi) - \dfrac{1}{1-\xi} \bigg ]_{+} 
& \quad 0 < \xi < 1 \\[0.5cm]
\bigg [ \dfrac{2}{1-\xi} \ln \dfrac{\xi -1}{\xi} - \dfrac{1}{1-\xi} + \dfrac{1}{1-\xi} \bigg ]_{+} 
& \quad \xi < 0 \,.
\end{cases}
\end{eqnarray}
%%%%%%%%%%%%%%%%%%%%%%
The transition from the $\rm{\overline{MS}}$ scheme to the M$\rm{\overline{MS}}$ scheme is given by the conversion function
%%%%%%%%%%%%%%%%%%%%%%
\begin{eqnarray}
Z^{\rm{M}\rm{\overline{MS}}} (\xi) &=& 1 - \frac{\alpha_{s} C_F}{2\pi}\left(- \frac{1}{\xi}\theta(\xi-1) + \frac{1}{\xi}\theta(-\xi - 1) \right)
+ {\alpha_s C_F \over 2\pi}\delta(1-\xi) \left( 1 - \ln \dfrac{1}{4} \right) \nonumber \\[0.2cm]
&-& \frac{\alpha_{s} C_F}{2\pi}\left(- \frac{1}{\xi}\theta(\xi-1) - \frac{1}{1-\xi}\theta(-\xi) \right)
- {\alpha_s C_F\over 2\pi}\delta(1-\xi) \left( 1 + \ln \dfrac{1}{4} \right) \,,
\label{e:conv_mom}
\end{eqnarray}
%%%%%%%%%%%%%%%%%%%%%%
where the first two terms of ${\cal O}(\alpha_s)$ are related to the singular contribution, and the remaining terms to the canonical contribution. 
The corresponding expression of the renormalization function in the position space has been given in the previous section; see Eq.~(\ref{e:ZMMS_pos_space}). 

By construction, the matching coefficients in Eqs.~(\ref{eq:singular_part}) and~(\ref{eq:canonical_part}) each integrate to zero.
This is obvious for $C^{(\rm{c})}_{\rm{M}\overline{\mathrm{MS}}}$ because of the plus-functions.
But one can also readily verify the same result for $C^{(\rm{s})}_{\rm{M}\overline{\mathrm{MS}}}$ by combining the contributions from the three different $\xi$-regions.
For the canonical term, this property guarantees that the norm of the corresponding contribution to the light-cone PDF $h_L(x)$ vanishes.
If the same would apply to the singular term, the norm of the full light-cone PDF and quasi-PDF would agree with each other based on the matching formula in Eq.~(\ref{e:def_matching}), as we expect from model-independent arguments~\cite{Bhattacharya:2019cme, Bhattacharya:2021boh}.
However, the singular matching coefficient does not lead to this property.
To illustrate this point, we take a simple model for the quasi-PDF, that is, 
%%%%%%%%%%%%%%%%%%%%%%
\begin{equation}
    \tilde{q}_{\rm mod} (x) = \dfrac{c}{x^2 + b} \,,
\label{e:hLQ_model}    
\end{equation}
%%%%%%%%%%%%%%%%%%%%%%
with $b > 0$. 
The expression in Eq.~(\ref{e:hLQ_model}), which is motivated by the large-$x$ behavior of the one-loop M$\rm{\overline{MS}}$ result for the quasi-PDF for the quark target, allows us to obtain an analytical result for the corresponding light-cone PDF.
By just considering the term $R_{0} (|\xi |)$ of the singular matching coefficient in Eq.~(\ref{eq:singular_part}) we find
\begin{eqnarray}
q_{\rm mod}^{R_0}(x) &=& \frac{2c}{\sqrt{b} x} \, {\rm{tan}}^{-1} \dfrac{\sqrt{b}}{x} \,,
\label{e:hL_model} 
\end{eqnarray}
which is well defined for all $x$ except $x = 0$.
Most importantly, the norm of this expression does not vanish.
Even worse, the norm is not defined, as the function in Eq.~(\ref{e:hL_model}) behaves like $1/|x|$ for $x \to 0$.
We actually expect that the same general result holds regardless of the specific form of the quasi-PDF. 
Furthermore, depending on the functional form of the quasi-PDFs, the contributions proportional to $\delta(1 - \xi)$ and $\delta(1 + \xi)$ in Eq.~(\ref{eq:singular_part}), which emerge when transitioning to the M$\rm{\overline{MS}}$ scheme, can give rise to contributions to the light-cone PDFs that have an undefined norm.
We repeat that the singular matching coefficient in Eq.~(\ref{eq:singular_part}), caused by zero-mode contributions, is a new feature at twist-3.
In Ref.~\cite{Bhattacharya:2020jfj} we have shown that, in principle, the quasi-PDF approach remains valid in the presence of such contributions.
However, we find that they can cause a problem when trying to compute the norm of the resulting light-cone PDF.
This point deserves further investigation which goes beyond the scope of the present work.

%%%%%%%%%%%%%%%%%%%%%%%%%%%

\begin{figure}[h!]
    \centering
    \includegraphics[scale=0.575]{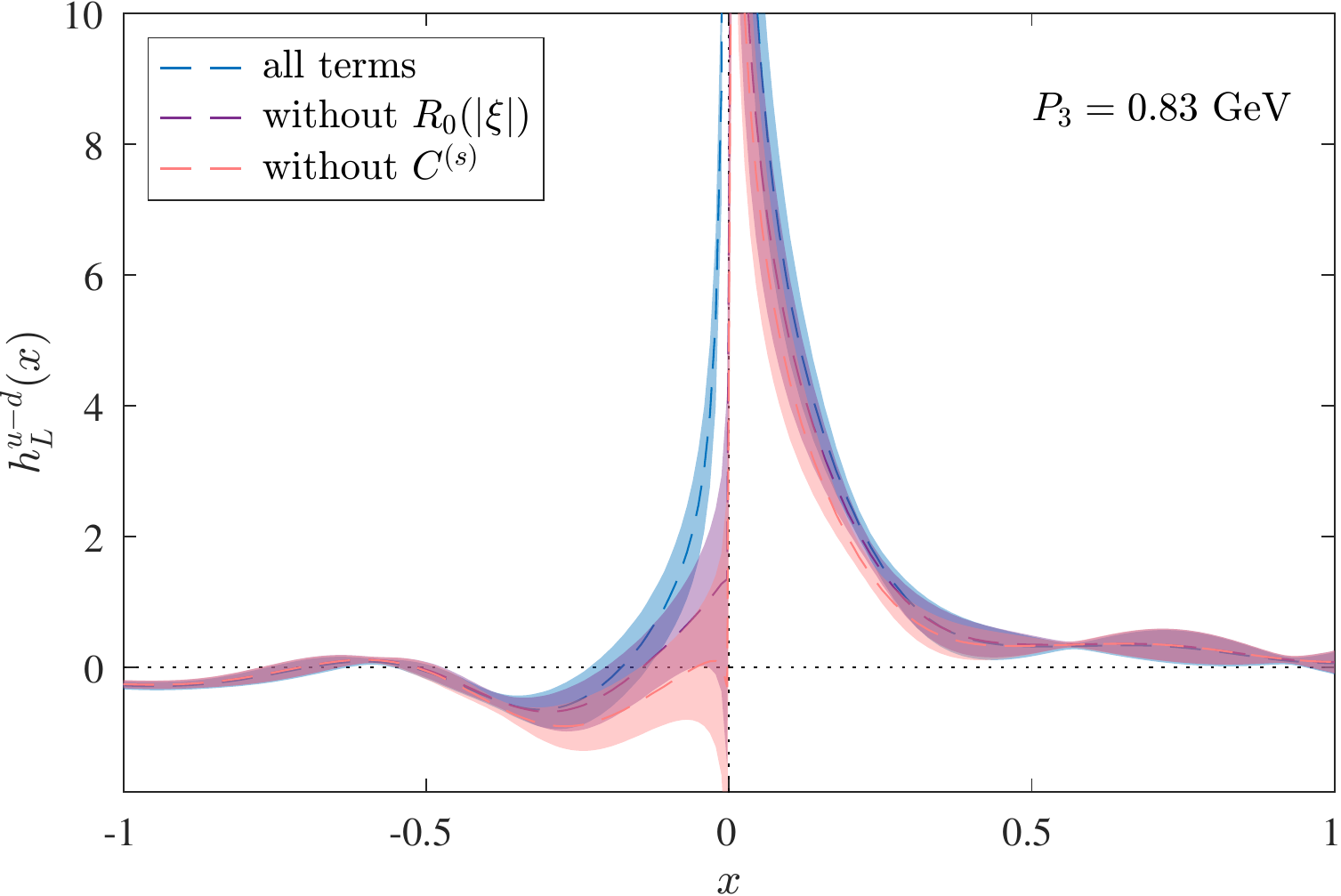}
    \includegraphics[scale=0.575]{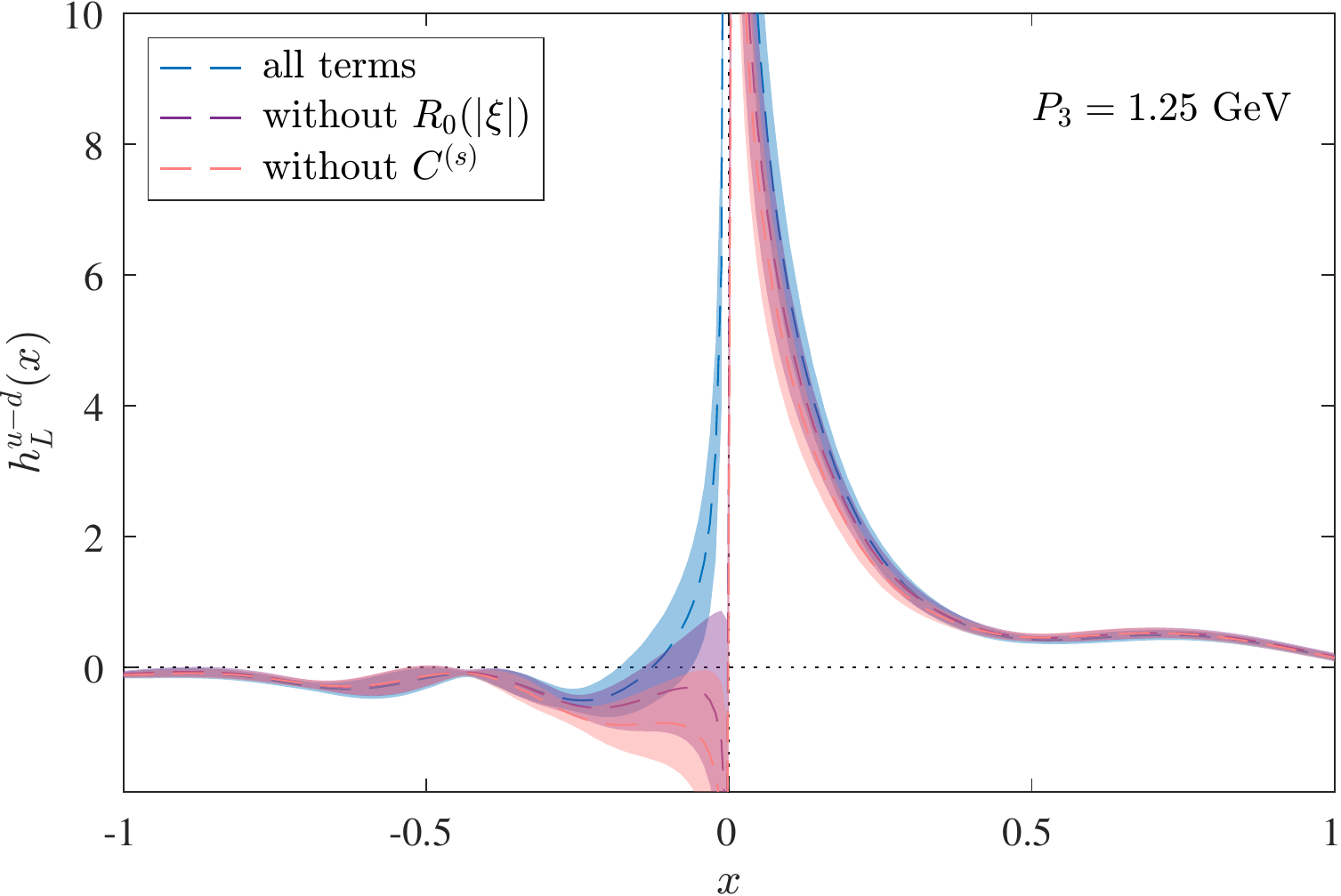} 
    \includegraphics[scale=0.575]{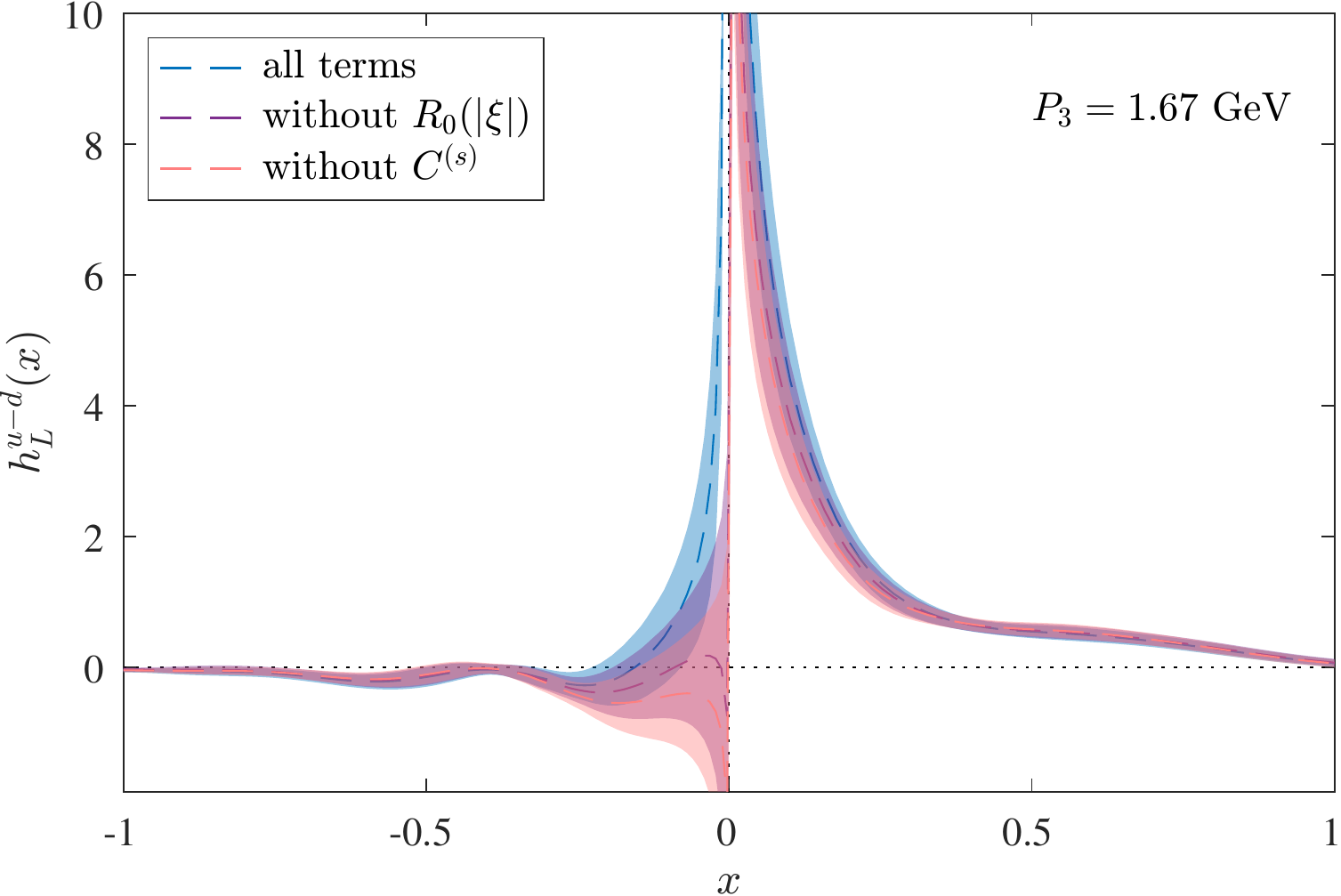}
    \caption{Matched $h_L(x)$ in the $\MMSb$-scheme at $P_3=\lbrace 0.83,1.25,1.67\rbrace$~GeV, including all terms in the matching formula of Eq.~(\ref{eq:matching_kernel}) (blue band), excluding the $R_0(|\xi|)$ term of Eq.(\ref{e:R0_def}) (violet band) and the singular part of Eq.~(\ref{eq:singular_part}) of the perturbative corrections (pink band).}
    \label{fig:matched_HL_options}
\end{figure}

\begin{figure}[h!]
    \centering
    \includegraphics[scale=0.575]{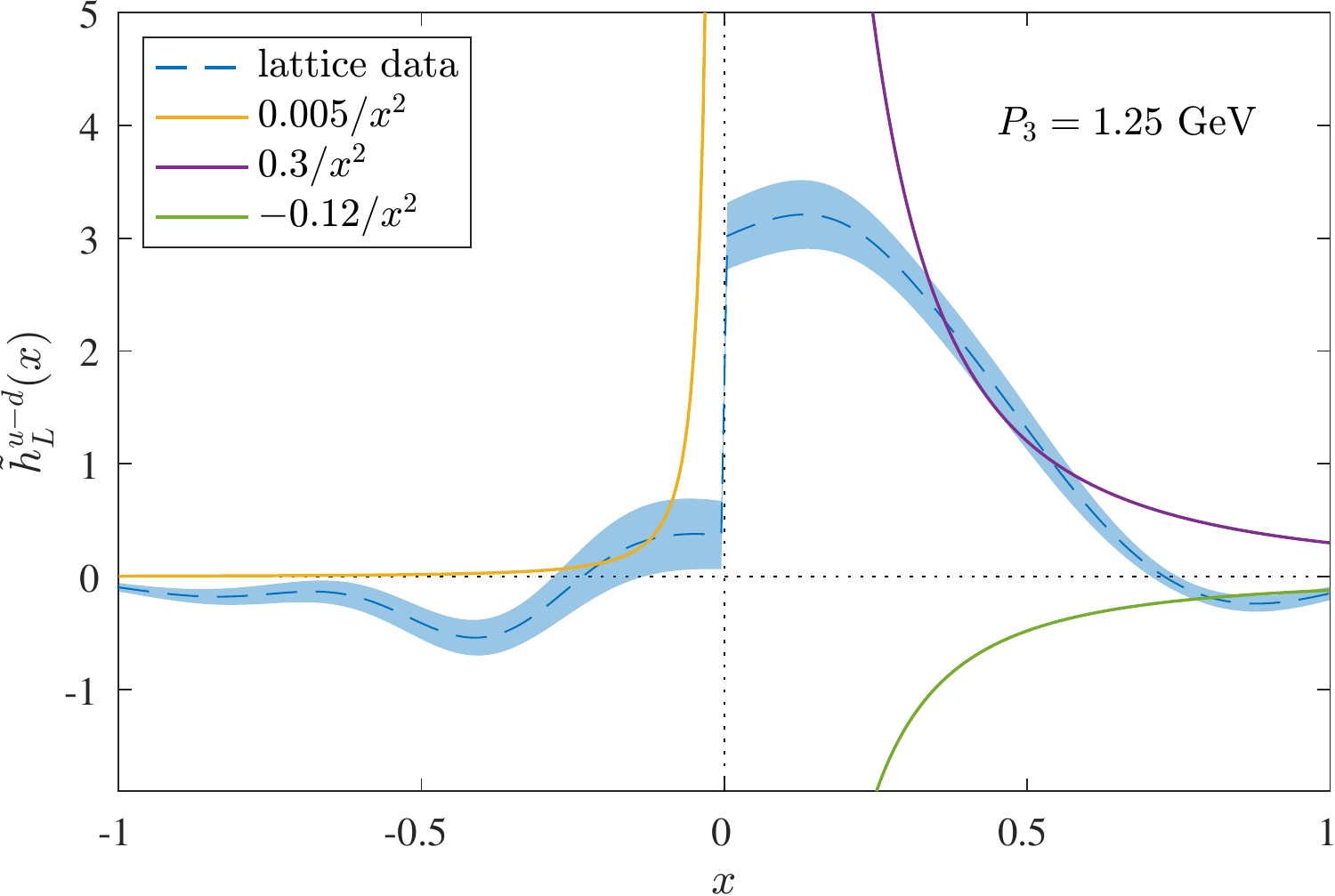}
   \includegraphics[scale=0.575]{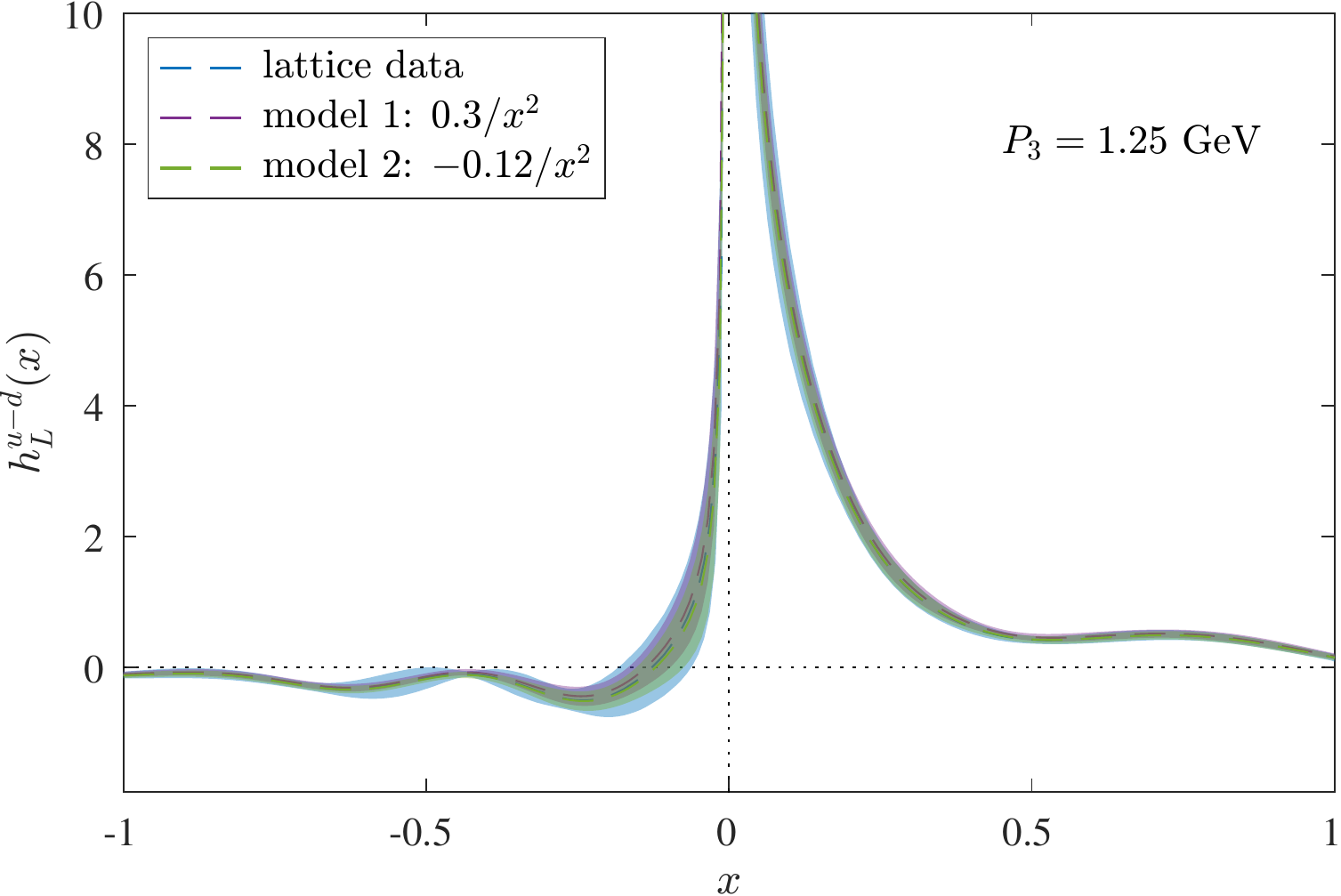} 
   
    \caption{Left: the quasi-$h_L$ function from our lattice data in the $\MMSb$-scheme at $P_3= 1.25$~GeV (blue band) and the considered models for the large-$x$ behavior (solid lines).
     Right: the matched $h_L$ function using only lattice data with the cutoff at $x_c=2$ (blue band) and with the large-$x$ contributions from the model quasi-$h_L$ function in the $R_0(|\xi|)$ term of Eq.~(\ref{e:R0_def}) (purple band for $0.3/x^2$ and green band for $-0.12/x^2$). }
    \label{fig:models}
\end{figure}

Instead, we explore now in detail the numerical impact of the singular matching coefficient in Eq.~(\ref{eq:singular_part}) on the $x$-dependence of the light-cone PDF $h_L(x)$.
In Fig.~\ref{fig:matched_HL_options}, we show the matched function $h_L(x)$ at our three nucleon boosts, with three options for the singular coefficient --- including the full $C^{(s)}$, excluding the $R_0(|\xi|)$ term and excluding $C^{(s)}$ altogether.
We note that in the quark part ($x>0$), the influence of the singular terms is smaller than the statistical errors, with positive contribution from the singular terms of increasing magnitude towards smaller $x$.
In the antiquark part ($x<0$), the contributions from the singular terms are larger and exceed statistical errors at $|x|\lesssim0.1$.
However, the numerical influence of the singular matching coefficient is overall insignificant for two reasons: this is not a precision calculation; the effect of the singular terms is basically limited to very small values of $x$, where the lattice-reconstructed distributions suffer anyway from uncontrolled higher-twist effects at the currently attained nucleon boosts.

We also test whether the behavior of the singular coefficient at large $x$ influences the numerical results, i.e.\ whether the cutoff of $|x_c|=2$ plays any role for the singular parts.
To this aim, we consider models 
of the form $\tilde{h}_L(x)=c/(x^2+b)$, as in Eq.~(\ref{e:hLQ_model}), but allowing for both signs of $b,c$.
In the left plot of Fig.~\ref{fig:models}, we show examples of models that we considered, along with the quasi-$h_L$ function from our lattice data at the intermediate nucleon boost.
We show the case where the coefficient $b=0$.
For the quark part, the positive value of $c$ is chosen such that the model matches the lattice $\tilde{h}_L$ function for intermediate $x\approx0.5$, which yields $c=0.3$ for $b=0$.
In turn, for $c<0$, we match the model and the lattice data at $x\approx1$, which gives $c=-0.12$.
The antiquark part is very much suppressed and we only consider positive $c=0.005$ that leads to agreement with the lattice quasi-$h_L$ function for $x\approx-0.2$. 
In the matching procedure, we take the lattice data for $\tilde{h}_L\in(-2,2)$ and outside of this interval, a model is used.
The ensuing matched distributions are shown in the right plot of Fig.~\ref{fig:models}.
We observe that the contribution from the models is much smaller than our statistical uncertainties and there is very little dependence on the form of the model, which holds also for nonzero values of $b$, taken to be up to 0.2.
This implies that the matched PDF is robustly determined by the available lattice data and there are no enhanced contributions from large $x$ via the $C^{(s)}$ part of the matching kernel.
This strengthens our confidence in the obtained results with respect to the role of the singular part of the matching, in the region of applicability of LaMET.

%%%%%%%%%%%%%%%%%%%%%%%%%%%%%%%%%%%%%%%%%%%%%%%%%%%%%%%%%%%%%%%%%%
\section{Final results for $h_{L}(x)$ and the Wandzura-Wilczek approximation}
\label{sec:numerics}
%%%%%%%%%%%%%%%%%%%%%%%%%%%%%%%%%%%%%%%%%%%%%%%%%%%%%%%%%%%%%%%%%%

\begin{figure}[h!]
    \centering
\includegraphics[scale=0.58]{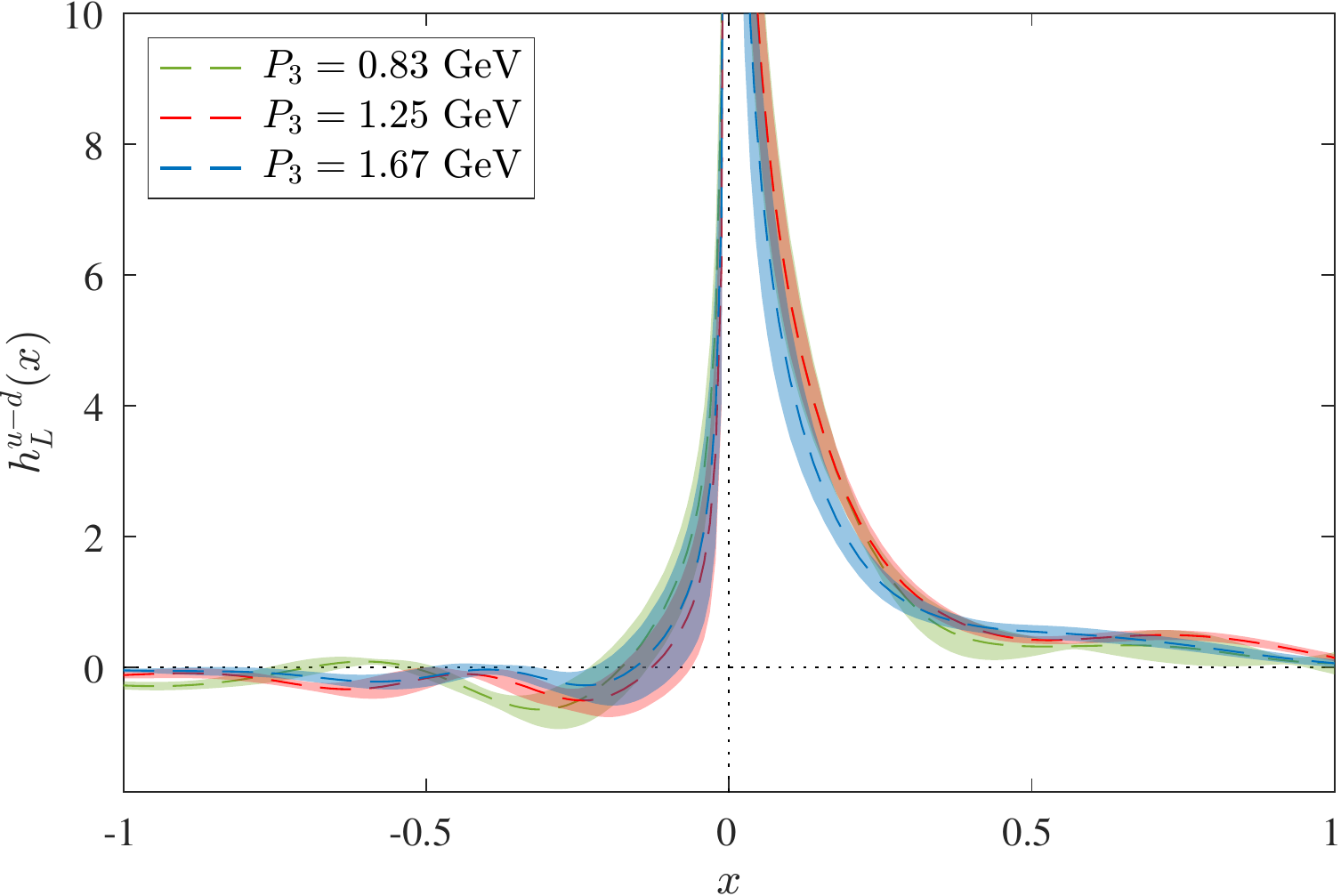}
\caption{Nucleon boost dependence for $h_L(x)$ in the $\MSb$ scheme at a scale of 2 GeV, using $P_3=0.83$~GeV (green curve), $P_3=1.25$~GeV (red curve) and $P_3=1.67$~GeV (blue curve).}
\label{fig:HL_mom_dep}
\end{figure}

In this section, we present the final results on the $x$-dependence of $h_L$ and show how they compare with the lattice extraction of the twist-2 transversity PDF, $h_1(x)$. The renormalized lattice data plotted in Fig.~\ref{fig:ME_R} are Fourier transformed to $x$-space, using the Backus-Gilbert approach, and finally matched to the light-cone $h_L(x)$ distribution using the $\MMSb$ matching kernel of Eq.~(\ref{eq:matching_kernel}). 
We apply this procedure for all values of the nucleon boost, and the results are shown in Fig.~\ref{fig:HL_mom_dep}. Green, red and blue bands correspond to $P_3=0.83$, $1.25$ and $1.67$~GeV, respectively, and include statistical errors as well as systematic uncertainties due to the choice of $z_{max}$ in the Fourier transform (see Eq.~(\ref{eq:syst_err})). 
We observe that for momentum boosts above 1 GeV, the distribution is insensitive to $P_3$ for almost all values of $x$. 
There is, however, a slight tendency of a steeper descent of the quark distribution for $x<0.4$ with increasing momentum.
Moreover, at $P_3=1.25$ and $1.67$~GeV, the distributions vanish at $|x|=1$ and are fully compatible with each other in the antiquark part, within the reported uncertainties.

\begin{figure}[h!]
    \centering
\includegraphics[scale=0.58]{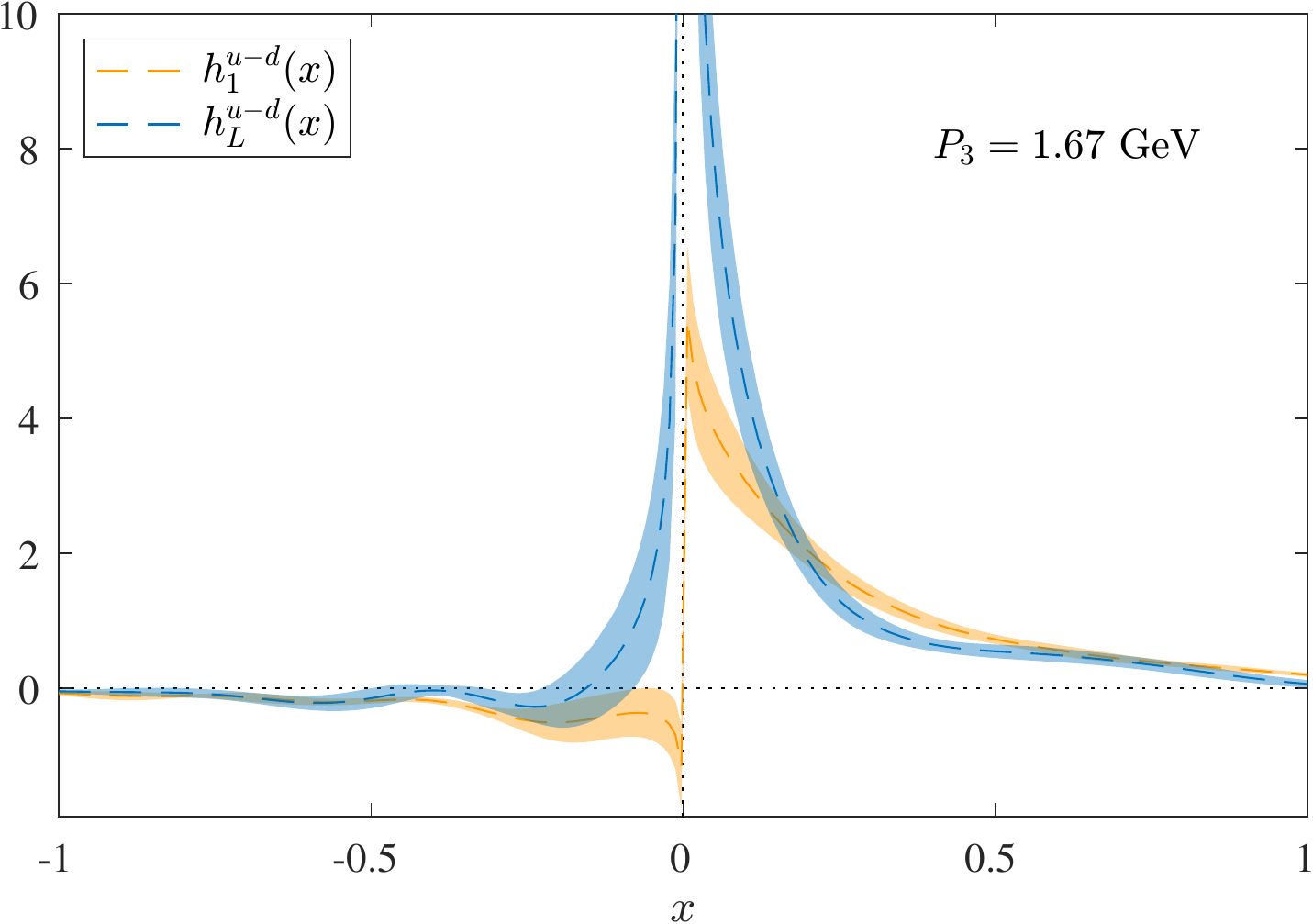}
\caption{Comparison of $x$-dependence of $h_L$ (blue band) and $h_1$ (orange band) for the nucleon boost $P_3=1.67$~GeV.}
\label{fig:HL_H1}
\end{figure}

We note that qualitative comparisons with phenomenological determinations of $h_L(x)$ cannot be included at this stage, as experimental data are not available. In fact, extracting $h_L(x)$ experimentally is complex because it is a chiral-odd function and, in addition, it enters the factorization theorems with a ${\cal O}(1/Q)$ suppression. However, despite the overall kinematical suppression, $h_L(x)$ at a given $x$ may be as sizeable as its twist-2 counterpart, $h_1(x)$. To see how the two compare, we also compute $h_1(x)$ on the same ensemble and with the same values for the momentum boost and source-sink separation. The comparison between the two functions for the largest boost is shown in Fig.~\ref{fig:HL_H1}.
It should be emphasized that the reconstruction of distribution functions in the region $|x| \lesssim 0.15$ is subject to large systematic uncertainties. Outside this region, $h_1(x)$ is dominant only for $0.2 \lesssim x \lesssim 0.5$. For the remaining $x$-values in the quark and antiquark regions, the two distributions are in agreement within uncertainties. 

Our numerical results allow us to also study the Burkhardt-Cottingham-type sum rule for the quasi-PDFs.
Generally, the Burkhardt-Cottingham sum rules~\cite{Burkhardt:1970ti, Tangerman:1994bb, Burkardt:1995ts} relate the integrals of twist-3 PDFs with their twist-2 counterparts. These sum rules have been known for quite some time, and are very useful tools in the qualitative understanding of twist-3 distribution functions. Recently, the equivalent sum rules for the quasi-distributions have been addressed in Ref.~\cite{Bhattacharya:2021boh}. In fact, it was shown that the sum rules also hold for the quasi-PDFs, that is,
\begin{equation}
\int^1_{-1} dx \,\tilde{h}_L(x,P_3) = \int^1_{-1} dx \,\tilde{h}_1(x,P_3) = g_T\,,
\end{equation}
where $g_T$ is the tensor charge\footnote{Not to be confused with the twist-3 PDF $g_T(x)$}. Note that $g_T$ is momentum-boost independent, and therefore, the relation holds for any value of $P_3$. 
We test this equality numerically using the results at all momenta, and we find 
\begin{eqnarray}
\label{eq:BC1}
\int dx\, \tilde{h}_L(x, 0.83\,{\rm GeV})=1.13(08)\,,&\quad&  \int dx\, \tilde{h}_1(x, 0.83\,{\rm GeV})=1.02(07)\,,   \\[1.5ex]
\int dx\, \tilde{h}_L(x, 1.25\,{\rm GeV})=1.09(10)\,,&\quad&  \int dx\, \tilde{h}_1(x, 1.25\,{\rm GeV})=1.07(08)\,,   \\[1.5ex]
\int dx\, \tilde{h}_L(x, 1.67\,{\rm GeV})=1.03(16)\,,&\quad&  \int dx\, \tilde{h}_1(x, 1.67\,{\rm GeV})=0.94(10)   \,.
\label{eq:BC3}
\end{eqnarray}
As can be seen, the sum rule is satisfied at each momentum within errors, as the integrals of $h_1(x)$ and $h_L(x)$ are compatible. Furthermore, we find that the BC sum rule is independent of the momentum, in accordance with the expectations of Ref.~\cite{Bhattacharya:2021boh}. We also note that the values of the tensor charge are within the range of values obtained directly from local operators; see, e.g., Ref.~\cite{PDFLat2020}.

\begin{figure}[h!]
    \centering
    \includegraphics[scale=0.61]{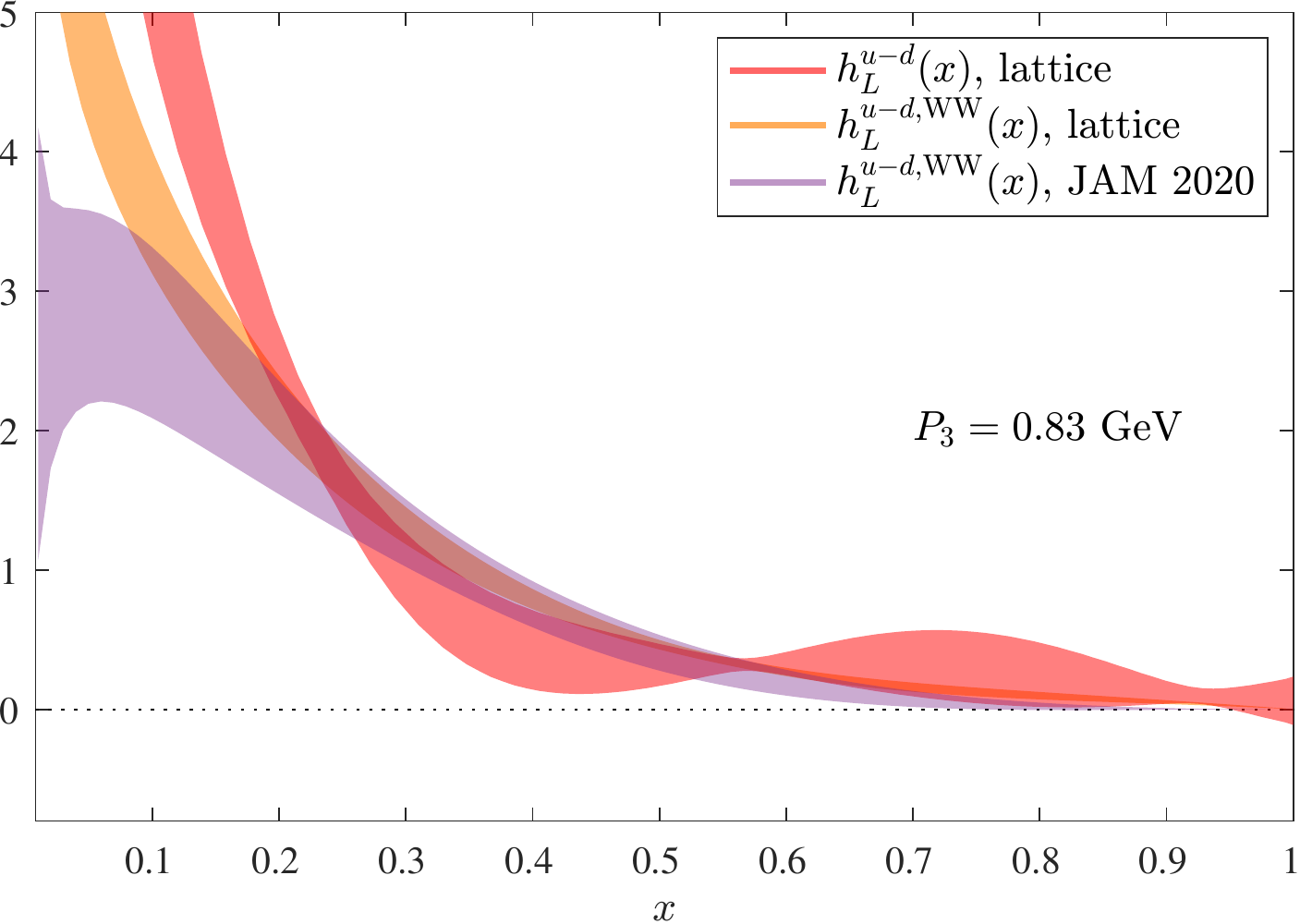}\hspace{0.1cm}
    \includegraphics[scale=0.61]{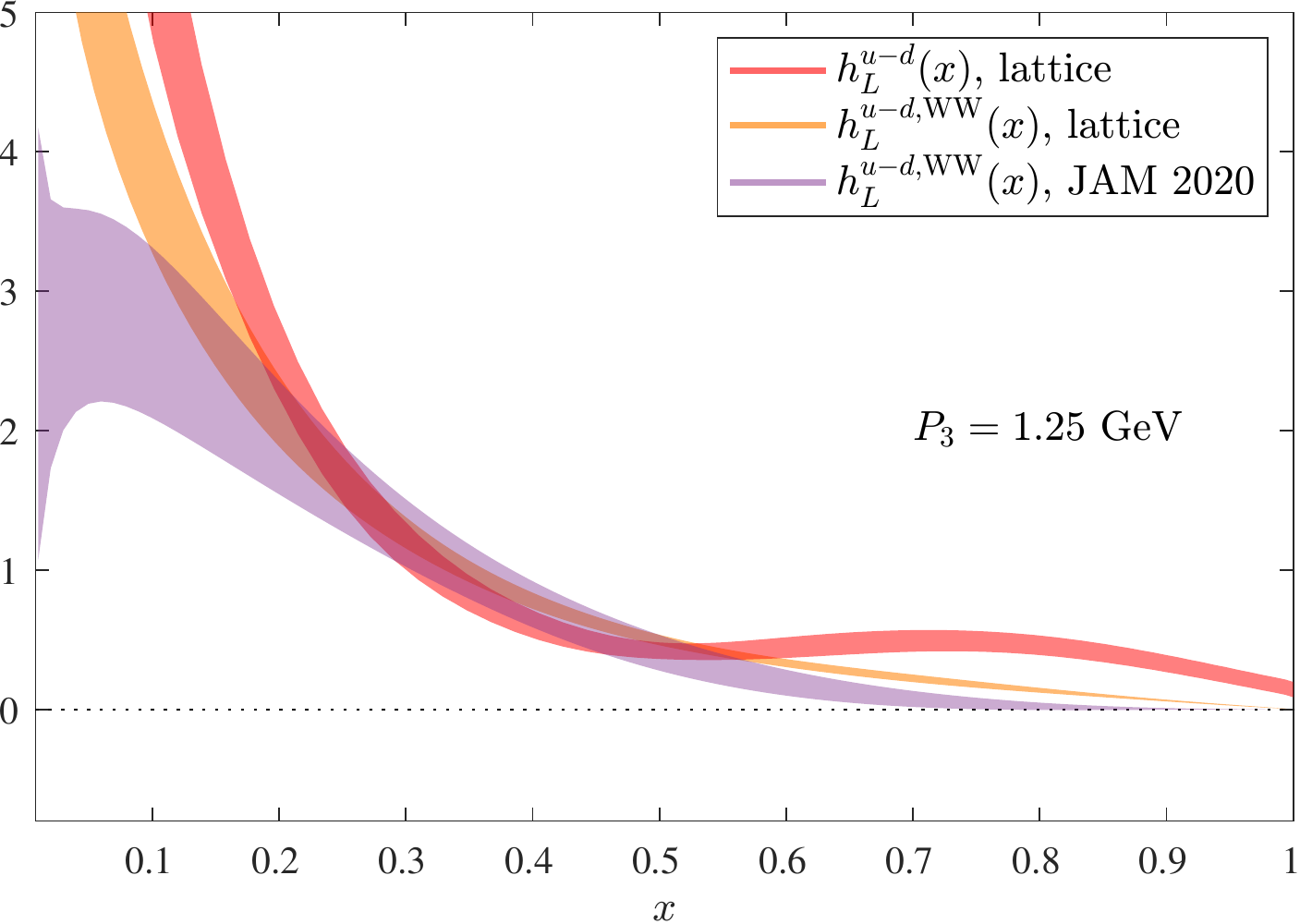}
    \includegraphics[scale=0.61]{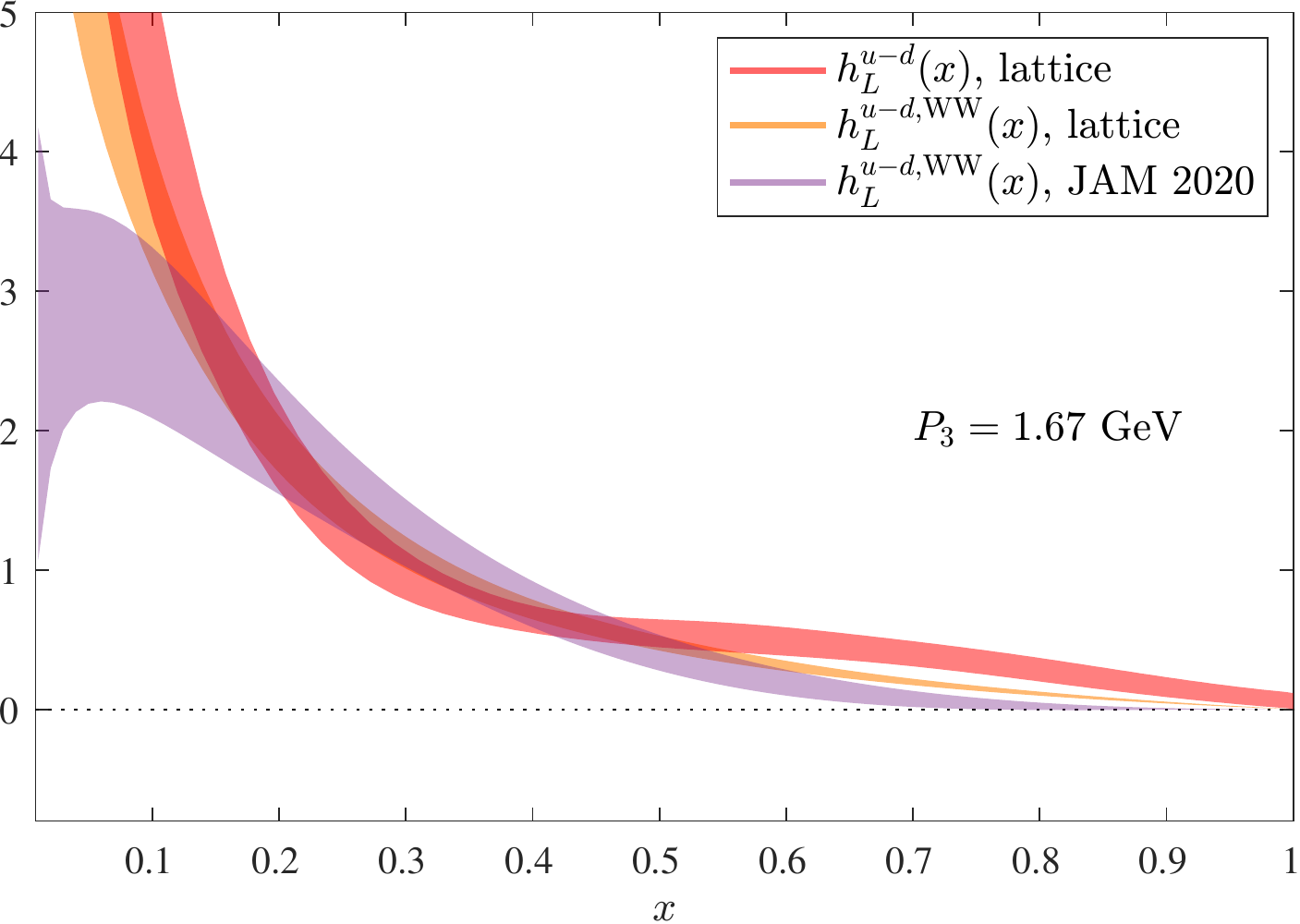}
    \caption{Test of the Wandzura-Wilczek approximation for $h_L(x)$, for the nucleon boosts $P_3=\lbrace 0.83,1.25,1.67\rbrace$~GeV. The lattice estimate of $h_L(x)$ (red band) is compared with its WW-approximation (orange band) extracted on the same gauge ensemble and the one obtained from global fits (violet band) from the JAM collaboration~\cite{Cammarota:2020qcw}.}
    \label{fig:WW_approx}
\end{figure}

The connection between $h_L(x)$ and $h_1(x)$, at a given $x$, can also be studied in more detail by using an analogous relation to the one that was derived by Wandzura and Wilczek for the helicity twist-3 $g_T(x)$ in Ref.~\cite{Wandzura:1977qf}. It is indeed known from Refs.~\cite{Jaffe:1991ra,Jaffe:1991kp} that also the Mellin moments of $h_L(x)$ can be split into twist-2 and twist-3 parts.
More specifically, in terms of distributions one has the relation
\begin{equation}
    h_L(x) = h_L^{\rm WW}(x) + h_L^{\rm twist-3}(x) 
    =  2x\int_x^1 dy \, \frac{h_1(y)}{y^2} + h_L^{\rm twist-3}(x) \,,
    \label{e:WW}
\end{equation}
where $h_L^{\rm twist-3}(x)$ is a genuine twist-3 contribution, which is given by quark-gluon correlations (and a current-quark mass term).
In the Wandzura-Wilczek (WW) approximation, one just keeps the term $h_L^{\rm WW}(x)$, which is determined by the transversity distribution.
It was found in the instanton model of the QCD vacuum that the lowest nontrivial moment of $h_L^{\rm twist-3}(x)$ is numerically very 
small~\cite{Dressler:1999hc}.
While this remains to be tested in experiments,
we explore here the significance of the contribution due to quark-gluon correlations in Eq.~(\ref{e:WW}) as a function of $x$ using our lattice data.
The results are shown in  Fig.~\ref{fig:WW_approx}, where  $h_L(x)$ and $h^{\rm WW}_L(x)$, which is computed through the lattice extracted $h_1(x)$ using Eq.~(\ref{e:WW}), are represented by the red and orange bands, respectively. 
We find that the agreement between $h_L(x)$ and $h^{\rm WW}_L(x)$ extends to a wider range of $x$ as the nucleon boost increases. In particular,  at $P_3=1.67$~GeV the distributions become consistent for $x\lesssim 0.55$. 
Moreover, in the region $0.15 \lesssim x\lesssim 0.55$, our lattice results are also in good agreement with $h^{\rm WW}_L(x)$ obtained from a global fit of the nucleon transversity by the JAM collaboration~\cite{Cammarota:2020qcw} (violet band in Fig.~\ref{fig:WW_approx}). Thus, our numerical findings seem to suggest that $h_L(x)$ could be determined by the twist-2 $h_1(x)$ for a considerable $x$-range. However, for more precise statements further investigations are needed. We repeat that the mixing with quark-gluon-quark operators has not been computed within this work, and other systematic effects have to be addressed as well, like those related to a finite lattice spacing and a non-physical light quark mass. We note that we expect a mild pion mass dependence on $h_1(x)$ and $h_L(x)$, as $h^{\rm WW}_L(x)$ extracted from this ensemble is compatible with the one obtained using simulations at the physical point from Ref.~\cite{Alexandrou:2018eet}. In addition, the tension observed between global fits and lattice data at small and large $x$ reveals that more control may be needed to constrain distributions in these regions.

Using the data at $h_1(x)$ we also extract the isovector tensor charge, for which we find $g_T^{u-d}=0.96(11)$, which is compatible with the value of the matrix element at $z=0$, i.e.\ $0.93(10)$, as well as the integrals of the quasi-PDFs of Eqs.~(\ref{eq:BC1})-(\ref{eq:BC3}). For completeness we provide the value extracted from the JAM collaboration, $g_T^{u-d}=0.87(11)$~\cite{Cammarota:2020qcw}. As can be seen, these values are compatible with each other, as well as other lattice calculations~\cite{PDFLat2020}.
We also remark that we can't obtain reliable numerical results for the lowest moment of $h_L(x)$ for the reasons explained in Sec.~\ref{sec:matching}.

%%%%%%%%%%%%%%%%%%%%%%%%%%%%%%%%%%%%%%%%%%%%%%%%%%%%%%%%%%%%%%%%%%
\section{Flavor decomposition}
\label{sec:up_down_PDFs}
%%%%%%%%%%%%%%%%%%%%%%%%%%%%%%%%%%%%%%%%%%%%%%%%%%%%%%%%%%%%%%%%%%
In the previous sections we focused on the isovector flavor combination $h_L^{u-d}(x)$. Within the same setup, we also extracted the isoscalar combination $h_L^{u+d}(x)$ for the connected diagram. We note that $h_L^{u+d}(x)$ receives a contribution from the disconnected diagram too. Ref.~\cite{Alexandrou:2021oih} reports the disconnected contributions to $h_1^{u+d}(x)$ using the same ensemble as this work. The finding is that the effect is very small for the tensor operator $\sigma^{3j}$ ($j=1,2$). We expect that the same applies for the operator $\sigma^{12}$ entering $h_L(x)$. Therefore, we proceed with the flavor decomposition of the up-quark and down-quark contributions using only the matrix elements extracted from the connected diagram. We also note that there neither exists a gluon transversity nor a twist-3 two-gluon matrix element for a longitudinally polarized target which could mix with $h_L(x)$~\cite{Mulders:2000sh}. Consequently, in the method of Ref.~\cite{Bhattacharya:2020jfj} used here, the one-loop matching kernel for the singlet $h_1(x)$ and $h_L(x)$ is the same as the one for the nonsinglet case. 

\begin{figure}[h!]
    \centering
    \includegraphics[scale=0.61]{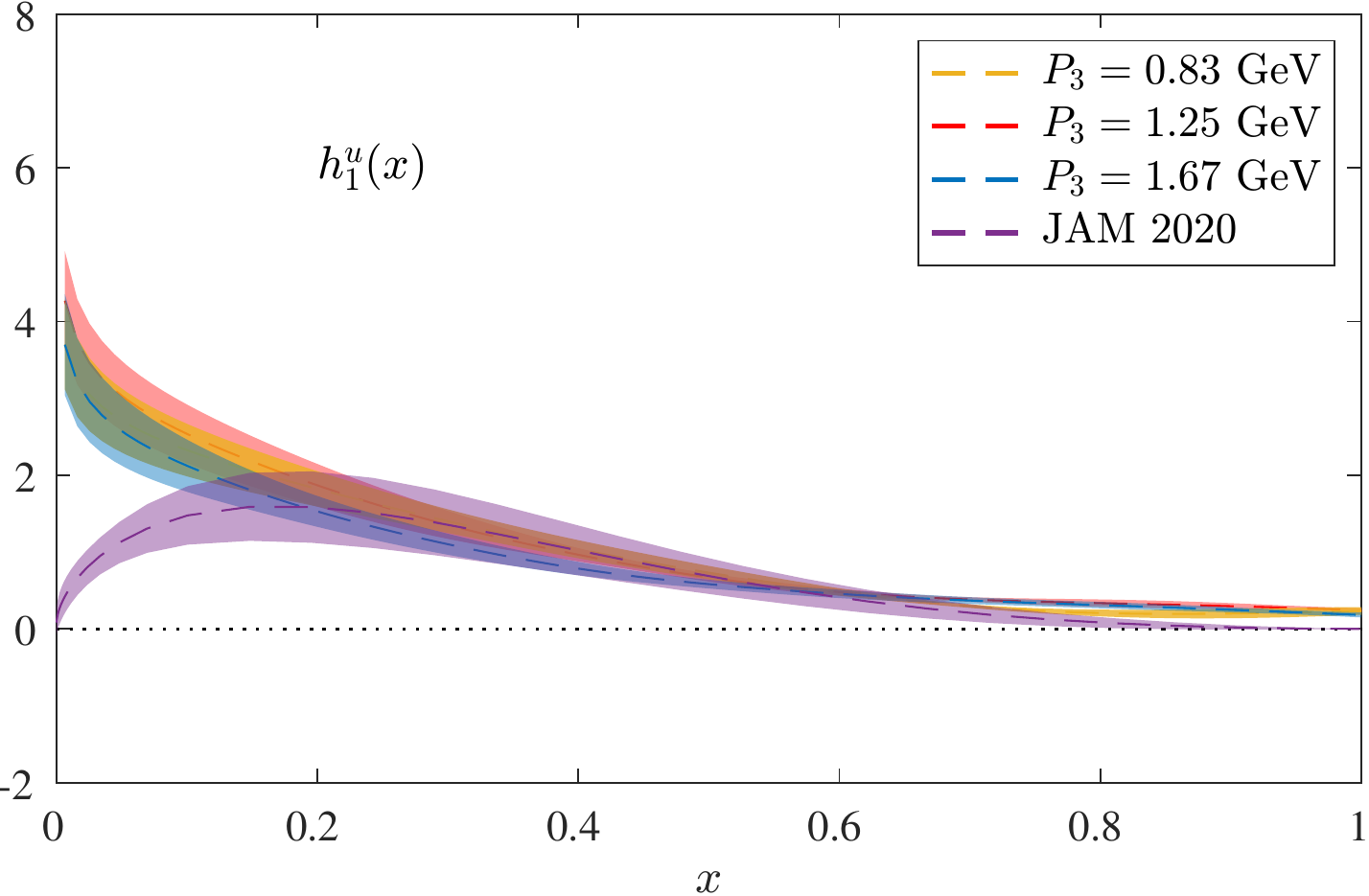}\hspace{0.1cm}
   \includegraphics[scale=0.61]{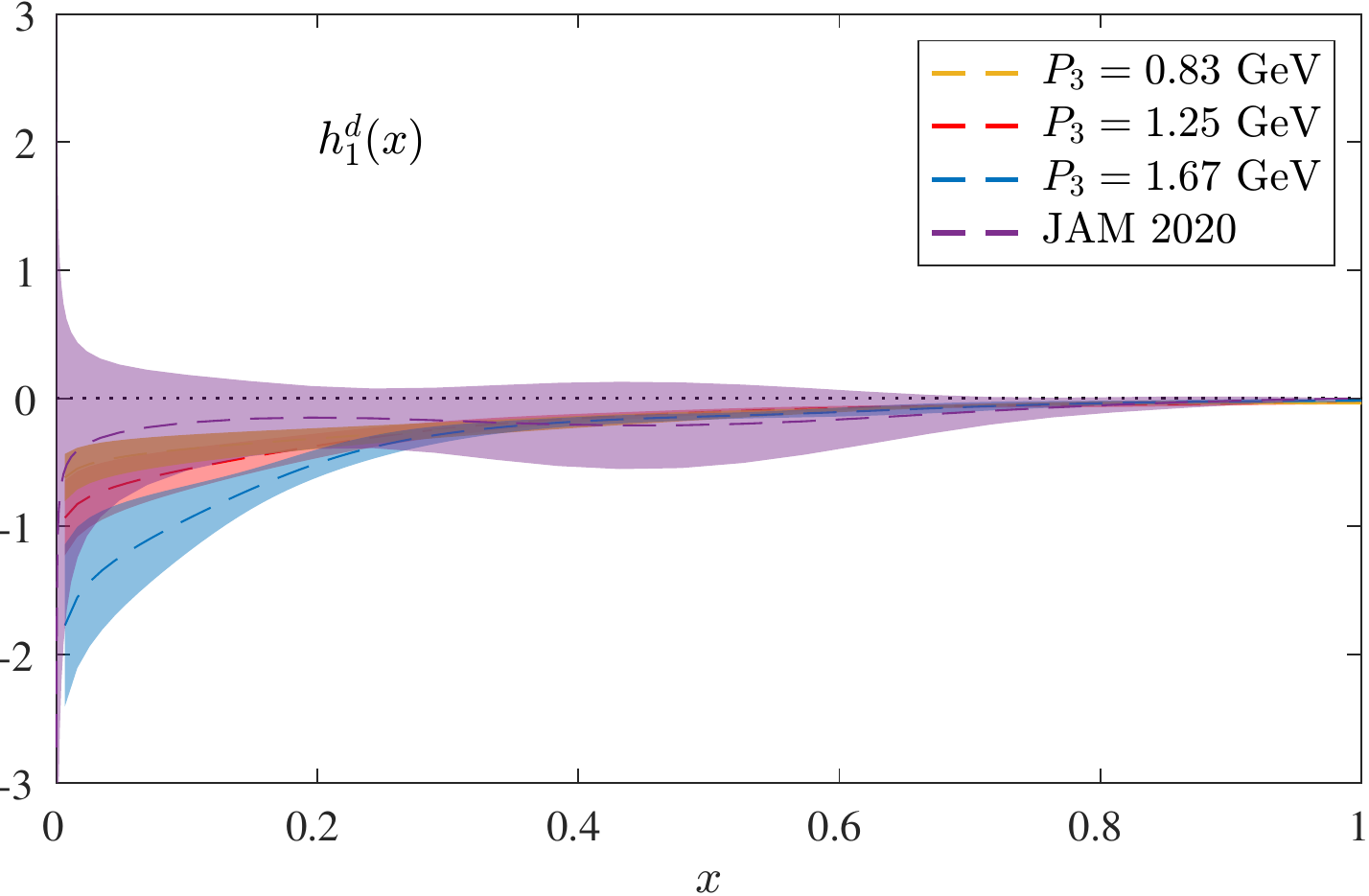} 
   \vspace{-0.25cm}
\caption{Twist-2 transversity $h_1(x)$ for up (left) and down (right) quarks, at nucleon boosts $P_3=0.83$~GeV (yellow), $P_3=1.25$~GeV (red) and $P_3=1.67$~GeV (blue). Results from the JAM collaboration~\cite{Cammarota:2020qcw} are shown with a violet band.}   
\label{h1_u_d}
   \end{figure}

\begin{figure}[h!]
    \centering
    \includegraphics[scale=0.61]{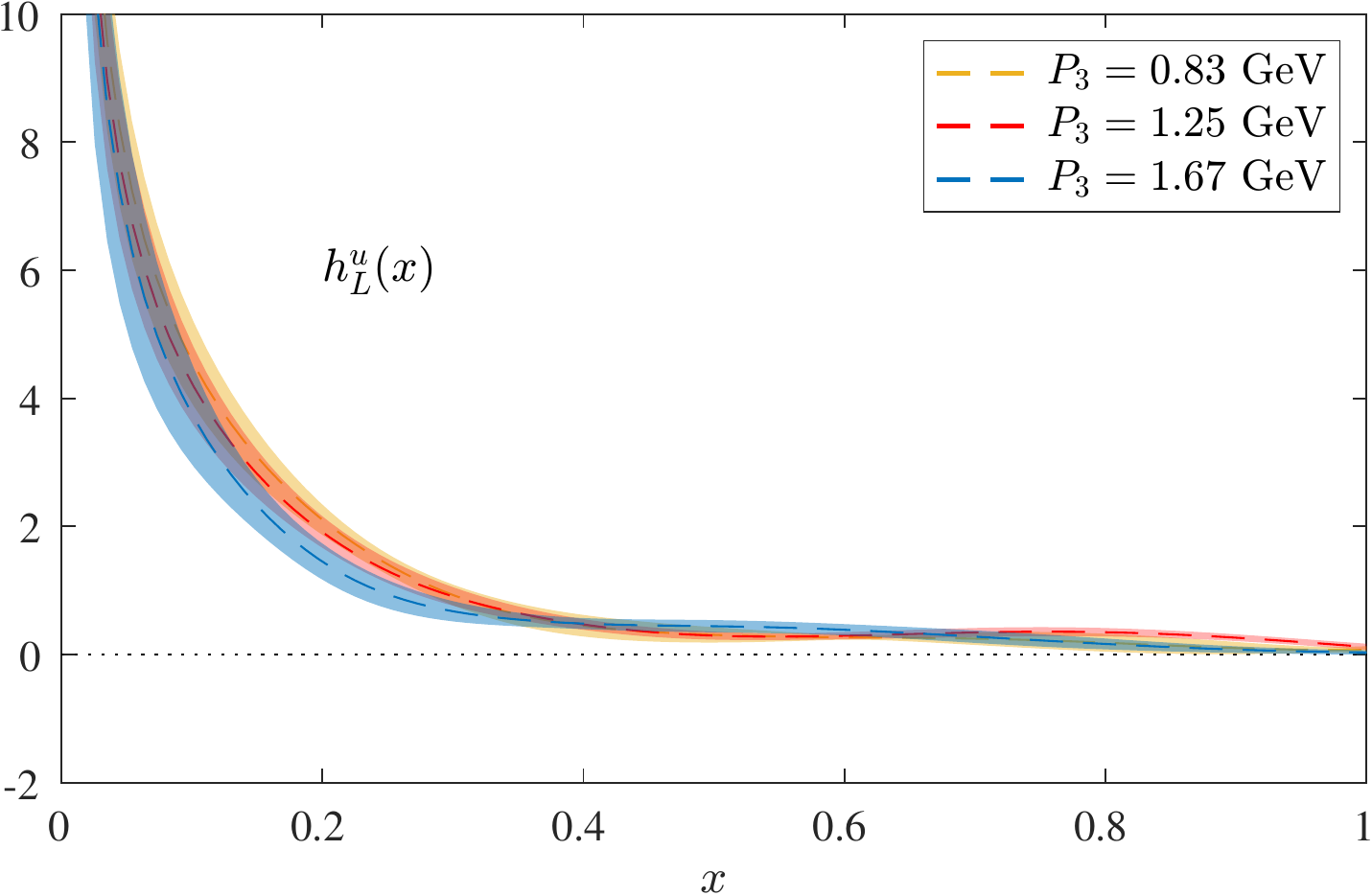}\hspace{0.1cm}
   \includegraphics[scale=0.61]{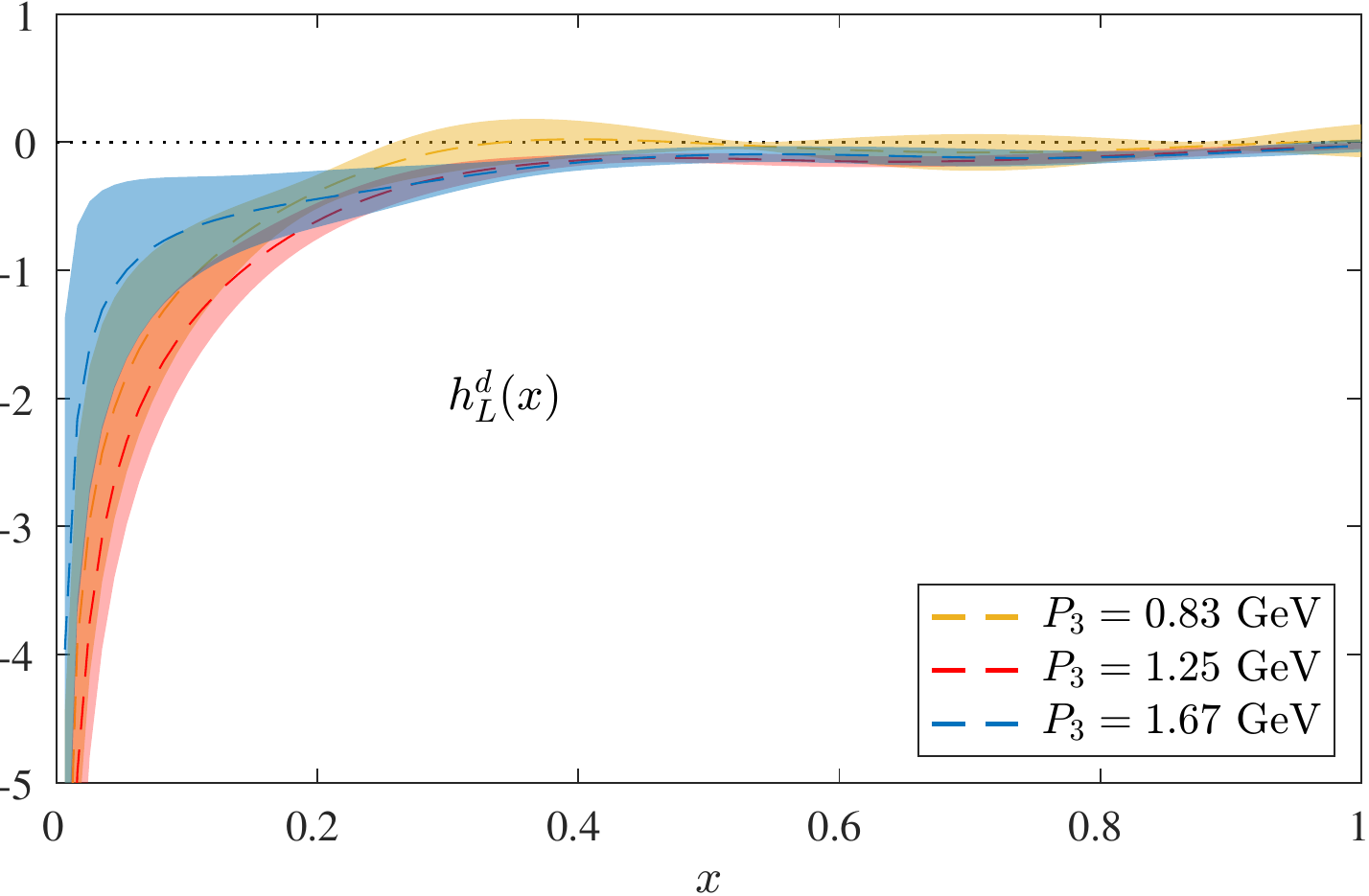} 
   \vspace{-0.25cm}
\caption{Twist-3 $h_L(x)$ for up (left) and down (right) quarks, at nucleon boosts $P_3=0.83$~GeV (yellow), $P_3=1.25$~GeV (red) and $P_3=1.67$~GeV (blue).}   
\label{hL_u_d}
   \end{figure}

Here, we focus on the individual quark contributions to $h_L(x)$ obtained from the isoscalar and isovector combination of connected contributions. In this discussion, we do not consider the antiquark contribution, as its extraction is sensitive to systematic effects~\cite{Cichy:2018mum}, and is suppressed compared to the quark part. For completeness, we present the momentum dependence for each flavor and for both $h_1(x)$ (Fig.~\ref{h1_u_d}) and $h_L(x)$ (Fig.~\ref{hL_u_d}).

The momentum dependence of $h^u_1(x)$ is very small for the three momenta we use in this work. The JAM20 data~\cite{Cammarota:2020qcw} show agreement with the lattice data in the region between $x=0.15$ and $x=0.7$. The down-quark transversity shows convergence between 1.25 GeV and 1.67 GeV, with reduced overlap in the region below $x=0.2$. For the case of $h_1^d(x)$ we find that 0.83 GeV is not large enough to achieve convergence. Unlike the case of $h_1(x)$, both $h_L^u(x)$ and $h_L^d(x)$ show convergence for all momenta. 

\begin{figure}[h!]
    \centering
    \includegraphics[scale=0.6]{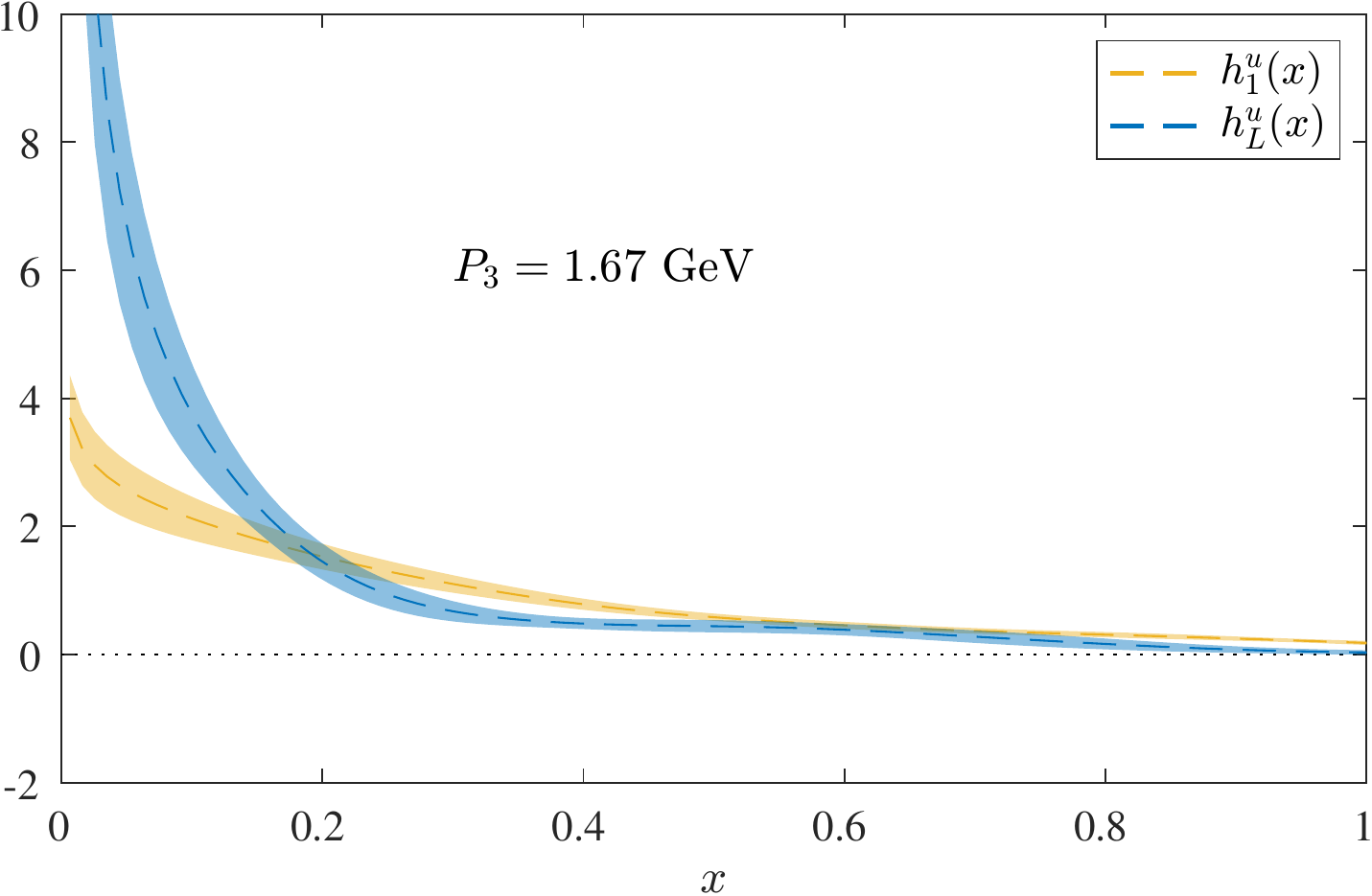}\hspace{0.1cm}
   \includegraphics[scale=0.6]{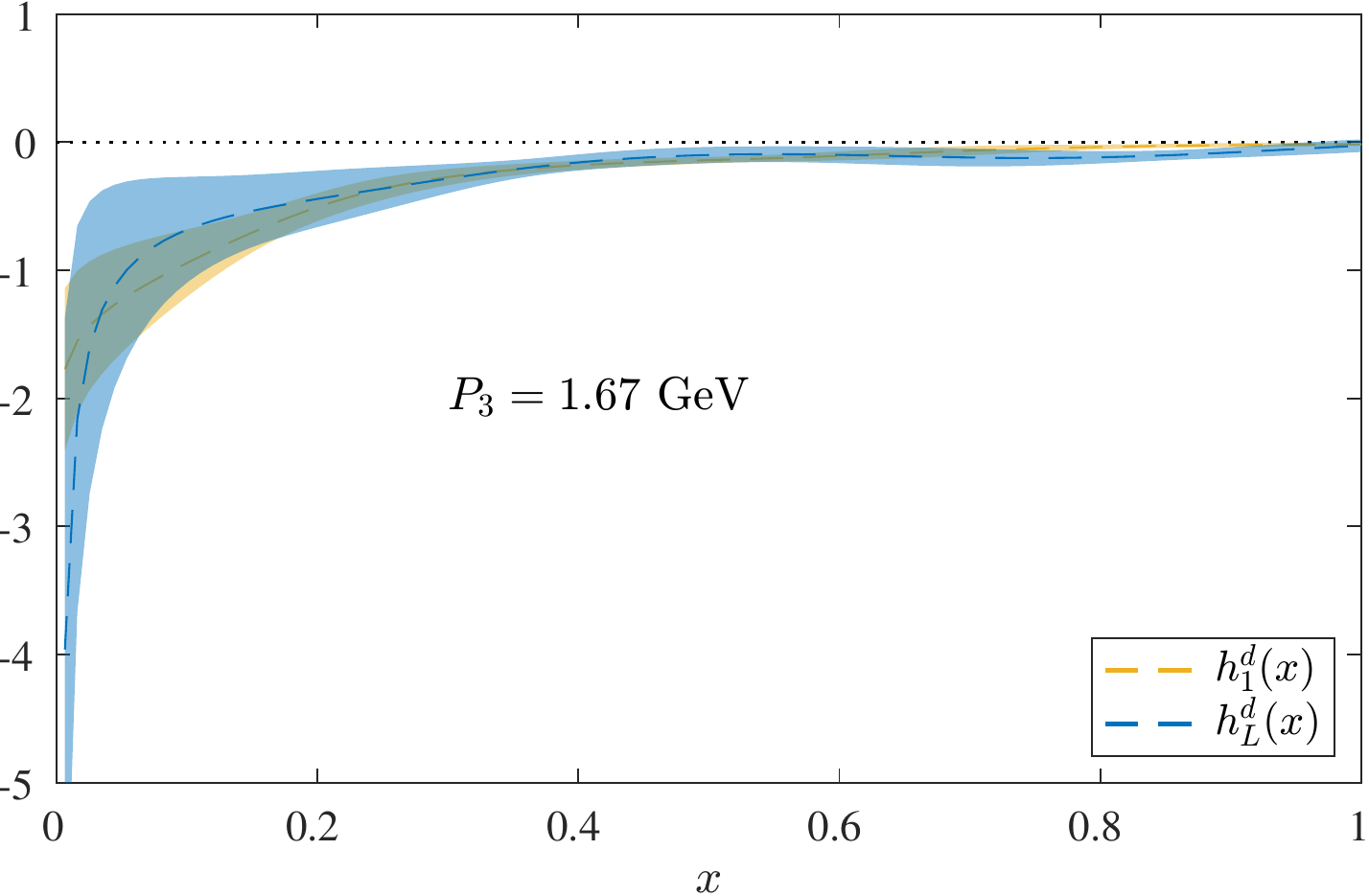} 
\caption{$x$-dependence of $h_1$ (yellow) and $h_L$ (blue) for up-quarks and down-quarks in the left and right plot, respectively. Results are shown at the largest nucleon boost, $P_3=1.67$~GeV.} 
\label{h1_hL_u_d}
   \end{figure}

An important question is about the role of the up- and down-quark in the proton. Furthermore, one may ask about the role of the quarks in twist-2 and twist-3 PDFs. To this end, we compare the individual-quark contributions to $h_1(x)$ and $h_L(x)$ in Fig.~\ref{h1_hL_u_d} using the lattice data at the momentum 1.67 GeV. There are a number of qualitative conclusions that one can draw. First, the up-quark is dominant in all regions of $x$, but the dominance is more apparent for $x<0.5$. While both $h_1^u(x)$ and $h_1^d(x)$ approach zero in the large-$x$ limit, $h_1^u(x)$ is typically twice larger. Second, similar conclusions are drawn in the comparison between $h_L^u(x)$ and $h_L^d(x)$, with the former being dominant. Third, the down-quark plays a similar role in $h_1(x)$ and $h_L(x)$ for all regions of $x$. In the case of the up-quark, we find similar magnitude between $h_1^u(x)$ and $h_L^u(x)$ for $x>0.2$. Forth and last, the statistical uncertainties are larger for the down-quark contributions. 
   
\begin{figure}[h!]
    \centering
    \includegraphics[scale=0.6]{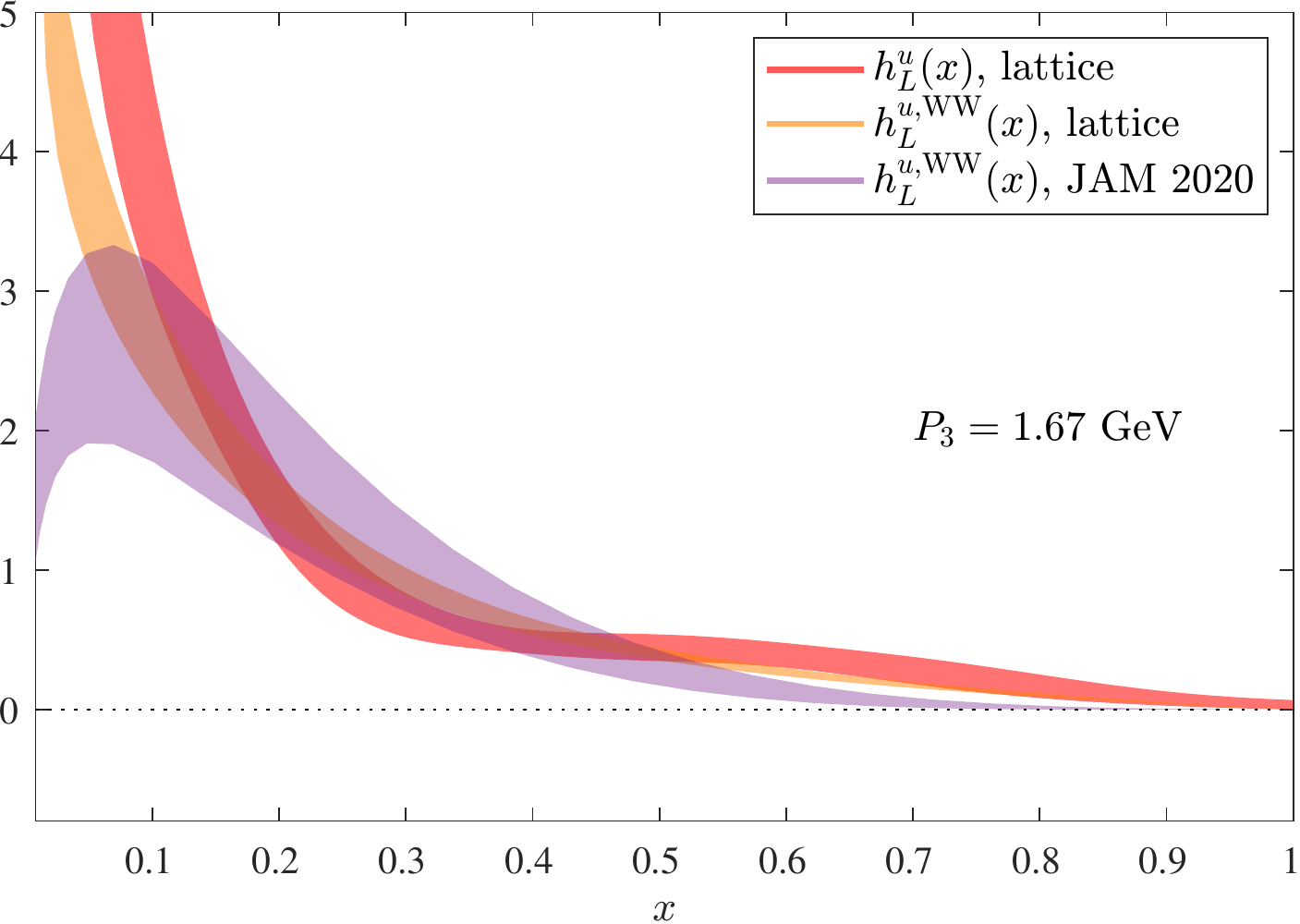}\hspace{0.1cm}
   \includegraphics[scale=0.6]{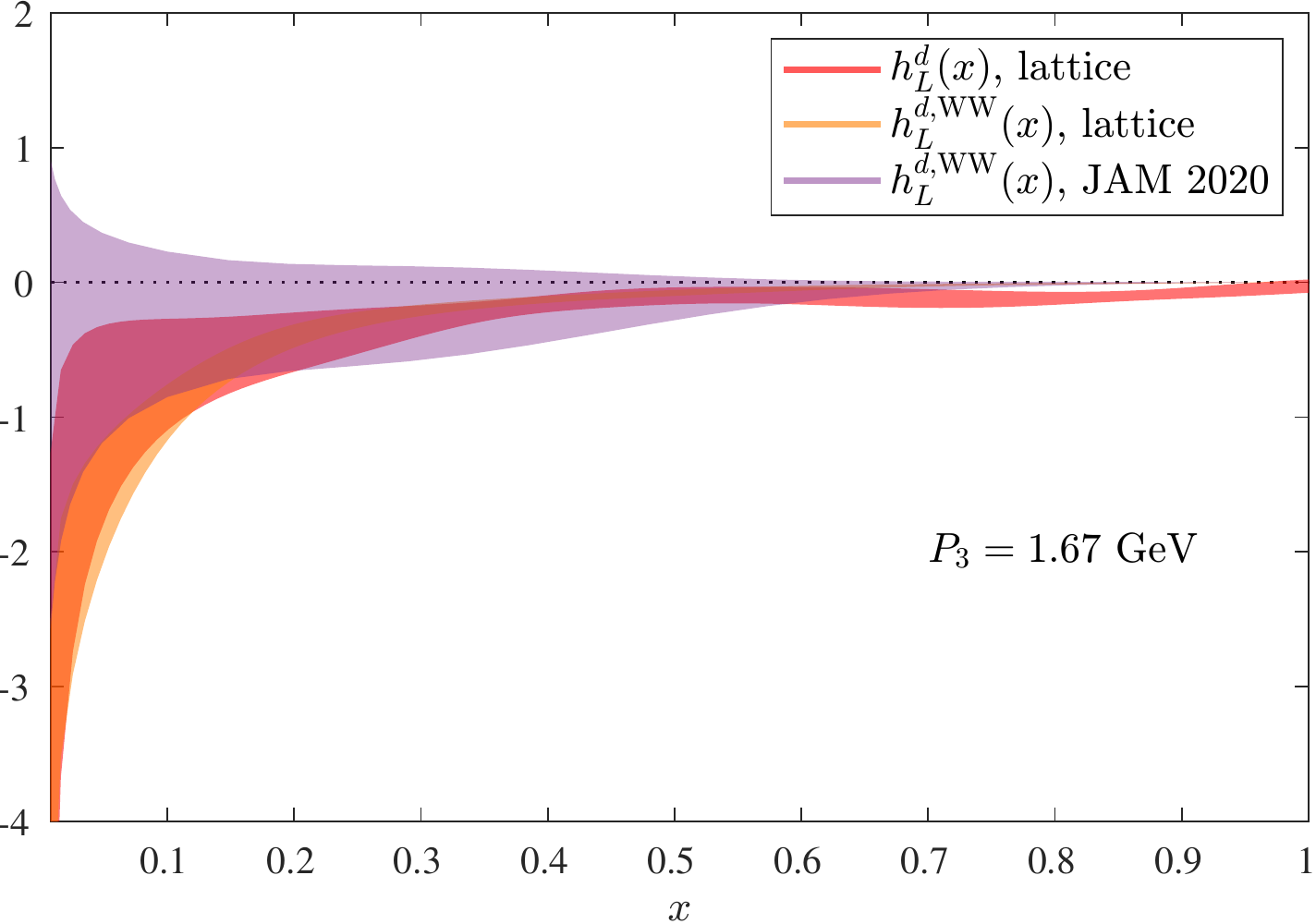} 
\caption{Test of the Wandzura-Wilczek approximation for up (left panel) and down (right panel) distributions, at the largest boost $P_3=1.67$~GeV. For the separate flavors we show $h_L(x)$ (red) with $h_L^{\rm WW}(x)$ (orange)  extracted from lattice QCD within this work. Results for $h_L^{\rm WW}(x)$ from the JAM collaboration~\cite{Cammarota:2020qcw} (violet) are also included for comparison.}   
\label{WW_u_d}
   \end{figure}

Finally, in Fig.~\ref{WW_u_d} we examine the WW approximation for each quark flavor. While the isovector flavor combination for $P_3=1.67$ GeV of Fig.~\ref{fig:WW_approx} shows an agreement within uncertainties for $x<0.55$, here we find a discrepancy for this region for the up-quark, even though $h_L^u(x)$ and $h_L^{u,\rm WW}(x)$ cross at around $x=0.2$. On the other hand, we find compatibility for the down-quark contributions up to $x=0.7$. The comparison of $h_L^{u,\rm WW}(x)$ obtained from the lattice data and JAM20 has the same features as in Fig.~\ref{h1_u_d}.

We emphasize once again that the data presented in this section neglect contributions from the disconnected diagram. However, as argued previously, these are expected to be within the reported uncertainties for these quantities~\cite{Alexandrou:2021oih}. 
%%%%%%%%%%%%%%%%%%%%%%%%%%%%%%%%%%%%%%%%%%%%%%%%%%%%%%%%%%%%%%%%%%

%%%%%%%%%%%%%%%%%%%%%%%%%%%%%%%%%%%%%%%%%%%%%%%%%%%%%%%%%%%%%%%%%%
\section{Summary}
\label{sec:summary}
%%%%%%%%%%%%%%%%%%%%%%%%%%%%%%%%%%%%%%%%%%%%%%%%%%%%%%%%%%%%%%%%%%

We report a pioneering lattice-QCD calculation of the chiral-odd twist-3 PDF $h_L(x)$ for the proton by making use of the quasi-PDF approach.
The lattice ensemble used in this work corresponds to a pion mass of 260~MeV.
In order to reconstruct the quasi-PDFs from the lattice results for the pertinent matrix elements in position space we used the Backus-Gilbert method. 
We performed the calculation for three different proton momenta, $P_3=0.83,\,1.25,\,1.67$ GeV, and generally observe a very good convergence of the final results for the PDFs as $P_3$ increases. 
While our main focus was on the isovector combination $h_L^{u - d}(x)$, we also obtain results for the isoscalar combination $h_L^{u + d}(x)$ by neglecting contributions from the disconnected diagram which are expected to be small~\cite{Alexandrou:2021oih}.
In order to relate the quasi-PDFs to the light-cone PDFs of interest we use the one-loop results for the matching coefficient of Ref.~\cite{Bhattacharya:2020jfj}, which represents an approximation in that quark-gluon correlations are not taken into account.
Furthermore, we compute both the isovector and isoscalar twist-2 transversity.
This allowed us, in particular, to explore the WW-approximation for $h_L$ which is determined through the transversity.
We also find that the Burkhardt-Cottingham-type sum rule for the quasi-PDFs is satisfied for all three proton momenta, which can be considered a consistency check of the numerics.
The tensor charge, which we computed for the quasi-PDFs and the light-cone PDF $h_1(x)$, agrees within errors with other lattice calculations and an extraction from experimental data by the JAM collaboration~\cite{Cammarota:2020qcw}.

It is well known that, generally, higher-twist PDFs can be as large as twist-2 PDFs.
This is indeed what we find when comparing $h_L^{u-d}(x)$ with the isovector transversity $h_1^{u-d}(x)$. 
More precisely, at intermediate values of $x$ around 0.4, the transversity is somewhat larger than $h_L^{u-d}(x)$, but the latter rises more rapidly towards smaller values of $x$ and exceeds the transversity for $x \lesssim 0.15$.
We also find that $h_L^u(x)$ for quarks is positive and $h_L^d(x)$ negative, like is the case for $h_1^u(x)$ and $h_1^d(x)$.
In the $x$-range $[0.1,0.5]$, for which we expect the systematics of the lattice data to be smallest, there is little difference between $h_L(x)$ and the WW-approximation $h_L^{\rm WW}(x)$ when considering for the latter both our lattice data and results from the JAM collaboration~\cite{Cammarota:2020qcw}. (An exception is $h_L^u(x)$ in the region around $x \sim 0.3$ where it noticeably differs from its WW approximation.) 
This finding is compatible with a result obtained in the instanton model of the QCD vacuum according to which the lowest nontrivial moment of the pure twist-3 term that breaks the WW-approximation is very small~\cite{Dressler:1999hc}.
However, we emphasize that it is too early for drawing a definite conclusion about the quality of the WW-approximation in the case of $h_L(x)$ as several aspects of our lattice calculation can be improved.
Apart from the (usual) sources of systematic errors of the lattice calculation, such as contamination due to excited states, errors in the reconstruction of the $x$-dependent quasi-PDFs, finite volume and discretization effects, and uncertainties from computing at unphysical quark masses, we want to mention again the approximation we used for the matching where quark-gluon correlations are neglected.
We plan to revisit the numerics once matching results along the lines of Ref.~\cite{Braun:2021aon} become available for $h_L(x)$.
We point out that a fully consistent calculation of $h_L(x)$ would also require lattice results for quark-gluon-quark correlations that depend on the parton momentum fractions.
While this requires a long-term dedicated program, efforts along those lines seem worthwhile given the importance of the topic and the unique opportunities for lattice QCD in view of the challenges to extract higher-twist PDFs from experimental data.

\begin{acknowledgements}
We thank D.~Pitonyak for providing us with the results for the transversity distribution from Ref.~\cite{Cammarota:2020qcw}.
The work of S.B.~and A.M.~has been supported by the National Science Foundation under grant number PHY-1812359.  A.M.~has also been supported by the U.S. Department of Energy, Office of Science, Office of Nuclear Physics, within the framework of the TMD Topical Collaboration. K.C.\ is supported by the National Science Centre (Poland) grant SONATA BIS no.\ 2016/22/E/ST2/00013. M.C.~and A.S.~acknowledge financial support by the U.S. Department of Energy, Office of Nuclear Physics, Early Career Award under Grant No.\ DE-SC0020405. F.S.\ was funded by by the NSFC and the Deutsche Forschungsgemeinschaft (DFG, German Research
Foundation) through the funds provided to the Sino-German Collaborative Research Center TRR110 “Symmetries and the Emergence of Structure in QCD” (NSFC Grant No. 12070131001, DFG Project-ID 196253076 - TRR 110). Computations for this work were carried out in part on facilities of the USQCD Collaboration, which are funded by the Office of Science of the U.S. Department of Energy. 
This research was supported in part by PLGrid Infrastructure (Prometheus supercomputer at AGH Cyfronet in Cracow).
Computations were also partially performed at the Poznan Supercomputing and Networking Center (Eagle supercomputer), the Interdisciplinary Centre for Mathematical and Computational Modelling of the Warsaw University (Okeanos supercomputer), and at the Academic Computer Centre in Gda\'nsk (Tryton supercomputer). The gauge configurations have been generated by the Extended Twisted Mass Collaboration on the KNL (A2) Partition of Marconi at CINECA, through the Prace project Pra13\_3304 ``SIMPHYS".
Inversions were performed using the DD-$\alpha$AMG solver~\cite{Frommer:2013fsa} with twisted mass
  support~\cite{Alexandrou:2016izb}.
\end{acknowledgements}

%\appendix

\bibliography{references}

\begin{thebibliography}{113}
\expandafter\ifx\csname natexlab\endcsname\relax\def\natexlab#1{#1}\fi
\expandafter\ifx\csname bibnamefont\endcsname\relax
  \def\bibnamefont#1{#1}\fi
\expandafter\ifx\csname bibfnamefont\endcsname\relax
  \def\bibfnamefont#1{#1}\fi
\expandafter\ifx\csname citenamefont\endcsname\relax
  \def\citenamefont#1{#1}\fi
\expandafter\ifx\csname url\endcsname\relax
  \def\url#1{\texttt{#1}}\fi
\expandafter\ifx\csname urlprefix\endcsname\relax\def\urlprefix{URL }\fi
\providecommand{\bibinfo}[2]{#2}
\providecommand{\eprint}[2][]{\url{#2}}

\bibitem[{\citenamefont{Collins and Soper}(1982)}]{Collins:1981uw}
\bibinfo{author}{\bibfnamefont{J.~C.} \bibnamefont{Collins}} \bibnamefont{and}
  \bibinfo{author}{\bibfnamefont{D.~E.} \bibnamefont{Soper}},
  \bibinfo{journal}{Nucl.\ Phys.\ B} \textbf{\bibinfo{volume}{194}},
  \bibinfo{pages}{445} (\bibinfo{year}{1982}).

\bibitem[{\citenamefont{Collins}(2011)}]{Collins:2011zzd}
\bibinfo{author}{\bibfnamefont{J.}~\bibnamefont{Collins}},
  \bibinfo{journal}{Camb. Monogr. Part. Phys. Nucl. Phys. Cosmol.}
  \textbf{\bibinfo{volume}{32}}, \bibinfo{pages}{1} (\bibinfo{year}{2011}).

\bibitem[{\citenamefont{Collins et~al.}(1989)\citenamefont{Collins, Soper, and
  Sterman}}]{Collins:1989gx}
\bibinfo{author}{\bibfnamefont{J.~C.} \bibnamefont{Collins}},
  \bibinfo{author}{\bibfnamefont{D.~E.} \bibnamefont{Soper}}, \bibnamefont{and}
  \bibinfo{author}{\bibfnamefont{G.~F.} \bibnamefont{Sterman}},
  \emph{\bibinfo{title}{{Factorization of Hard Processes in QCD}}}
  (\bibinfo{year}{1989}), vol.~\bibinfo{volume}{5}, pp. \bibinfo{pages}{1--91},
  \eprint{hep-ph/0409313}.

\bibitem[{\citenamefont{Jaffe}(1996)}]{Jaffe:1996zw}
\bibinfo{author}{\bibfnamefont{R.~L.} \bibnamefont{Jaffe}}, in
  \emph{\bibinfo{booktitle}{{The spin structure of the nucleon (1995)}}}
  (\bibinfo{year}{1996}), pp. \bibinfo{pages}{42--129},
  \eprint{hep-ph/9602236}.

\bibitem[{\citenamefont{Balitsky and Braun}(1989)}]{Balitsky:1987bk}
\bibinfo{author}{\bibfnamefont{I.}~\bibnamefont{Balitsky}} \bibnamefont{and}
  \bibinfo{author}{\bibfnamefont{V.~M.} \bibnamefont{Braun}},
  \bibinfo{journal}{Nucl.\ Phys.\ B} \textbf{\bibinfo{volume}{311}},
  \bibinfo{pages}{541} (\bibinfo{year}{1989}).

\bibitem[{\citenamefont{Kanazawa et~al.}(2016)\citenamefont{Kanazawa, Koike,
  Metz, Pitonyak, and Schlegel}}]{Kanazawa:2015ajw}
\bibinfo{author}{\bibfnamefont{K.}~\bibnamefont{Kanazawa}},
  \bibinfo{author}{\bibfnamefont{Y.}~\bibnamefont{Koike}},
  \bibinfo{author}{\bibfnamefont{A.}~\bibnamefont{Metz}},
  \bibinfo{author}{\bibfnamefont{D.}~\bibnamefont{Pitonyak}}, \bibnamefont{and}
  \bibinfo{author}{\bibfnamefont{M.}~\bibnamefont{Schlegel}},
  \bibinfo{journal}{Phys.\ Rev.\ D} \textbf{\bibinfo{volume}{93}},
  \bibinfo{pages}{054024} (\bibinfo{year}{2016}), \eprint{1512.07233}.

\bibitem[{\citenamefont{Boer et~al.}(2003)\citenamefont{Boer, Mulders, and
  Pijlman}}]{Boer:2003cm}
\bibinfo{author}{\bibfnamefont{D.}~\bibnamefont{Boer}},
  \bibinfo{author}{\bibfnamefont{P.~J.} \bibnamefont{Mulders}},
  \bibnamefont{and} \bibinfo{author}{\bibfnamefont{F.}~\bibnamefont{Pijlman}},
  \bibinfo{journal}{Nucl. Phys. B} \textbf{\bibinfo{volume}{667}},
  \bibinfo{pages}{201} (\bibinfo{year}{2003}), \eprint{hep-ph/0303034}.

\bibitem[{\citenamefont{Accardi et~al.}(2009)\citenamefont{Accardi, Bacchetta,
  Melnitchouk, and Schlegel}}]{Accardi:2009au}
\bibinfo{author}{\bibfnamefont{A.}~\bibnamefont{Accardi}},
  \bibinfo{author}{\bibfnamefont{A.}~\bibnamefont{Bacchetta}},
  \bibinfo{author}{\bibfnamefont{W.}~\bibnamefont{Melnitchouk}},
  \bibnamefont{and} \bibinfo{author}{\bibfnamefont{M.}~\bibnamefont{Schlegel}},
  \bibinfo{journal}{JHEP} \textbf{\bibinfo{volume}{11}}, \bibinfo{pages}{093}
  (\bibinfo{year}{2009}), \eprint{0907.2942}.

\bibitem[{\citenamefont{Gamberg et~al.}(2018)\citenamefont{Gamberg, Metz,
  Pitonyak, and Prokudin}}]{Gamberg:2017jha}
\bibinfo{author}{\bibfnamefont{L.}~\bibnamefont{Gamberg}},
  \bibinfo{author}{\bibfnamefont{A.}~\bibnamefont{Metz}},
  \bibinfo{author}{\bibfnamefont{D.}~\bibnamefont{Pitonyak}}, \bibnamefont{and}
  \bibinfo{author}{\bibfnamefont{A.}~\bibnamefont{Prokudin}},
  \bibinfo{journal}{Phys. Lett. B} \textbf{\bibinfo{volume}{781}},
  \bibinfo{pages}{443} (\bibinfo{year}{2018}), \eprint{1712.08116}.

\bibitem[{\citenamefont{Cammarota et~al.}(2020)\citenamefont{Cammarota,
  Gamberg, Kang, Miller, Pitonyak, Prokudin, Rogers, and
  Sato}}]{Cammarota:2020qcw}
\bibinfo{author}{\bibfnamefont{J.}~\bibnamefont{Cammarota}},
  \bibinfo{author}{\bibfnamefont{L.}~\bibnamefont{Gamberg}},
  \bibinfo{author}{\bibfnamefont{Z.-B.} \bibnamefont{Kang}},
  \bibinfo{author}{\bibfnamefont{J.~A.} \bibnamefont{Miller}},
  \bibinfo{author}{\bibfnamefont{D.}~\bibnamefont{Pitonyak}},
  \bibinfo{author}{\bibfnamefont{A.}~\bibnamefont{Prokudin}},
  \bibinfo{author}{\bibfnamefont{T.~C.} \bibnamefont{Rogers}},
  \bibnamefont{and} \bibinfo{author}{\bibfnamefont{N.}~\bibnamefont{Sato}}
  (\bibinfo{collaboration}{Jefferson Lab Angular Momentum}),
  \bibinfo{journal}{Phys. Rev. D} \textbf{\bibinfo{volume}{102}},
  \bibinfo{pages}{054002} (\bibinfo{year}{2020}), \eprint{2002.08384}.

\bibitem[{\citenamefont{Burkardt}(2013)}]{Burkardt:2008ps}
\bibinfo{author}{\bibfnamefont{M.}~\bibnamefont{Burkardt}},
  \bibinfo{journal}{Phys. Rev.} \textbf{\bibinfo{volume}{D88}},
  \bibinfo{pages}{114502} (\bibinfo{year}{2013}), \eprint{0810.3589}.

\bibitem[{\citenamefont{Jaffe and Ji}(1991)}]{Jaffe:1991kp}
\bibinfo{author}{\bibfnamefont{R.~L.} \bibnamefont{Jaffe}} \bibnamefont{and}
  \bibinfo{author}{\bibfnamefont{X.-D.} \bibnamefont{Ji}},
  \bibinfo{journal}{Phys. Rev. Lett.} \textbf{\bibinfo{volume}{67}},
  \bibinfo{pages}{552} (\bibinfo{year}{1991}).

\bibitem[{\citenamefont{Jaffe and Ji}(1992)}]{Jaffe:1991ra}
\bibinfo{author}{\bibfnamefont{R.~L.} \bibnamefont{Jaffe}} \bibnamefont{and}
  \bibinfo{author}{\bibfnamefont{X.-D.} \bibnamefont{Ji}},
  \bibinfo{journal}{Nucl. Phys. B} \textbf{\bibinfo{volume}{375}},
  \bibinfo{pages}{527} (\bibinfo{year}{1992}).

\bibitem[{\citenamefont{Flay et~al.}(2016)}]{Flay:2016wie}
\bibinfo{author}{\bibfnamefont{D.}~\bibnamefont{Flay}} \bibnamefont{et~al.}
  (\bibinfo{collaboration}{Jefferson Lab Hall A}), \bibinfo{journal}{Phys.
  Rev.} \textbf{\bibinfo{volume}{D94}}, \bibinfo{pages}{052003}
  (\bibinfo{year}{2016}), \eprint{1603.03612}.

\bibitem[{\citenamefont{Armstrong et~al.}(2019)}]{Armstrong:2018xgk}
\bibinfo{author}{\bibfnamefont{W.}~\bibnamefont{Armstrong}}
  \bibnamefont{et~al.} (\bibinfo{collaboration}{SANE}), \bibinfo{journal}{Phys.
  Rev. Lett.} \textbf{\bibinfo{volume}{122}}, \bibinfo{pages}{022002}
  (\bibinfo{year}{2019}), \eprint{1805.08835}.

\bibitem[{\citenamefont{Koike et~al.}(2008)\citenamefont{Koike, Tanaka, and
  Yoshida}}]{Koike:2008du}
\bibinfo{author}{\bibfnamefont{Y.}~\bibnamefont{Koike}},
  \bibinfo{author}{\bibfnamefont{K.}~\bibnamefont{Tanaka}}, \bibnamefont{and}
  \bibinfo{author}{\bibfnamefont{S.}~\bibnamefont{Yoshida}},
  \bibinfo{journal}{Phys. Lett. B} \textbf{\bibinfo{volume}{668}},
  \bibinfo{pages}{286} (\bibinfo{year}{2008}), \eprint{0805.2289}.

\bibitem[{\citenamefont{Koike et~al.}(2016)\citenamefont{Koike, Pitonyak, and
  Yoshida}}]{Koike:2016ura}
\bibinfo{author}{\bibfnamefont{Y.}~\bibnamefont{Koike}},
  \bibinfo{author}{\bibfnamefont{D.}~\bibnamefont{Pitonyak}}, \bibnamefont{and}
  \bibinfo{author}{\bibfnamefont{S.}~\bibnamefont{Yoshida}},
  \bibinfo{journal}{Phys. Lett. B} \textbf{\bibinfo{volume}{759}},
  \bibinfo{pages}{75} (\bibinfo{year}{2016}), \eprint{1603.07908}.

\bibitem[{\citenamefont{Jakob et~al.}(1997)\citenamefont{Jakob, Mulders, and
  Rodrigues}}]{Jakob:1997wg}
\bibinfo{author}{\bibfnamefont{R.}~\bibnamefont{Jakob}},
  \bibinfo{author}{\bibfnamefont{P.~J.} \bibnamefont{Mulders}},
  \bibnamefont{and}
  \bibinfo{author}{\bibfnamefont{J.}~\bibnamefont{Rodrigues}},
  \bibinfo{journal}{Nucl. Phys. A} \textbf{\bibinfo{volume}{626}},
  \bibinfo{pages}{937} (\bibinfo{year}{1997}), \eprint{hep-ph/9704335}.

\bibitem[{\citenamefont{Bastami et~al.}(2021)\citenamefont{Bastami, Efremov,
  Schweitzer, Teryaev, and Zavada}}]{Bastami:2020rxn}
\bibinfo{author}{\bibfnamefont{S.}~\bibnamefont{Bastami}},
  \bibinfo{author}{\bibfnamefont{A.~V.} \bibnamefont{Efremov}},
  \bibinfo{author}{\bibfnamefont{P.}~\bibnamefont{Schweitzer}},
  \bibinfo{author}{\bibfnamefont{O.~V.} \bibnamefont{Teryaev}},
  \bibnamefont{and} \bibinfo{author}{\bibfnamefont{P.}~\bibnamefont{Zavada}},
  \bibinfo{journal}{Phys. Rev. D} \textbf{\bibinfo{volume}{103}},
  \bibinfo{pages}{014024} (\bibinfo{year}{2021}), \eprint{2011.06203}.

\bibitem[{\citenamefont{Bhattacharya
  et~al.}(2020{\natexlab{a}})\citenamefont{Bhattacharya, Cichy, Constantinou,
  Metz, Scapellato, and Steffens}}]{Bhattacharya:2020cen}
\bibinfo{author}{\bibfnamefont{S.}~\bibnamefont{Bhattacharya}},
  \bibinfo{author}{\bibfnamefont{K.}~\bibnamefont{Cichy}},
  \bibinfo{author}{\bibfnamefont{M.}~\bibnamefont{Constantinou}},
  \bibinfo{author}{\bibfnamefont{A.}~\bibnamefont{Metz}},
  \bibinfo{author}{\bibfnamefont{A.}~\bibnamefont{Scapellato}},
  \bibnamefont{and} \bibinfo{author}{\bibfnamefont{F.}~\bibnamefont{Steffens}},
  \bibinfo{journal}{Phys. Rev. D} \textbf{\bibinfo{volume}{102}},
  \bibinfo{pages}{111501} (\bibinfo{year}{2020}{\natexlab{a}}),
  \eprint{2004.04130}.

\bibitem[{\citenamefont{Ji}(2013)}]{Ji:2013dva}
\bibinfo{author}{\bibfnamefont{X.}~\bibnamefont{Ji}}, \bibinfo{journal}{Phys.
  Rev. Lett.} \textbf{\bibinfo{volume}{110}}, \bibinfo{pages}{262002}
  (\bibinfo{year}{2013}), \eprint{1305.1539}.

\bibitem[{\citenamefont{Ji}(2014)}]{Ji:2014gla}
\bibinfo{author}{\bibfnamefont{X.}~\bibnamefont{Ji}}, \bibinfo{journal}{Sci.
  China Phys. Mech. Astron.} \textbf{\bibinfo{volume}{57}},
  \bibinfo{pages}{1407} (\bibinfo{year}{2014}), \eprint{1404.6680}.

\bibitem[{\citenamefont{Braun and Mueller}(2008)}]{Braun:2007wv}
\bibinfo{author}{\bibfnamefont{V.}~\bibnamefont{Braun}} \bibnamefont{and}
  \bibinfo{author}{\bibfnamefont{D.}~\bibnamefont{Mueller}},
  \bibinfo{journal}{Eur. Phys. J.} \textbf{\bibinfo{volume}{C55}},
  \bibinfo{pages}{349} (\bibinfo{year}{2008}), \eprint{0709.1348}.

\bibitem[{\citenamefont{Radyushkin}(2017)}]{Radyushkin:2017cyf}
\bibinfo{author}{\bibfnamefont{A.~V.} \bibnamefont{Radyushkin}},
  \bibinfo{journal}{Phys. Rev.} \textbf{\bibinfo{volume}{D96}},
  \bibinfo{pages}{034025} (\bibinfo{year}{2017}), \eprint{1705.01488}.

\bibitem[{\citenamefont{Ma and Qiu}(2018{\natexlab{a}})}]{Ma:2017pxb}
\bibinfo{author}{\bibfnamefont{Y.-Q.} \bibnamefont{Ma}} \bibnamefont{and}
  \bibinfo{author}{\bibfnamefont{J.-W.} \bibnamefont{Qiu}},
  \bibinfo{journal}{Phys. Rev. Lett.} \textbf{\bibinfo{volume}{120}},
  \bibinfo{pages}{022003} (\bibinfo{year}{2018}{\natexlab{a}}),
  \eprint{1709.03018}.

\bibitem[{\citenamefont{Lin et~al.}(2015)\citenamefont{Lin, Chen, Cohen, and
  Ji}}]{Lin:2014zya}
\bibinfo{author}{\bibfnamefont{H.-W.} \bibnamefont{Lin}},
  \bibinfo{author}{\bibfnamefont{J.-W.} \bibnamefont{Chen}},
  \bibinfo{author}{\bibfnamefont{S.~D.} \bibnamefont{Cohen}}, \bibnamefont{and}
  \bibinfo{author}{\bibfnamefont{X.}~\bibnamefont{Ji}}, \bibinfo{journal}{Phys.
  Rev.} \textbf{\bibinfo{volume}{D91}}, \bibinfo{pages}{054510}
  (\bibinfo{year}{2015}), \eprint{1402.1462}.

\bibitem[{\citenamefont{Alexandrou et~al.}(2015)\citenamefont{Alexandrou,
  Cichy, Drach, Garcia-Ramos, Hadjiyiannakou, Jansen, Steffens, and
  Wiese}}]{Alexandrou:2015rja}
\bibinfo{author}{\bibfnamefont{C.}~\bibnamefont{Alexandrou}},
  \bibinfo{author}{\bibfnamefont{K.}~\bibnamefont{Cichy}},
  \bibinfo{author}{\bibfnamefont{V.}~\bibnamefont{Drach}},
  \bibinfo{author}{\bibfnamefont{E.}~\bibnamefont{Garcia-Ramos}},
  \bibinfo{author}{\bibfnamefont{K.}~\bibnamefont{Hadjiyiannakou}},
  \bibinfo{author}{\bibfnamefont{K.}~\bibnamefont{Jansen}},
  \bibinfo{author}{\bibfnamefont{F.}~\bibnamefont{Steffens}}, \bibnamefont{and}
  \bibinfo{author}{\bibfnamefont{C.}~\bibnamefont{Wiese}},
  \bibinfo{journal}{Phys. Rev.} \textbf{\bibinfo{volume}{D92}},
  \bibinfo{pages}{014502} (\bibinfo{year}{2015}), \eprint{1504.07455}.

\bibitem[{\citenamefont{Chen et~al.}(2016)\citenamefont{Chen, Cohen, Ji, Lin,
  and Zhang}}]{Chen:2016utp}
\bibinfo{author}{\bibfnamefont{J.-W.} \bibnamefont{Chen}},
  \bibinfo{author}{\bibfnamefont{S.~D.} \bibnamefont{Cohen}},
  \bibinfo{author}{\bibfnamefont{X.}~\bibnamefont{Ji}},
  \bibinfo{author}{\bibfnamefont{H.-W.} \bibnamefont{Lin}}, \bibnamefont{and}
  \bibinfo{author}{\bibfnamefont{J.-H.} \bibnamefont{Zhang}},
  \bibinfo{journal}{Nucl. Phys.} \textbf{\bibinfo{volume}{B911}},
  \bibinfo{pages}{246} (\bibinfo{year}{2016}), \eprint{1603.06664}.

\bibitem[{\citenamefont{Alexandrou
  et~al.}(2017{\natexlab{a}})\citenamefont{Alexandrou, Cichy, Constantinou,
  Hadjiyiannakou, Jansen, Steffens, and Wiese}}]{Alexandrou:2016jqi}
\bibinfo{author}{\bibfnamefont{C.}~\bibnamefont{Alexandrou}},
  \bibinfo{author}{\bibfnamefont{K.}~\bibnamefont{Cichy}},
  \bibinfo{author}{\bibfnamefont{M.}~\bibnamefont{Constantinou}},
  \bibinfo{author}{\bibfnamefont{K.}~\bibnamefont{Hadjiyiannakou}},
  \bibinfo{author}{\bibfnamefont{K.}~\bibnamefont{Jansen}},
  \bibinfo{author}{\bibfnamefont{F.}~\bibnamefont{Steffens}}, \bibnamefont{and}
  \bibinfo{author}{\bibfnamefont{C.}~\bibnamefont{Wiese}},
  \bibinfo{journal}{Phys. Rev.} \textbf{\bibinfo{volume}{D96}},
  \bibinfo{pages}{014513} (\bibinfo{year}{2017}{\natexlab{a}}),
  \eprint{1610.03689}.

\bibitem[{\citenamefont{Chambers et~al.}(2017)\citenamefont{Chambers, Horsley,
  Nakamura, Perlt, Rakow, Schierholz, Schiller, Somfleth, Young, and
  Zanotti}}]{Chambers:2017dov}
\bibinfo{author}{\bibfnamefont{A.~J.} \bibnamefont{Chambers}},
  \bibinfo{author}{\bibfnamefont{R.}~\bibnamefont{Horsley}},
  \bibinfo{author}{\bibfnamefont{Y.}~\bibnamefont{Nakamura}},
  \bibinfo{author}{\bibfnamefont{H.}~\bibnamefont{Perlt}},
  \bibinfo{author}{\bibfnamefont{P.~E.~L.} \bibnamefont{Rakow}},
  \bibinfo{author}{\bibfnamefont{G.}~\bibnamefont{Schierholz}},
  \bibinfo{author}{\bibfnamefont{A.}~\bibnamefont{Schiller}},
  \bibinfo{author}{\bibfnamefont{K.}~\bibnamefont{Somfleth}},
  \bibinfo{author}{\bibfnamefont{R.~D.} \bibnamefont{Young}}, \bibnamefont{and}
  \bibinfo{author}{\bibfnamefont{J.~M.} \bibnamefont{Zanotti}},
  \bibinfo{journal}{Phys. Rev. Lett.} \textbf{\bibinfo{volume}{118}},
  \bibinfo{pages}{242001} (\bibinfo{year}{2017}), \eprint{1703.01153}.

\bibitem[{\citenamefont{Alexandrou
  et~al.}(2017{\natexlab{b}})\citenamefont{Alexandrou, Cichy, Constantinou,
  Hadjiyiannakou, Jansen, Panagopoulos, and Steffens}}]{Alexandrou:2017huk}
\bibinfo{author}{\bibfnamefont{C.}~\bibnamefont{Alexandrou}},
  \bibinfo{author}{\bibfnamefont{K.}~\bibnamefont{Cichy}},
  \bibinfo{author}{\bibfnamefont{M.}~\bibnamefont{Constantinou}},
  \bibinfo{author}{\bibfnamefont{K.}~\bibnamefont{Hadjiyiannakou}},
  \bibinfo{author}{\bibfnamefont{K.}~\bibnamefont{Jansen}},
  \bibinfo{author}{\bibfnamefont{H.}~\bibnamefont{Panagopoulos}},
  \bibnamefont{and} \bibinfo{author}{\bibfnamefont{F.}~\bibnamefont{Steffens}},
  \bibinfo{journal}{Nucl. Phys.} \textbf{\bibinfo{volume}{B923}},
  \bibinfo{pages}{394} (\bibinfo{year}{2017}{\natexlab{b}}),
  \eprint{1706.00265}.

\bibitem[{\citenamefont{Orginos et~al.}(2017)\citenamefont{Orginos, Radyushkin,
  Karpie, and Zafeiropoulos}}]{Orginos:2017kos}
\bibinfo{author}{\bibfnamefont{K.}~\bibnamefont{Orginos}},
  \bibinfo{author}{\bibfnamefont{A.}~\bibnamefont{Radyushkin}},
  \bibinfo{author}{\bibfnamefont{J.}~\bibnamefont{Karpie}}, \bibnamefont{and}
  \bibinfo{author}{\bibfnamefont{S.}~\bibnamefont{Zafeiropoulos}},
  \bibinfo{journal}{Phys. Rev.} \textbf{\bibinfo{volume}{D96}},
  \bibinfo{pages}{094503} (\bibinfo{year}{2017}), \eprint{1706.05373}.

\bibitem[{\citenamefont{Ishikawa et~al.}(2017)\citenamefont{Ishikawa, Ma, Qiu,
  and Yoshida}}]{Ishikawa:2017faj}
\bibinfo{author}{\bibfnamefont{T.}~\bibnamefont{Ishikawa}},
  \bibinfo{author}{\bibfnamefont{Y.-Q.} \bibnamefont{Ma}},
  \bibinfo{author}{\bibfnamefont{J.-W.} \bibnamefont{Qiu}}, \bibnamefont{and}
  \bibinfo{author}{\bibfnamefont{S.}~\bibnamefont{Yoshida}},
  \bibinfo{journal}{Phys. Rev.} \textbf{\bibinfo{volume}{D96}},
  \bibinfo{pages}{094019} (\bibinfo{year}{2017}), \eprint{1707.03107}.

\bibitem[{\citenamefont{Ji et~al.}(2018)\citenamefont{Ji, Zhang, and
  Zhao}}]{Ji:2017oey}
\bibinfo{author}{\bibfnamefont{X.}~\bibnamefont{Ji}},
  \bibinfo{author}{\bibfnamefont{J.-H.} \bibnamefont{Zhang}}, \bibnamefont{and}
  \bibinfo{author}{\bibfnamefont{Y.}~\bibnamefont{Zhao}},
  \bibinfo{journal}{Phys. Rev. Lett.} \textbf{\bibinfo{volume}{120}},
  \bibinfo{pages}{112001} (\bibinfo{year}{2018}), \eprint{1706.08962}.

\bibitem[{\citenamefont{Radyushkin}(2018)}]{Radyushkin:2018cvn}
\bibinfo{author}{\bibfnamefont{A.}~\bibnamefont{Radyushkin}},
  \bibinfo{journal}{Phys. Rev.} \textbf{\bibinfo{volume}{D98}},
  \bibinfo{pages}{014019} (\bibinfo{year}{2018}), \eprint{1801.02427}.

\bibitem[{\citenamefont{Alexandrou
  et~al.}(2018{\natexlab{a}})\citenamefont{Alexandrou, Cichy, Constantinou,
  Jansen, Scapellato, and Steffens}}]{Alexandrou:2018pbm}
\bibinfo{author}{\bibfnamefont{C.}~\bibnamefont{Alexandrou}},
  \bibinfo{author}{\bibfnamefont{K.}~\bibnamefont{Cichy}},
  \bibinfo{author}{\bibfnamefont{M.}~\bibnamefont{Constantinou}},
  \bibinfo{author}{\bibfnamefont{K.}~\bibnamefont{Jansen}},
  \bibinfo{author}{\bibfnamefont{A.}~\bibnamefont{Scapellato}},
  \bibnamefont{and} \bibinfo{author}{\bibfnamefont{F.}~\bibnamefont{Steffens}},
  \bibinfo{journal}{Phys. Rev. Lett.} \textbf{\bibinfo{volume}{121}},
  \bibinfo{pages}{112001} (\bibinfo{year}{2018}{\natexlab{a}}),
  \eprint{1803.02685}.

\bibitem[{\citenamefont{Zhang et~al.}(2019{\natexlab{a}})\citenamefont{Zhang,
  Chen, Jin, Lin, Schäfer, and Zhao}}]{Chen:2018fwa}
\bibinfo{author}{\bibfnamefont{J.-H.} \bibnamefont{Zhang}},
  \bibinfo{author}{\bibfnamefont{J.-W.} \bibnamefont{Chen}},
  \bibinfo{author}{\bibfnamefont{L.}~\bibnamefont{Jin}},
  \bibinfo{author}{\bibfnamefont{H.-W.} \bibnamefont{Lin}},
  \bibinfo{author}{\bibfnamefont{A.}~\bibnamefont{Schäfer}}, \bibnamefont{and}
  \bibinfo{author}{\bibfnamefont{Y.}~\bibnamefont{Zhao}},
  \bibinfo{journal}{Phys. Rev.} \textbf{\bibinfo{volume}{D100}},
  \bibinfo{pages}{034505} (\bibinfo{year}{2019}{\natexlab{a}}),
  \eprint{1804.01483}.

\bibitem[{\citenamefont{Alexandrou
  et~al.}(2018{\natexlab{b}})\citenamefont{Alexandrou, Cichy, Constantinou,
  Jansen, Scapellato, and Steffens}}]{Alexandrou:2018eet}
\bibinfo{author}{\bibfnamefont{C.}~\bibnamefont{Alexandrou}},
  \bibinfo{author}{\bibfnamefont{K.}~\bibnamefont{Cichy}},
  \bibinfo{author}{\bibfnamefont{M.}~\bibnamefont{Constantinou}},
  \bibinfo{author}{\bibfnamefont{K.}~\bibnamefont{Jansen}},
  \bibinfo{author}{\bibfnamefont{A.}~\bibnamefont{Scapellato}},
  \bibnamefont{and} \bibinfo{author}{\bibfnamefont{F.}~\bibnamefont{Steffens}},
  \bibinfo{journal}{Phys. Rev.} \textbf{\bibinfo{volume}{D98}},
  \bibinfo{pages}{091503} (\bibinfo{year}{2018}{\natexlab{b}}),
  \eprint{1807.00232}.

\bibitem[{\citenamefont{Liu et~al.}(2020)}]{Liu:2018uuj}
\bibinfo{author}{\bibfnamefont{Y.-S.} \bibnamefont{Liu}} \bibnamefont{et~al.}
  (\bibinfo{collaboration}{Lattice Parton}), \bibinfo{journal}{Phys. Rev. D}
  \textbf{\bibinfo{volume}{101}}, \bibinfo{pages}{034020}
  (\bibinfo{year}{2020}), \eprint{1807.06566}.

\bibitem[{\citenamefont{Karpie et~al.}(2018)\citenamefont{Karpie, Orginos, and
  Zafeiropoulos}}]{Karpie:2018zaz}
\bibinfo{author}{\bibfnamefont{J.}~\bibnamefont{Karpie}},
  \bibinfo{author}{\bibfnamefont{K.}~\bibnamefont{Orginos}}, \bibnamefont{and}
  \bibinfo{author}{\bibfnamefont{S.}~\bibnamefont{Zafeiropoulos}},
  \bibinfo{journal}{JHEP} \textbf{\bibinfo{volume}{11}}, \bibinfo{pages}{178}
  (\bibinfo{year}{2018}), \eprint{1807.10933}.

\bibitem[{\citenamefont{Zhang et~al.}(2019{\natexlab{b}})\citenamefont{Zhang,
  Ji, Schäfer, Wang, and Zhao}}]{Zhang:2018diq}
\bibinfo{author}{\bibfnamefont{J.-H.} \bibnamefont{Zhang}},
  \bibinfo{author}{\bibfnamefont{X.}~\bibnamefont{Ji}},
  \bibinfo{author}{\bibfnamefont{A.}~\bibnamefont{Schäfer}},
  \bibinfo{author}{\bibfnamefont{W.}~\bibnamefont{Wang}}, \bibnamefont{and}
  \bibinfo{author}{\bibfnamefont{S.}~\bibnamefont{Zhao}},
  \bibinfo{journal}{Phys. Rev. Lett.} \textbf{\bibinfo{volume}{122}},
  \bibinfo{pages}{142001} (\bibinfo{year}{2019}{\natexlab{b}}),
  \eprint{1808.10824}.

\bibitem[{\citenamefont{Bhattacharya et~al.}(2019)\citenamefont{Bhattacharya,
  Cocuzza, and Metz}}]{Bhattacharya:2018zxi}
\bibinfo{author}{\bibfnamefont{S.}~\bibnamefont{Bhattacharya}},
  \bibinfo{author}{\bibfnamefont{C.}~\bibnamefont{Cocuzza}}, \bibnamefont{and}
  \bibinfo{author}{\bibfnamefont{A.}~\bibnamefont{Metz}},
  \bibinfo{journal}{Phys. Lett. B} \textbf{\bibinfo{volume}{788}},
  \bibinfo{pages}{453} (\bibinfo{year}{2019}), \eprint{1808.01437}.

\bibitem[{\citenamefont{Li et~al.}(2019)\citenamefont{Li, Ma, and
  Qiu}}]{Li:2018tpe}
\bibinfo{author}{\bibfnamefont{Z.-Y.} \bibnamefont{Li}},
  \bibinfo{author}{\bibfnamefont{Y.-Q.} \bibnamefont{Ma}}, \bibnamefont{and}
  \bibinfo{author}{\bibfnamefont{J.-W.} \bibnamefont{Qiu}},
  \bibinfo{journal}{Phys. Rev. Lett.} \textbf{\bibinfo{volume}{122}},
  \bibinfo{pages}{062002} (\bibinfo{year}{2019}), \eprint{1809.01836}.

\bibitem[{\citenamefont{Sufian et~al.}(2019)\citenamefont{Sufian, Karpie,
  Egerer, Orginos, Qiu, and Richards}}]{Sufian:2019bol}
\bibinfo{author}{\bibfnamefont{R.~S.} \bibnamefont{Sufian}},
  \bibinfo{author}{\bibfnamefont{J.}~\bibnamefont{Karpie}},
  \bibinfo{author}{\bibfnamefont{C.}~\bibnamefont{Egerer}},
  \bibinfo{author}{\bibfnamefont{K.}~\bibnamefont{Orginos}},
  \bibinfo{author}{\bibfnamefont{J.-W.} \bibnamefont{Qiu}}, \bibnamefont{and}
  \bibinfo{author}{\bibfnamefont{D.~G.} \bibnamefont{Richards}},
  \bibinfo{journal}{Phys. Rev.} \textbf{\bibinfo{volume}{D99}},
  \bibinfo{pages}{074507} (\bibinfo{year}{2019}), \eprint{1901.03921}.

\bibitem[{\citenamefont{Karpie et~al.}(2019)\citenamefont{Karpie, Orginos,
  Rothkopf, and Zafeiropoulos}}]{Karpie:2019eiq}
\bibinfo{author}{\bibfnamefont{J.}~\bibnamefont{Karpie}},
  \bibinfo{author}{\bibfnamefont{K.}~\bibnamefont{Orginos}},
  \bibinfo{author}{\bibfnamefont{A.}~\bibnamefont{Rothkopf}}, \bibnamefont{and}
  \bibinfo{author}{\bibfnamefont{S.}~\bibnamefont{Zafeiropoulos}},
  \bibinfo{journal}{JHEP} \textbf{\bibinfo{volume}{04}}, \bibinfo{pages}{057}
  (\bibinfo{year}{2019}), \eprint{1901.05408}.

\bibitem[{\citenamefont{Alexandrou et~al.}(2019)\citenamefont{Alexandrou,
  Cichy, Constantinou, Hadjiyiannakou, Jansen, Scapellato, and
  Steffens}}]{Alexandrou:2019lfo}
\bibinfo{author}{\bibfnamefont{C.}~\bibnamefont{Alexandrou}},
  \bibinfo{author}{\bibfnamefont{K.}~\bibnamefont{Cichy}},
  \bibinfo{author}{\bibfnamefont{M.}~\bibnamefont{Constantinou}},
  \bibinfo{author}{\bibfnamefont{K.}~\bibnamefont{Hadjiyiannakou}},
  \bibinfo{author}{\bibfnamefont{K.}~\bibnamefont{Jansen}},
  \bibinfo{author}{\bibfnamefont{A.}~\bibnamefont{Scapellato}},
  \bibnamefont{and} \bibinfo{author}{\bibfnamefont{F.}~\bibnamefont{Steffens}},
  \bibinfo{journal}{Phys. Rev.} \textbf{\bibinfo{volume}{D99}},
  \bibinfo{pages}{114504} (\bibinfo{year}{2019}), \eprint{1902.00587}.

\bibitem[{\citenamefont{Izubuchi et~al.}(2019)\citenamefont{Izubuchi, Jin,
  Kallidonis, Karthik, Mukherjee, Petreczky, Shugert, and
  Syritsyn}}]{Izubuchi:2019lyk}
\bibinfo{author}{\bibfnamefont{T.}~\bibnamefont{Izubuchi}},
  \bibinfo{author}{\bibfnamefont{L.}~\bibnamefont{Jin}},
  \bibinfo{author}{\bibfnamefont{C.}~\bibnamefont{Kallidonis}},
  \bibinfo{author}{\bibfnamefont{N.}~\bibnamefont{Karthik}},
  \bibinfo{author}{\bibfnamefont{S.}~\bibnamefont{Mukherjee}},
  \bibinfo{author}{\bibfnamefont{P.}~\bibnamefont{Petreczky}},
  \bibinfo{author}{\bibfnamefont{C.}~\bibnamefont{Shugert}}, \bibnamefont{and}
  \bibinfo{author}{\bibfnamefont{S.}~\bibnamefont{Syritsyn}},
  \bibinfo{journal}{Phys. Rev.} \textbf{\bibinfo{volume}{D100}},
  \bibinfo{pages}{034516} (\bibinfo{year}{2019}), \eprint{1905.06349}.

\bibitem[{\citenamefont{Cichy et~al.}(2019)\citenamefont{Cichy, Del~Debbio, and
  Giani}}]{Cichy:2019ebf}
\bibinfo{author}{\bibfnamefont{K.}~\bibnamefont{Cichy}},
  \bibinfo{author}{\bibfnamefont{L.}~\bibnamefont{Del~Debbio}},
  \bibnamefont{and} \bibinfo{author}{\bibfnamefont{T.}~\bibnamefont{Giani}},
  \bibinfo{journal}{JHEP} \textbf{\bibinfo{volume}{10}}, \bibinfo{pages}{137}
  (\bibinfo{year}{2019}), \eprint{1907.06037}.

\bibitem[{\citenamefont{Joó et~al.}(2019{\natexlab{a}})\citenamefont{Joó,
  Karpie, Orginos, Radyushkin, Richards, and Zafeiropoulos}}]{Joo:2019jct}
\bibinfo{author}{\bibfnamefont{B.}~\bibnamefont{Joó}},
  \bibinfo{author}{\bibfnamefont{J.}~\bibnamefont{Karpie}},
  \bibinfo{author}{\bibfnamefont{K.}~\bibnamefont{Orginos}},
  \bibinfo{author}{\bibfnamefont{A.}~\bibnamefont{Radyushkin}},
  \bibinfo{author}{\bibfnamefont{D.}~\bibnamefont{Richards}}, \bibnamefont{and}
  \bibinfo{author}{\bibfnamefont{S.}~\bibnamefont{Zafeiropoulos}},
  \bibinfo{journal}{JHEP} \textbf{\bibinfo{volume}{12}}, \bibinfo{pages}{081}
  (\bibinfo{year}{2019}{\natexlab{a}}), \eprint{1908.09771}.

\bibitem[{\citenamefont{Radyushkin}(2019)}]{Radyushkin:2019owq}
\bibinfo{author}{\bibfnamefont{A.~V.} \bibnamefont{Radyushkin}},
  \bibinfo{journal}{Phys. Rev. D} \textbf{\bibinfo{volume}{100}},
  \bibinfo{pages}{116011} (\bibinfo{year}{2019}), \eprint{1909.08474}.

\bibitem[{\citenamefont{Joó et~al.}(2019{\natexlab{b}})\citenamefont{Joó,
  Karpie, Orginos, Radyushkin, Richards, Sufian, and
  Zafeiropoulos}}]{Joo:2019bzr}
\bibinfo{author}{\bibfnamefont{B.}~\bibnamefont{Joó}},
  \bibinfo{author}{\bibfnamefont{J.}~\bibnamefont{Karpie}},
  \bibinfo{author}{\bibfnamefont{K.}~\bibnamefont{Orginos}},
  \bibinfo{author}{\bibfnamefont{A.~V.} \bibnamefont{Radyushkin}},
  \bibinfo{author}{\bibfnamefont{D.~G.} \bibnamefont{Richards}},
  \bibinfo{author}{\bibfnamefont{R.~S.} \bibnamefont{Sufian}},
  \bibnamefont{and}
  \bibinfo{author}{\bibfnamefont{S.}~\bibnamefont{Zafeiropoulos}},
  \bibinfo{journal}{Phys. Rev.} \textbf{\bibinfo{volume}{D100}},
  \bibinfo{pages}{114512} (\bibinfo{year}{2019}{\natexlab{b}}),
  \eprint{1909.08517}.

\bibitem[{\citenamefont{Chai et~al.}(2020)}]{Chai:2020nxw}
\bibinfo{author}{\bibfnamefont{Y.}~\bibnamefont{Chai}} \bibnamefont{et~al.},
  \bibinfo{journal}{Phys. Rev. D} \textbf{\bibinfo{volume}{102}},
  \bibinfo{pages}{014508} (\bibinfo{year}{2020}), \eprint{2002.12044}.

\bibitem[{\citenamefont{Ji}(2020)}]{Ji:2020baz}
\bibinfo{author}{\bibfnamefont{X.}~\bibnamefont{Ji}}, \bibinfo{journal}{Nucl.
  Phys.} \textbf{\bibinfo{volume}{B}}, \bibinfo{pages}{115181}
  (\bibinfo{year}{2020}), \eprint{2003.04478}.

\bibitem[{\citenamefont{Braun et~al.}(2020)\citenamefont{Braun, Chetyrkin, and
  Kniehl}}]{Braun:2020ymy}
\bibinfo{author}{\bibfnamefont{V.}~\bibnamefont{Braun}},
  \bibinfo{author}{\bibfnamefont{K.}~\bibnamefont{Chetyrkin}},
  \bibnamefont{and} \bibinfo{author}{\bibfnamefont{B.}~\bibnamefont{Kniehl}},
  \bibinfo{journal}{JHEP} \textbf{\bibinfo{volume}{07}}, \bibinfo{pages}{161}
  (\bibinfo{year}{2020}), \eprint{2004.01043}.

\bibitem[{\citenamefont{Bhat et~al.}(2021)\citenamefont{Bhat, Cichy,
  Constantinou, and Scapellato}}]{Bhat:2020ktg}
\bibinfo{author}{\bibfnamefont{M.}~\bibnamefont{Bhat}},
  \bibinfo{author}{\bibfnamefont{K.}~\bibnamefont{Cichy}},
  \bibinfo{author}{\bibfnamefont{M.}~\bibnamefont{Constantinou}},
  \bibnamefont{and}
  \bibinfo{author}{\bibfnamefont{A.}~\bibnamefont{Scapellato}},
  \bibinfo{journal}{Phys. Rev. D} \textbf{\bibinfo{volume}{103}},
  \bibinfo{pages}{034510} (\bibinfo{year}{2021}), \eprint{2005.02102}.

\bibitem[{\citenamefont{Alexandrou
  et~al.}(2020{\natexlab{a}})\citenamefont{Alexandrou, Cichy, Constantinou,
  Hadjiyiannakou, Jansen, Scapellato, and Steffens}}]{Alexandrou:2020zbe}
\bibinfo{author}{\bibfnamefont{C.}~\bibnamefont{Alexandrou}},
  \bibinfo{author}{\bibfnamefont{K.}~\bibnamefont{Cichy}},
  \bibinfo{author}{\bibfnamefont{M.}~\bibnamefont{Constantinou}},
  \bibinfo{author}{\bibfnamefont{K.}~\bibnamefont{Hadjiyiannakou}},
  \bibinfo{author}{\bibfnamefont{K.}~\bibnamefont{Jansen}},
  \bibinfo{author}{\bibfnamefont{A.}~\bibnamefont{Scapellato}},
  \bibnamefont{and} \bibinfo{author}{\bibfnamefont{F.}~\bibnamefont{Steffens}},
  \bibinfo{journal}{Phys. Rev. Lett.} \textbf{\bibinfo{volume}{125}},
  \bibinfo{pages}{262001} (\bibinfo{year}{2020}{\natexlab{a}}),
  \eprint{2008.10573}.

\bibitem[{\citenamefont{Alexandrou
  et~al.}(2021{\natexlab{a}})\citenamefont{Alexandrou, Constantinou,
  Hadjiyiannakou, Jansen, and Manigrasso}}]{Alexandrou:2020uyt}
\bibinfo{author}{\bibfnamefont{C.}~\bibnamefont{Alexandrou}},
  \bibinfo{author}{\bibfnamefont{M.}~\bibnamefont{Constantinou}},
  \bibinfo{author}{\bibfnamefont{K.}~\bibnamefont{Hadjiyiannakou}},
  \bibinfo{author}{\bibfnamefont{K.}~\bibnamefont{Jansen}}, \bibnamefont{and}
  \bibinfo{author}{\bibfnamefont{F.}~\bibnamefont{Manigrasso}},
  \bibinfo{journal}{Phys. Rev. Lett.} \textbf{\bibinfo{volume}{126}},
  \bibinfo{pages}{102003} (\bibinfo{year}{2021}{\natexlab{a}}),
  \eprint{2009.13061}.

\bibitem[{\citenamefont{Bringewatt et~al.}(2021)\citenamefont{Bringewatt, Sato,
  Melnitchouk, Qiu, Steffens, and Constantinou}}]{Bringewatt:2020ixn}
\bibinfo{author}{\bibfnamefont{J.}~\bibnamefont{Bringewatt}},
  \bibinfo{author}{\bibfnamefont{N.}~\bibnamefont{Sato}},
  \bibinfo{author}{\bibfnamefont{W.}~\bibnamefont{Melnitchouk}},
  \bibinfo{author}{\bibfnamefont{J.-W.} \bibnamefont{Qiu}},
  \bibinfo{author}{\bibfnamefont{F.}~\bibnamefont{Steffens}}, \bibnamefont{and}
  \bibinfo{author}{\bibfnamefont{M.}~\bibnamefont{Constantinou}},
  \bibinfo{journal}{Phys. Rev. D} \textbf{\bibinfo{volume}{103}},
  \bibinfo{pages}{016003} (\bibinfo{year}{2021}), \eprint{2010.00548}.

\bibitem[{\citenamefont{Liu and Chen}(2020{\natexlab{a}})}]{Liu:2020rqi}
\bibinfo{author}{\bibfnamefont{W.-Y.} \bibnamefont{Liu}} \bibnamefont{and}
  \bibinfo{author}{\bibfnamefont{J.-W.} \bibnamefont{Chen}}
  (\bibinfo{year}{2020}{\natexlab{a}}), \eprint{2010.06623}.

\bibitem[{\citenamefont{Del~Debbio et~al.}(2021)\citenamefont{Del~Debbio,
  Giani, Karpie, Orginos, Radyushkin, and Zafeiropoulos}}]{DelDebbio:2020rgv}
\bibinfo{author}{\bibfnamefont{L.}~\bibnamefont{Del~Debbio}},
  \bibinfo{author}{\bibfnamefont{T.}~\bibnamefont{Giani}},
  \bibinfo{author}{\bibfnamefont{J.}~\bibnamefont{Karpie}},
  \bibinfo{author}{\bibfnamefont{K.}~\bibnamefont{Orginos}},
  \bibinfo{author}{\bibfnamefont{A.}~\bibnamefont{Radyushkin}},
  \bibnamefont{and}
  \bibinfo{author}{\bibfnamefont{S.}~\bibnamefont{Zafeiropoulos}},
  \bibinfo{journal}{JHEP} \textbf{\bibinfo{volume}{02}}, \bibinfo{pages}{138}
  (\bibinfo{year}{2021}), \eprint{2010.03996}.

\bibitem[{\citenamefont{Alexandrou
  et~al.}(2021{\natexlab{b}})\citenamefont{Alexandrou, Cichy, Constantinou,
  Green, Hadjiyiannakou, Jansen, Manigrasso, Scapellato, and
  Steffens}}]{Alexandrou:2020qtt}
\bibinfo{author}{\bibfnamefont{C.}~\bibnamefont{Alexandrou}},
  \bibinfo{author}{\bibfnamefont{K.}~\bibnamefont{Cichy}},
  \bibinfo{author}{\bibfnamefont{M.}~\bibnamefont{Constantinou}},
  \bibinfo{author}{\bibfnamefont{J.~R.} \bibnamefont{Green}},
  \bibinfo{author}{\bibfnamefont{K.}~\bibnamefont{Hadjiyiannakou}},
  \bibinfo{author}{\bibfnamefont{K.}~\bibnamefont{Jansen}},
  \bibinfo{author}{\bibfnamefont{F.}~\bibnamefont{Manigrasso}},
  \bibinfo{author}{\bibfnamefont{A.}~\bibnamefont{Scapellato}},
  \bibnamefont{and} \bibinfo{author}{\bibfnamefont{F.}~\bibnamefont{Steffens}},
  \bibinfo{journal}{Phys. Rev. D} \textbf{\bibinfo{volume}{103}},
  \bibinfo{pages}{094512} (\bibinfo{year}{2021}{\natexlab{b}}),
  \eprint{2011.00964}.

\bibitem[{\citenamefont{Liu and Chen}(2020{\natexlab{b}})}]{Liu:2020krc}
\bibinfo{author}{\bibfnamefont{W.-Y.} \bibnamefont{Liu}} \bibnamefont{and}
  \bibinfo{author}{\bibfnamefont{J.-W.} \bibnamefont{Chen}}
  (\bibinfo{year}{2020}{\natexlab{b}}), \eprint{2011.13536}.

\bibitem[{\citenamefont{Huo et~al.}(2021)}]{Huo:2021rpe}
\bibinfo{author}{\bibfnamefont{Y.-K.} \bibnamefont{Huo}} \bibnamefont{et~al.}
  (\bibinfo{collaboration}{Lattice Parton}) (\bibinfo{year}{2021}),
  \eprint{2103.02965}.

\bibitem[{\citenamefont{Detmold et~al.}(2021)\citenamefont{Detmold, Grebe,
  Kanamori, Lin, Perry, and Zhao}}]{Detmold:2021uru}
\bibinfo{author}{\bibfnamefont{W.}~\bibnamefont{Detmold}},
  \bibinfo{author}{\bibfnamefont{A.~V.} \bibnamefont{Grebe}},
  \bibinfo{author}{\bibfnamefont{I.}~\bibnamefont{Kanamori}},
  \bibinfo{author}{\bibfnamefont{C.~J.~D.} \bibnamefont{Lin}},
  \bibinfo{author}{\bibfnamefont{R.~J.} \bibnamefont{Perry}}, \bibnamefont{and}
  \bibinfo{author}{\bibfnamefont{Y.}~\bibnamefont{Zhao}}
  (\bibinfo{year}{2021}), \eprint{2103.09529}.

\bibitem[{\citenamefont{Karpie et~al.}(2021)\citenamefont{Karpie, Orginos,
  Radyushkin, and Zafeiropoulos}}]{Karpie:2021pap}
\bibinfo{author}{\bibfnamefont{J.}~\bibnamefont{Karpie}},
  \bibinfo{author}{\bibfnamefont{K.}~\bibnamefont{Orginos}},
  \bibinfo{author}{\bibfnamefont{A.}~\bibnamefont{Radyushkin}},
  \bibnamefont{and}
  \bibinfo{author}{\bibfnamefont{S.}~\bibnamefont{Zafeiropoulos}}
  (\bibinfo{year}{2021}), \eprint{2105.13313}.

\bibitem[{\citenamefont{Alexandrou
  et~al.}(2021{\natexlab{c}})\citenamefont{Alexandrou, Constantinou,
  Hadjiyiannakou, Jansen, and Manigrasso}}]{Alexandrou:2021oih}
\bibinfo{author}{\bibfnamefont{C.}~\bibnamefont{Alexandrou}},
  \bibinfo{author}{\bibfnamefont{M.}~\bibnamefont{Constantinou}},
  \bibinfo{author}{\bibfnamefont{K.}~\bibnamefont{Hadjiyiannakou}},
  \bibinfo{author}{\bibfnamefont{K.}~\bibnamefont{Jansen}}, \bibnamefont{and}
  \bibinfo{author}{\bibfnamefont{F.}~\bibnamefont{Manigrasso}}
  (\bibinfo{year}{2021}{\natexlab{c}}), \eprint{2106.16065}.

\bibitem[{\citenamefont{Cichy and Constantinou}(2019)}]{Cichy:2018mum}
\bibinfo{author}{\bibfnamefont{K.}~\bibnamefont{Cichy}} \bibnamefont{and}
  \bibinfo{author}{\bibfnamefont{M.}~\bibnamefont{Constantinou}},
  \bibinfo{journal}{Adv. High Energy Phys.} \textbf{\bibinfo{volume}{2019}},
  \bibinfo{pages}{3036904} (\bibinfo{year}{2019}), \eprint{1811.07248}.

\bibitem[{\citenamefont{Ji et~al.}(2020)\citenamefont{Ji, Liu, Liu, Zhang, and
  Zhao}}]{Ji:2020ect}
\bibinfo{author}{\bibfnamefont{X.}~\bibnamefont{Ji}},
  \bibinfo{author}{\bibfnamefont{Y.-S.} \bibnamefont{Liu}},
  \bibinfo{author}{\bibfnamefont{Y.}~\bibnamefont{Liu}},
  \bibinfo{author}{\bibfnamefont{J.-H.} \bibnamefont{Zhang}}, \bibnamefont{and}
  \bibinfo{author}{\bibfnamefont{Y.}~\bibnamefont{Zhao}}
  (\bibinfo{year}{2020}), \eprint{2004.03543}.

\bibitem[{\citenamefont{Constantinou}(2021)}]{Constantinou:2020pek}
\bibinfo{author}{\bibfnamefont{M.}~\bibnamefont{Constantinou}},
  \bibinfo{journal}{Eur. Phys. J. A} \textbf{\bibinfo{volume}{57}},
  \bibinfo{pages}{77} (\bibinfo{year}{2021}), \eprint{2010.02445}.

\bibitem[{\citenamefont{Brice{\~n}o et~al.}(2017)\citenamefont{Brice{\~n}o,
  Hansen, and Monahan}}]{Briceno:2017cpo}
\bibinfo{author}{\bibfnamefont{R.~A.} \bibnamefont{Brice{\~n}o}},
  \bibinfo{author}{\bibfnamefont{M.~T.} \bibnamefont{Hansen}},
  \bibnamefont{and} \bibinfo{author}{\bibfnamefont{C.~J.}
  \bibnamefont{Monahan}}, \bibinfo{journal}{Phys. Rev.}
  \textbf{\bibinfo{volume}{D96}}, \bibinfo{pages}{014502}
  (\bibinfo{year}{2017}), \eprint{1703.06072}.

\bibitem[{\citenamefont{Xiong et~al.}(2014)\citenamefont{Xiong, Ji, Zhang, and
  Zhao}}]{Xiong:2013bka}
\bibinfo{author}{\bibfnamefont{X.}~\bibnamefont{Xiong}},
  \bibinfo{author}{\bibfnamefont{X.}~\bibnamefont{Ji}},
  \bibinfo{author}{\bibfnamefont{J.-H.} \bibnamefont{Zhang}}, \bibnamefont{and}
  \bibinfo{author}{\bibfnamefont{Y.}~\bibnamefont{Zhao}},
  \bibinfo{journal}{Phys.Rev.} \textbf{\bibinfo{volume}{D90}},
  \bibinfo{pages}{014051} (\bibinfo{year}{2014}), \eprint{1310.7471}.

\bibitem[{\citenamefont{Ma and Qiu}(2018{\natexlab{b}})}]{Ma:2014jla}
\bibinfo{author}{\bibfnamefont{Y.-Q.} \bibnamefont{Ma}} \bibnamefont{and}
  \bibinfo{author}{\bibfnamefont{J.-W.} \bibnamefont{Qiu}},
  \bibinfo{journal}{Phys. Rev.} \textbf{\bibinfo{volume}{D98}},
  \bibinfo{pages}{074021} (\bibinfo{year}{2018}{\natexlab{b}}),
  \eprint{1404.6860}.

\bibitem[{\citenamefont{Wang et~al.}(2018)\citenamefont{Wang, Zhao, and
  Zhu}}]{Wang:2017qyg}
\bibinfo{author}{\bibfnamefont{W.}~\bibnamefont{Wang}},
  \bibinfo{author}{\bibfnamefont{S.}~\bibnamefont{Zhao}}, \bibnamefont{and}
  \bibinfo{author}{\bibfnamefont{R.}~\bibnamefont{Zhu}}, \bibinfo{journal}{Eur.
  Phys. J.} \textbf{\bibinfo{volume}{C78}}, \bibinfo{pages}{147}
  (\bibinfo{year}{2018}), \eprint{1708.02458}.

\bibitem[{\citenamefont{Stewart and Zhao}(2018)}]{Stewart:2017tvs}
\bibinfo{author}{\bibfnamefont{I.~W.} \bibnamefont{Stewart}} \bibnamefont{and}
  \bibinfo{author}{\bibfnamefont{Y.}~\bibnamefont{Zhao}},
  \bibinfo{journal}{Phys. Rev.} \textbf{\bibinfo{volume}{D97}},
  \bibinfo{pages}{054512} (\bibinfo{year}{2018}), \eprint{1709.04933}.

\bibitem[{\citenamefont{Izubuchi et~al.}(2018)\citenamefont{Izubuchi, Ji, Jin,
  Stewart, and Zhao}}]{Izubuchi:2018srq}
\bibinfo{author}{\bibfnamefont{T.}~\bibnamefont{Izubuchi}},
  \bibinfo{author}{\bibfnamefont{X.}~\bibnamefont{Ji}},
  \bibinfo{author}{\bibfnamefont{L.}~\bibnamefont{Jin}},
  \bibinfo{author}{\bibfnamefont{I.~W.} \bibnamefont{Stewart}},
  \bibnamefont{and} \bibinfo{author}{\bibfnamefont{Y.}~\bibnamefont{Zhao}},
  \bibinfo{journal}{Phys. Rev.} \textbf{\bibinfo{volume}{D98}},
  \bibinfo{pages}{056004} (\bibinfo{year}{2018}), \eprint{1801.03917}.

\bibitem[{\citenamefont{Balitsky et~al.}(2020)\citenamefont{Balitsky, Morris,
  and Radyushkin}}]{Balitsky:2019krf}
\bibinfo{author}{\bibfnamefont{I.}~\bibnamefont{Balitsky}},
  \bibinfo{author}{\bibfnamefont{W.}~\bibnamefont{Morris}}, \bibnamefont{and}
  \bibinfo{author}{\bibfnamefont{A.}~\bibnamefont{Radyushkin}},
  \bibinfo{journal}{Phys. Lett. B} \textbf{\bibinfo{volume}{808}},
  \bibinfo{pages}{135621} (\bibinfo{year}{2020}), \eprint{1910.13963}.

\bibitem[{\citenamefont{Bhattacharya
  et~al.}(2020{\natexlab{b}})\citenamefont{Bhattacharya, Cichy, Constantinou,
  Metz, Scapellato, and Steffens}}]{Bhattacharya:2020xlt}
\bibinfo{author}{\bibfnamefont{S.}~\bibnamefont{Bhattacharya}},
  \bibinfo{author}{\bibfnamefont{K.}~\bibnamefont{Cichy}},
  \bibinfo{author}{\bibfnamefont{M.}~\bibnamefont{Constantinou}},
  \bibinfo{author}{\bibfnamefont{A.}~\bibnamefont{Metz}},
  \bibinfo{author}{\bibfnamefont{A.}~\bibnamefont{Scapellato}},
  \bibnamefont{and} \bibinfo{author}{\bibfnamefont{F.}~\bibnamefont{Steffens}},
  \bibinfo{journal}{Phys. Rev. D} \textbf{\bibinfo{volume}{102}},
  \bibinfo{pages}{034005} (\bibinfo{year}{2020}{\natexlab{b}}),
  \eprint{2005.10939}.

\bibitem[{\citenamefont{Bhattacharya
  et~al.}(2020{\natexlab{c}})\citenamefont{Bhattacharya, Cichy, Constantinou,
  Metz, Scapellato, and Steffens}}]{Bhattacharya:2020jfj}
\bibinfo{author}{\bibfnamefont{S.}~\bibnamefont{Bhattacharya}},
  \bibinfo{author}{\bibfnamefont{K.}~\bibnamefont{Cichy}},
  \bibinfo{author}{\bibfnamefont{M.}~\bibnamefont{Constantinou}},
  \bibinfo{author}{\bibfnamefont{A.}~\bibnamefont{Metz}},
  \bibinfo{author}{\bibfnamefont{A.}~\bibnamefont{Scapellato}},
  \bibnamefont{and} \bibinfo{author}{\bibfnamefont{F.}~\bibnamefont{Steffens}},
  \bibinfo{journal}{Phys. Rev. D} \textbf{\bibinfo{volume}{102}},
  \bibinfo{pages}{114025} (\bibinfo{year}{2020}{\natexlab{c}}),
  \eprint{2006.12347}.

\bibitem[{\citenamefont{Li et~al.}(2021)\citenamefont{Li, Ma, and
  Qiu}}]{Li:2020xml}
\bibinfo{author}{\bibfnamefont{Z.-Y.} \bibnamefont{Li}},
  \bibinfo{author}{\bibfnamefont{Y.-Q.} \bibnamefont{Ma}}, \bibnamefont{and}
  \bibinfo{author}{\bibfnamefont{J.-W.} \bibnamefont{Qiu}},
  \bibinfo{journal}{Phys. Rev. Lett.} \textbf{\bibinfo{volume}{126}},
  \bibinfo{pages}{072001} (\bibinfo{year}{2021}), \eprint{2006.12370}.

\bibitem[{\citenamefont{Chen et~al.}(2021)\citenamefont{Chen, Wang, and
  Zhu}}]{Chen:2020ody}
\bibinfo{author}{\bibfnamefont{L.-B.} \bibnamefont{Chen}},
  \bibinfo{author}{\bibfnamefont{W.}~\bibnamefont{Wang}}, \bibnamefont{and}
  \bibinfo{author}{\bibfnamefont{R.}~\bibnamefont{Zhu}},
  \bibinfo{journal}{Phys. Rev. Lett.} \textbf{\bibinfo{volume}{126}},
  \bibinfo{pages}{072002} (\bibinfo{year}{2021}), \eprint{2006.14825}.

\bibitem[{\citenamefont{Braun et~al.}(2021)\citenamefont{Braun, Ji, and
  Vladimirov}}]{Braun:2021aon}
\bibinfo{author}{\bibfnamefont{V.~M.} \bibnamefont{Braun}},
  \bibinfo{author}{\bibfnamefont{Y.}~\bibnamefont{Ji}}, \bibnamefont{and}
  \bibinfo{author}{\bibfnamefont{A.}~\bibnamefont{Vladimirov}},
  \bibinfo{journal}{JHEP} \textbf{\bibinfo{volume}{05}}, \bibinfo{pages}{086}
  (\bibinfo{year}{2021}), \eprint{2103.12105}.

\bibitem[{\citenamefont{Burkardt}(1995)}]{Burkardt:1995ts}
\bibinfo{author}{\bibfnamefont{M.}~\bibnamefont{Burkardt}},
  \bibinfo{journal}{Phys. Rev. D} \textbf{\bibinfo{volume}{52}},
  \bibinfo{pages}{3841} (\bibinfo{year}{1995}), \eprint{hep-ph/9505226}.

\bibitem[{\citenamefont{Burkardt and Koike}(2002)}]{Burkardt:2001iy}
\bibinfo{author}{\bibfnamefont{M.}~\bibnamefont{Burkardt}} \bibnamefont{and}
  \bibinfo{author}{\bibfnamefont{Y.}~\bibnamefont{Koike}},
  \bibinfo{journal}{Nucl. Phys. B} \textbf{\bibinfo{volume}{632}},
  \bibinfo{pages}{311} (\bibinfo{year}{2002}), \eprint{hep-ph/0111343}.

\bibitem[{\citenamefont{Efremov and Schweitzer}(2003)}]{Efremov:2002qh}
\bibinfo{author}{\bibfnamefont{A.~V.} \bibnamefont{Efremov}} \bibnamefont{and}
  \bibinfo{author}{\bibfnamefont{P.}~\bibnamefont{Schweitzer}},
  \bibinfo{journal}{JHEP} \textbf{\bibinfo{volume}{08}}, \bibinfo{pages}{006}
  (\bibinfo{year}{2003}), \eprint{hep-ph/0212044}.

\bibitem[{\citenamefont{Wakamatsu and Ohnishi}(2003)}]{Wakamatsu:2003uu}
\bibinfo{author}{\bibfnamefont{M.}~\bibnamefont{Wakamatsu}} \bibnamefont{and}
  \bibinfo{author}{\bibfnamefont{Y.}~\bibnamefont{Ohnishi}},
  \bibinfo{journal}{Phys. Rev. D} \textbf{\bibinfo{volume}{67}},
  \bibinfo{pages}{114011} (\bibinfo{year}{2003}), \eprint{hep-ph/0303007}.

\bibitem[{\citenamefont{Pasquini and Rodini}(2019)}]{Pasquini:2018oyz}
\bibinfo{author}{\bibfnamefont{B.}~\bibnamefont{Pasquini}} \bibnamefont{and}
  \bibinfo{author}{\bibfnamefont{S.}~\bibnamefont{Rodini}},
  \bibinfo{journal}{Phys. Lett. B} \textbf{\bibinfo{volume}{788}},
  \bibinfo{pages}{414} (\bibinfo{year}{2019}), \eprint{1806.10932}.

\bibitem[{\citenamefont{Aslan et~al.}(2018)\citenamefont{Aslan, Burkardt,
  Lorc\'e, Metz, and Pasquini}}]{Aslan:2018zzk}
\bibinfo{author}{\bibfnamefont{F.}~\bibnamefont{Aslan}},
  \bibinfo{author}{\bibfnamefont{M.}~\bibnamefont{Burkardt}},
  \bibinfo{author}{\bibfnamefont{C.}~\bibnamefont{Lorc\'e}},
  \bibinfo{author}{\bibfnamefont{A.}~\bibnamefont{Metz}}, \bibnamefont{and}
  \bibinfo{author}{\bibfnamefont{B.}~\bibnamefont{Pasquini}},
  \bibinfo{journal}{Phys. Rev. D} \textbf{\bibinfo{volume}{98}},
  \bibinfo{pages}{014038} (\bibinfo{year}{2018}), \eprint{1802.06243}.

\bibitem[{\citenamefont{Aslan and Burkardt}(2020)}]{Aslan:2018tff}
\bibinfo{author}{\bibfnamefont{F.}~\bibnamefont{Aslan}} \bibnamefont{and}
  \bibinfo{author}{\bibfnamefont{M.}~\bibnamefont{Burkardt}},
  \bibinfo{journal}{Phys. Rev. D} \textbf{\bibinfo{volume}{101}},
  \bibinfo{pages}{016010} (\bibinfo{year}{2020}), \eprint{1811.00938}.

\bibitem[{\citenamefont{Bhattacharya and Metz}(2021)}]{Bhattacharya:2021boh}
\bibinfo{author}{\bibfnamefont{S.}~\bibnamefont{Bhattacharya}}
  \bibnamefont{and} \bibinfo{author}{\bibfnamefont{A.}~\bibnamefont{Metz}}
  (\bibinfo{year}{2021}), \eprint{2105.07282}.

\bibitem[{\citenamefont{Wandzura and Wilczek}(1977)}]{Wandzura:1977qf}
\bibinfo{author}{\bibfnamefont{S.}~\bibnamefont{Wandzura}} \bibnamefont{and}
  \bibinfo{author}{\bibfnamefont{F.}~\bibnamefont{Wilczek}},
  \bibinfo{journal}{Phys. Lett.} \textbf{\bibinfo{volume}{72B}},
  \bibinfo{pages}{195} (\bibinfo{year}{1977}).

\bibitem[{\citenamefont{Ralston and Soper}(1979)}]{Ralston:1979ys}
\bibinfo{author}{\bibfnamefont{J.~P.} \bibnamefont{Ralston}} \bibnamefont{and}
  \bibinfo{author}{\bibfnamefont{D.~E.} \bibnamefont{Soper}},
  \bibinfo{journal}{Nucl. Phys. B} \textbf{\bibinfo{volume}{152}},
  \bibinfo{pages}{109} (\bibinfo{year}{1979}).

\bibitem[{\citenamefont{Alexandrou
  et~al.}(2021{\natexlab{d}})}]{Alexandrou:2021gqw}
\bibinfo{author}{\bibfnamefont{C.}~\bibnamefont{Alexandrou}}
  \bibnamefont{et~al.} (\bibinfo{year}{2021}{\natexlab{d}}),
  \eprint{2104.13408}.

\bibitem[{\citenamefont{Iwasaki}(1985)}]{Iwasaki:1985we}
\bibinfo{author}{\bibfnamefont{Y.}~\bibnamefont{Iwasaki}},
  \bibinfo{journal}{Nucl. Phys. B} \textbf{\bibinfo{volume}{258}},
  \bibinfo{pages}{141} (\bibinfo{year}{1985}).

\bibitem[{\citenamefont{Sheikholeslami and
  Wohlert}(1985)}]{Sheikholeslami:1985ij}
\bibinfo{author}{\bibfnamefont{B.}~\bibnamefont{Sheikholeslami}}
  \bibnamefont{and} \bibinfo{author}{\bibfnamefont{R.}~\bibnamefont{Wohlert}},
  \bibinfo{journal}{Nucl. Phys.} \textbf{\bibinfo{volume}{B259}},
  \bibinfo{pages}{572} (\bibinfo{year}{1985}).

\bibitem[{\citenamefont{Bali et~al.}(2016)\citenamefont{Bali, Lang, Musch, and
  Schäfer}}]{Bali:2016lva}
\bibinfo{author}{\bibfnamefont{G.~S.} \bibnamefont{Bali}},
  \bibinfo{author}{\bibfnamefont{B.}~\bibnamefont{Lang}},
  \bibinfo{author}{\bibfnamefont{B.~U.} \bibnamefont{Musch}}, \bibnamefont{and}
  \bibinfo{author}{\bibfnamefont{A.}~\bibnamefont{Schäfer}},
  \bibinfo{journal}{Phys. Rev.} \textbf{\bibinfo{volume}{D93}},
  \bibinfo{pages}{094515} (\bibinfo{year}{2016}), \eprint{1602.05525}.

\bibitem[{\citenamefont{Albanese et~al.}(1987)}]{Albanese:1987ds}
\bibinfo{author}{\bibfnamefont{M.}~\bibnamefont{Albanese}} \bibnamefont{et~al.}
  (\bibinfo{collaboration}{APE}), \bibinfo{journal}{Phys. Lett.}
  \textbf{\bibinfo{volume}{B192}}, \bibinfo{pages}{163} (\bibinfo{year}{1987}).

\bibitem[{\citenamefont{Morningstar and Peardon}(2004)}]{Morningstar:2003gk}
\bibinfo{author}{\bibfnamefont{C.}~\bibnamefont{Morningstar}} \bibnamefont{and}
  \bibinfo{author}{\bibfnamefont{M.~J.} \bibnamefont{Peardon}},
  \bibinfo{journal}{Phys. Rev.} \textbf{\bibinfo{volume}{D69}},
  \bibinfo{pages}{054501} (\bibinfo{year}{2004}), \eprint{hep-lat/0311018}.

\bibitem[{\citenamefont{Alexandrou
  et~al.}(2017{\natexlab{c}})\citenamefont{Alexandrou, Constantinou,
  Hadjiyiannakou, Jansen, Panagopoulos, and Wiese}}]{Alexandrou:2016ekb}
\bibinfo{author}{\bibfnamefont{C.}~\bibnamefont{Alexandrou}},
  \bibinfo{author}{\bibfnamefont{M.}~\bibnamefont{Constantinou}},
  \bibinfo{author}{\bibfnamefont{K.}~\bibnamefont{Hadjiyiannakou}},
  \bibinfo{author}{\bibfnamefont{K.}~\bibnamefont{Jansen}},
  \bibinfo{author}{\bibfnamefont{H.}~\bibnamefont{Panagopoulos}},
  \bibnamefont{and} \bibinfo{author}{\bibfnamefont{C.}~\bibnamefont{Wiese}},
  \bibinfo{journal}{Phys. Rev.} \textbf{\bibinfo{volume}{D96}},
  \bibinfo{pages}{054503} (\bibinfo{year}{2017}{\natexlab{c}}),
  \eprint{1611.06901}.

\bibitem[{\citenamefont{Alexandrou
  et~al.}(2020{\natexlab{b}})\citenamefont{Alexandrou, Bacchio, Constantinou,
  Finkenrath, Hadjiyiannakou, Jansen, Koutsou, Panagopoulos, and
  Spanoudes}}]{Alexandrou:2020sml}
\bibinfo{author}{\bibfnamefont{C.}~\bibnamefont{Alexandrou}},
  \bibinfo{author}{\bibfnamefont{S.}~\bibnamefont{Bacchio}},
  \bibinfo{author}{\bibfnamefont{M.}~\bibnamefont{Constantinou}},
  \bibinfo{author}{\bibfnamefont{J.}~\bibnamefont{Finkenrath}},
  \bibinfo{author}{\bibfnamefont{K.}~\bibnamefont{Hadjiyiannakou}},
  \bibinfo{author}{\bibfnamefont{K.}~\bibnamefont{Jansen}},
  \bibinfo{author}{\bibfnamefont{G.}~\bibnamefont{Koutsou}},
  \bibinfo{author}{\bibfnamefont{H.}~\bibnamefont{Panagopoulos}},
  \bibnamefont{and} \bibinfo{author}{\bibfnamefont{G.}~\bibnamefont{Spanoudes}}
  (\bibinfo{year}{2020}{\natexlab{b}}), \eprint{2003.08486}.

\bibitem[{\citenamefont{Constantinou and
  Panagopoulos}(2017)}]{Constantinou:2017sej}
\bibinfo{author}{\bibfnamefont{M.}~\bibnamefont{Constantinou}}
  \bibnamefont{and}
  \bibinfo{author}{\bibfnamefont{H.}~\bibnamefont{Panagopoulos}},
  \bibinfo{journal}{Phys. Rev.} \textbf{\bibinfo{volume}{D96}},
  \bibinfo{pages}{054506} (\bibinfo{year}{2017}), \eprint{1705.11193}.

\bibitem[{\citenamefont{Martinelli et~al.}(1995)\citenamefont{Martinelli,
  Pittori, Sachrajda, Testa, and Vladikas}}]{Martinelli:1994ty}
\bibinfo{author}{\bibfnamefont{G.}~\bibnamefont{Martinelli}},
  \bibinfo{author}{\bibfnamefont{C.}~\bibnamefont{Pittori}},
  \bibinfo{author}{\bibfnamefont{C.~T.} \bibnamefont{Sachrajda}},
  \bibinfo{author}{\bibfnamefont{M.}~\bibnamefont{Testa}}, \bibnamefont{and}
  \bibinfo{author}{\bibfnamefont{A.}~\bibnamefont{Vladikas}},
  \bibinfo{journal}{Nucl. Phys.} \textbf{\bibinfo{volume}{B445}},
  \bibinfo{pages}{81} (\bibinfo{year}{1995}), \eprint{hep-lat/9411010}.

\bibitem[{\citenamefont{Alexandrou
  et~al.}(2017{\natexlab{d}})\citenamefont{Alexandrou, Constantinou, and
  Panagopoulos}}]{Alexandrou:2015sea}
\bibinfo{author}{\bibfnamefont{C.}~\bibnamefont{Alexandrou}},
  \bibinfo{author}{\bibfnamefont{M.}~\bibnamefont{Constantinou}},
  \bibnamefont{and}
  \bibinfo{author}{\bibfnamefont{H.}~\bibnamefont{Panagopoulos}}
  (\bibinfo{collaboration}{ETM}), \bibinfo{journal}{Phys. Rev.}
  \textbf{\bibinfo{volume}{D95}}, \bibinfo{pages}{034505}
  (\bibinfo{year}{2017}{\natexlab{d}}), \eprint{1509.00213}.

\bibitem[{\citenamefont{Constantinou et~al.}(2010)}]{Constantinou:2010gr}
\bibinfo{author}{\bibfnamefont{M.}~\bibnamefont{Constantinou}}
  \bibnamefont{et~al.} (\bibinfo{collaboration}{ETM}), \bibinfo{journal}{JHEP}
  \textbf{\bibinfo{volume}{08}}, \bibinfo{pages}{068} (\bibinfo{year}{2010}),
  \eprint{1004.1115}.

\bibitem[{\citenamefont{Backus and Gilbert}(1968)}]{BackusGilbert}
\bibinfo{author}{\bibfnamefont{G.}~\bibnamefont{Backus}} \bibnamefont{and}
  \bibinfo{author}{\bibfnamefont{F.}~\bibnamefont{Gilbert}},
  \bibinfo{journal}{Geophysical Journal International}
  \textbf{\bibinfo{volume}{16}}, \bibinfo{pages}{169} (\bibinfo{year}{1968}),
  \urlprefix\url{http://dx.doi.org/10.1111/j.1365-246X.1968.tb00216.x}.

\bibitem[{\citenamefont{Tikhonov}(1963)}]{Tikhonov:1963}
\bibinfo{author}{\bibfnamefont{A.~N.} \bibnamefont{Tikhonov}},
  \bibinfo{journal}{Soviet Math. Dokl.} \textbf{\bibinfo{volume}{4}},
  \bibinfo{pages}{1035} (\bibinfo{year}{1963}).

\bibitem[{\citenamefont{Bhattacharya
  et~al.}(2020{\natexlab{d}})\citenamefont{Bhattacharya, Cocuzza, and
  Metz}}]{Bhattacharya:2019cme}
\bibinfo{author}{\bibfnamefont{S.}~\bibnamefont{Bhattacharya}},
  \bibinfo{author}{\bibfnamefont{C.}~\bibnamefont{Cocuzza}}, \bibnamefont{and}
  \bibinfo{author}{\bibfnamefont{A.}~\bibnamefont{Metz}},
  \bibinfo{journal}{Phys. Rev. D} \textbf{\bibinfo{volume}{102}},
  \bibinfo{pages}{054021} (\bibinfo{year}{2020}{\natexlab{d}}),
  \eprint{1903.05721}.

\bibitem[{\citenamefont{Burkhardt and Cottingham}(1970)}]{Burkhardt:1970ti}
\bibinfo{author}{\bibfnamefont{H.}~\bibnamefont{Burkhardt}} \bibnamefont{and}
  \bibinfo{author}{\bibfnamefont{W.~N.} \bibnamefont{Cottingham}},
  \bibinfo{journal}{Annals Phys.} \textbf{\bibinfo{volume}{56}},
  \bibinfo{pages}{453} (\bibinfo{year}{1970}).

\bibitem[{\citenamefont{Tangerman and Mulders}(1994)}]{Tangerman:1994bb}
\bibinfo{author}{\bibfnamefont{R.~D.} \bibnamefont{Tangerman}}
  \bibnamefont{and} \bibinfo{author}{\bibfnamefont{P.~J.}
  \bibnamefont{Mulders}} (\bibinfo{year}{1994}), \eprint{hep-ph/9408305}.

\bibitem[{\citenamefont{Constantinou et~al.}(2020)}]{PDFLat2020}
\bibinfo{author}{\bibfnamefont{M.}~\bibnamefont{Constantinou}}
  \bibnamefont{et~al.} (\bibinfo{year}{2020}), \eprint{2006.08636}.

\bibitem[{\citenamefont{Dressler and Polyakov}(2000)}]{Dressler:1999hc}
\bibinfo{author}{\bibfnamefont{B.}~\bibnamefont{Dressler}} \bibnamefont{and}
  \bibinfo{author}{\bibfnamefont{M.~V.} \bibnamefont{Polyakov}},
  \bibinfo{journal}{Phys. Rev. D} \textbf{\bibinfo{volume}{61}},
  \bibinfo{pages}{097501} (\bibinfo{year}{2000}), \eprint{hep-ph/9912376}.

\bibitem[{\citenamefont{Mulders and Rodrigues}(2001)}]{Mulders:2000sh}
\bibinfo{author}{\bibfnamefont{P.~J.} \bibnamefont{Mulders}} \bibnamefont{and}
  \bibinfo{author}{\bibfnamefont{J.}~\bibnamefont{Rodrigues}},
  \bibinfo{journal}{Phys. Rev. D} \textbf{\bibinfo{volume}{63}},
  \bibinfo{pages}{094021} (\bibinfo{year}{2001}), \eprint{hep-ph/0009343}.

\bibitem[{\citenamefont{Frommer et~al.}(2014)\citenamefont{Frommer, Kahl,
  Krieg, Leder, and Rottmann}}]{Frommer:2013fsa}
\bibinfo{author}{\bibfnamefont{A.}~\bibnamefont{Frommer}},
  \bibinfo{author}{\bibfnamefont{K.}~\bibnamefont{Kahl}},
  \bibinfo{author}{\bibfnamefont{S.}~\bibnamefont{Krieg}},
  \bibinfo{author}{\bibfnamefont{B.}~\bibnamefont{Leder}}, \bibnamefont{and}
  \bibinfo{author}{\bibfnamefont{M.}~\bibnamefont{Rottmann}},
  \bibinfo{journal}{SIAM J. Sci. Comput.} \textbf{\bibinfo{volume}{36}},
  \bibinfo{pages}{A1581} (\bibinfo{year}{2014}), \eprint{1303.1377}.

\bibitem[{\citenamefont{Alexandrou et~al.}(2016)\citenamefont{Alexandrou,
  Bacchio, Finkenrath, Frommer, Kahl, and Rottmann}}]{Alexandrou:2016izb}
\bibinfo{author}{\bibfnamefont{C.}~\bibnamefont{Alexandrou}},
  \bibinfo{author}{\bibfnamefont{S.}~\bibnamefont{Bacchio}},
  \bibinfo{author}{\bibfnamefont{J.}~\bibnamefont{Finkenrath}},
  \bibinfo{author}{\bibfnamefont{A.}~\bibnamefont{Frommer}},
  \bibinfo{author}{\bibfnamefont{K.}~\bibnamefont{Kahl}}, \bibnamefont{and}
  \bibinfo{author}{\bibfnamefont{M.}~\bibnamefont{Rottmann}},
  \bibinfo{journal}{Phys. Rev. D} \textbf{\bibinfo{volume}{94}},
  \bibinfo{pages}{114509} (\bibinfo{year}{2016}), \eprint{1610.02370}.

\end{thebibliography}
\end{document}